\documentclass[preprint,11pt]{elsarticle}

\makeatletter
\def\ps@pprintTitle{%
  \let\@oddhead\@empty
  \let\@evenhead\@empty
  \let\@oddfoot\@empty
  \let\@evenfoot\@empty
}
\makeatother

\usepackage{geometry}
\usepackage{xcolor}
\usepackage{siunitx}
\usepackage[export]{adjustbox}
\usepackage{tabularx}
\usepackage{graphicx}
\usepackage{paralist}
\usepackage{url}
\usepackage{wrapfig}
\usepackage{amsmath}
\usepackage{amssymb}
\usepackage{mathrsfs}
\usepackage{placeins}
\usepackage{caption}
\usepackage{subcaption}
\usepackage{ifthen,tikz}
\usepackage{pgfplots}
\pgfplotsset{compat=newest}
\usepackage{comment}
\usepackage{enumerate}
\usepackage{algorithm}
\usepackage{algpseudocode}
\usepackage{enumitem}
\usepackage{booktabs}
\usepackage{nicefrac}
\usepackage{hyperref} 
\usepackage{physics}

\usetikzlibrary{shapes,positioning,intersections,quotes, calc, angles}
\usepackage{color}

\newcommand{\commentout}[1]{}

\usepackage{soul} 

\newcommand{\AddA}[1]{{{#1}}}

\newcommand{\sfrac}[2]{\mathchoice
  {\kern0em\raise.5ex\hbox{\the\scriptfont0 #1}\kern-.15em/
   \kern-.15em\lower.25ex\hbox{\the\scriptfont0 #2}}
  {\kern0em\raise.5ex\hbox{\the\scriptfont0 #1}\kern-.15em/
   \kern-.15em\lower.25ex\hbox{\the\scriptfont0 #2}}
  {\kern0em\raise.5ex\hbox{\the\scriptscriptfont0 #1}\kern-.2em/
   \kern-.15em\lower.25ex\hbox{\the\scriptscriptfont0 #2}}
  {#1\!/#2}}

\newcommand{\vol}{V}

\DeclareMathSymbol{\shortminus}{\mathbin}{AMSa}{"39}
\def\dt     {\Delta t}
\def\dx     {\Delta x}
\def\dy     {\Delta y}
\def\dz     {\Delta z}

\def\EB     {\text{EB}}

\def\U       {{U}}
\def\ub      {{\bf u}}

\def\iph    {i+\frac{1}{2}}

\def\imh    {i-\frac{1}{2}}
\def\iph    {i+\frac{1}{2}}

\def\iph    {{i+\frac{1}{2},j,k}}
\def\ijph   {{i,j+\frac{1}{2},k}}
\def\ijkph  {{i,j,k+\frac{1}{2}}}
\def\imh    {{i-\frac{1}{2},j,k}}
\def\ijmh   {{i,j-\frac{1}{2},k}}
\def\ijkmh  {{i,j,k-\frac{1}{2}}}
\def\ivec   {{i,j,k}}

\def\Fb      {\mathbf{F}}

\def\Sb      {\mathbf{S}}
\def\Fb      {\mathbf{F}}

\def\Ib      {\mathbf{I}}

\def\Ub      {\mathbf{U}}
\def\nb      {\mathbf{n}}
\def\xb      {\mathbf{x}}

\def\taub      {\boldsymbol{\tau}}

\journal{Journal of Computational Physics}

\begin{document}

\begin{frontmatter}

\title{An Embedded Boundary Scheme for Three-Dimensional Flow Over Terrain on a  Staggered Mesh}

\author[inst1]{Soonpil Kang\corref{cor1}}
\ead{kang18@llnl.gov}
\cortext[cor1]{Corresponding author}
\author[inst2]{Ann S. Almgren}
\author[inst2]{Mahesh Natarajan}
\author[inst2]{Aaron M. Lattanzi}
\author[inst1]{Jeffrey D. Mirocha}
\author[inst1]{Katie Lundquist}
\author[inst3]{Jordan Musser}
\author[inst2]{Weiqun Zhang}

\affiliation[inst1]{organization={Lawrence Livermore National Laboratory},
            addressline={7000 East Ave}, 
            city={Livermore},
            postcode={94550}, 
            state={CA},
            country={USA}}

\affiliation[inst2]{organization={Lawrence Berkeley National Laboratory},
            addressline={1 Cyclotron Rd}, 
            city={Berkeley},
            postcode={94720}, 
            state={CA},
            country={USA}}

\affiliation[inst3]{organization={Independent Researcher},
            country={USA}}
%

\begin{abstract}
This paper describes an embedded boundary (EB) approach for simulating three-dimensional fluid flow on a staggered mesh where the velocity components are defined on cell faces and the thermodynamic state is defined on cell centers. Most EB approaches assume that all components of the solution, including the velocity, are co-located. To compute solution quantities on faces as well as cell centers, we construct and store multiple instances of the geometric information, one for the quantities stored at cell centers and one for each velocity component. In addition, we extend the weighted state redistribution (WSRD) scheme to staggered meshes to address the small-cell instability issue. This new approach is implemented in the Energy Research and Forecasting (ERF) model that provides performance portability and adaptive mesh refinement. We validate the new EB method by comparing EB simulations to those computed using terrain-following coordinates.
\end{abstract}


\end{frontmatter}
\newpage

\newpage

\newpage
\section{Introduction}

Numerical solution of partial differential equations in complex geometric domains arises in a wide range of important scientific and engineering applications. Traditional boundary-fitted methods require the computational grid to match the complex domain boundaries, whereas embedded boundary (EB) approaches, also known as \textit{cut cell} methods, treat the complex domain boundaries as surfaces that intersect a regular Cartesian grid. This geometric flexibility reduces the problem of mesh generation to the problem of constructing cut cells that capture these intersections, and offers the computational convenience of using a Cartesian grid on most of the domain. 

In the context of finite volume methods, EB methods modify the control volume stencils near the boundary to directly impose Dirichlet or Neumann boundary conditions on surfaces that are not aligned with the computational grid~\cite{johansen1998cartesian, graves2013cartesian}.
A numerical difficulty arising in this process is that cells cut by the interface may have arbitrarily small volumes, which could lead to a requirement for arbitrarily small timestep size when using standard explicit temporal discretizations. This is referred to as the \textit{small cell problem}, and developing stable methodology to avoid severe time step restrictions is one of the major challenges in EB methods.

A number of different approaches have been proposed to address the small cell problem. A novel dimensionally split method was developed to compute stabilized cut cell fluxes for hyperbolic system \cite{gokhaleNikosKlein:2018}. The $h$-box method \cite{mjb-hel-rjl:hbox, berger_simplified_2012} shows nice stability properties, but has not been extended to three dimensions. Cell merging, in which small cells are merged with neighboring larger cells, is an intuitive approach that has been successfully used to mitigate the small cell problem, and has seen significant recent progress~\cite{MURALIDHARAN2016, saye2017implicit, saye2017part2, GULIZZI2022}.

An alternative to cell merging, called flux redistribution (FRD), has been successfully used for simulations of compressible \cite{pember1995adaptive, colella2006cartesian, hu2006conservative, klein2009well, graves2013cartesian, SCHNEIDERS2013786} and incompressible flows \cite{almgren1997cartesian, trebotich:2015}. In FRD schemes, a conservative but potentially unstable update for the solution is first computed using the difference of area-weighted fluxes divided by the cell volume. The solution in cut cells is then stabilized or regularized by adjusting the update to the solution, and the remainder of the update is redistributed to neighboring cells to maintain global conservation. This approach is attractive due to its simple implementation as a postprocessing step; however, it suffers a loss of accuracy in cut cells and does not preserve linearity.

Recently, a new approach called state redistribution (SRD) was proposed for finite volume methods on two-dimensional grids as an alternative to FRD \cite{berger2021state}. Subsequently, the method was extended to three dimensions using a weighted version of the algorithm, weighted state redistribution (WSRD)~\cite{giuliani2022weighted}. WSRD is a minimally invasive stabilization technique that is linearity preserving, conservative, and straightforward to implement for hyperbolic conservation laws. Inspired by SRD, WSRD postprocesses the numerical solution by accurately redistributing the solution states in a conservative manner.

The goal of this paper is to extend the EB methodology using WSRD for co-located cell-centered grids to a staggered grid. We employ the Arakawa C-grid, in which scalar quantities are located at cell centers and velocity components are located on cell faces. The accuracy and stability of the algorithm will be demonstrated for flows governed by both the compressible Navier-Stokes equations and the anelastic equations.

The proposed method is implemented in the Energy Research and Forecasting (ERF) model~\cite{lattanzi2025erf}. ERF, built on the AMReX framework~\cite{AMReX:IJHPCA}, simulates atmospheric flows and offers adaptive mesh refinement (AMR) capability and performance portability across modern high-performance computing architectures, including graphics processing units (GPUs).



The paper is organized as follows. In Section~\ref{sec:goveqs}, we introduce the governing equations in ERF. Section~\ref{sec:eb} describes the EB representation within the finite volume method. Section~\ref{sec:numeth} presents the numerical method for spatial discretization with embedded boundaries. Section~\ref{sec:WSRD} reviews the WSRD algorithm and describes its incorporation into the time integration procedure. In Section~\ref{sec:num}, we validate our numerical scheme using canonical two- and three-dimensional example problems. Finally, conclusions are summarized in Section~\ref{sec:conclusion}.

\section{Governing equations} \label{sec:goveqs}

In this study, we consider two different sets of equations that govern the dry dynamics of atmospheric flows. The first set is the compressible Navier-Stokes equations, and the second set is the anelastic equations. In both formulations, for numerical convenience the pressure, density, and potential temperature are split into reference (also called background) and perturbation parts, $p(\xb,t)=p_0(z) + p'(\xb,t)$, $\rho(\xb,t)=\rho_0(z) + \rho'(\xb,t)$, and $\theta(\xb,t)=\theta_0(z) + \theta'(\xb,t)$. We assume that the reference state satisfies the equation of state (see below) and is hydrostatically balanced and dependent only on the vertical coordinate $z$, i.e., $\pdv{p_0}{z}=-\rho_0 g$ where $g$ is the gravitational acceleration.


\subsection{Compressible flow model} \label{subsec:goveqs_comp}

The fully compressible equations are written as follows:
\begin{align}
  \pdv{\rho}{t} &= - \nabla \cdot (\rho \ub), \label{eq:cons} \\
  \pdv{(\rho \ub)}{t} &= - \nabla \cdot (\rho \ub \otimes\ub) - \nabla p' +\mathbf{B} +\nabla \cdot \taub, \label{eq:mom-comp}\\
  \pdv{(\rho \theta)}{t} &= - \nabla \cdot (\rho \ub \theta) + \nabla \cdot ( \rho \alpha_{\theta}\ \nabla \theta), \label{eq:theta-comp}
\end{align}
where the conserved variables are $(\rho,\; \rho\ub,\; \rho\theta)$, $\ub = ( u,\; v ,\; w )$ is the fluid velocity vector, $\theta$ is the potential temperature, and $\alpha_\theta$ is the thermal diffusivity. The buoyancy term $\mathbf{B}$ is defined as $\mathbf{B}=-\rho' \bf g$ with the gravitational acceleration vector $\bf g$. Equations~\eqref{eq:cons}--\eqref{eq:theta-comp} represent local mass conservation, momentum balance, and thermodynamics, respectively. The viscous stress tensor $\taub$ is defined as
\begin{equation}
  \taub = \lambda(\nabla\cdot\ub)\Ib + \mu\qty(\nabla\ub+\nabla\ub^T)
\end{equation}
where $\mu$ and $\lambda=\tfrac{2}{3}\mu$ are the viscosity coefficients, and $\Ib$ is the identity tensor. The pressure is obtained from the diagnostic equation of state:
\begin{align}
p = P_{00} \qty( \frac{ R \; \rho \; \theta }{P_{00}} )^{\gamma},
\end{align}
where $\gamma = c_p / c_v$ is the specific heat ratio for dry air, $R$ is the gas constant for dry air, and $P_{00}=1\times 10^{5}$ Pa is the reference pressure.


\subsection{Anelastic flow model} \label{subsec:Anelastic} 

The anelastic system replaces the equation of state with a divergence constraint and assumes that the density $\rho$ is close to the constant reference value $\rho_0$. As a result, we solve 
\begin{align}
  \pdv{(\rho_0 \ub)}{t} &= - \nabla \cdot (\rho_0 \ub \otimes\ub) - \nabla p'  + {\mathbf B}+ \nabla \cdot \taub, \label{eq:mom-anelastic}\\
  \pdv{(\rho_0 \theta)}{t} &= - \nabla \cdot (\rho_0 \ub \theta) + \nabla \cdot ( \rho_0 \alpha_{\theta}\ \nabla \theta), \label{eq:theta-anelastic} \\
    \nabla \cdot (\rho_0 \ub) &= 0,
\end{align}
%
%
 where the prognostic variables are $(\rho_0\ub,\; \rho_0\theta)$, the buoyancy term is defined as $\mathbf{B} = \rho_0 \theta' / \theta_0 \; \mathbf{g}$, and the other variables are as defined for the compressible formulation. In this model, the perturbation pressure, $p'$, acts as a Lagrange multiplier enforcing the divergence constraint and is computed by solving a Poisson equation that arises in the temporal discretization applying the divergence constraint.
 


\section{Embedded boundary scheme for finite volume methods}\label{sec:eb}

In this section, we present the finite volume approach for spatial discretization of cut cells and describe the construction of the geometric data to represent them on a staggered grid. In the EB approach, the control volume of each cut cell is modified to match the physical (fluid) portion of the cell, and the governing equations are integrated over this modified control volume. As a result, the embedded boundary is explicitly represented in the discrete formulation. The following sections discuss this process in detail.


\subsection{Finite volume discretization for cut cells}\label{sec:FV}

\begin{figure}[h!]
    \begin{center}
        \includegraphics[width=0.5\textwidth]{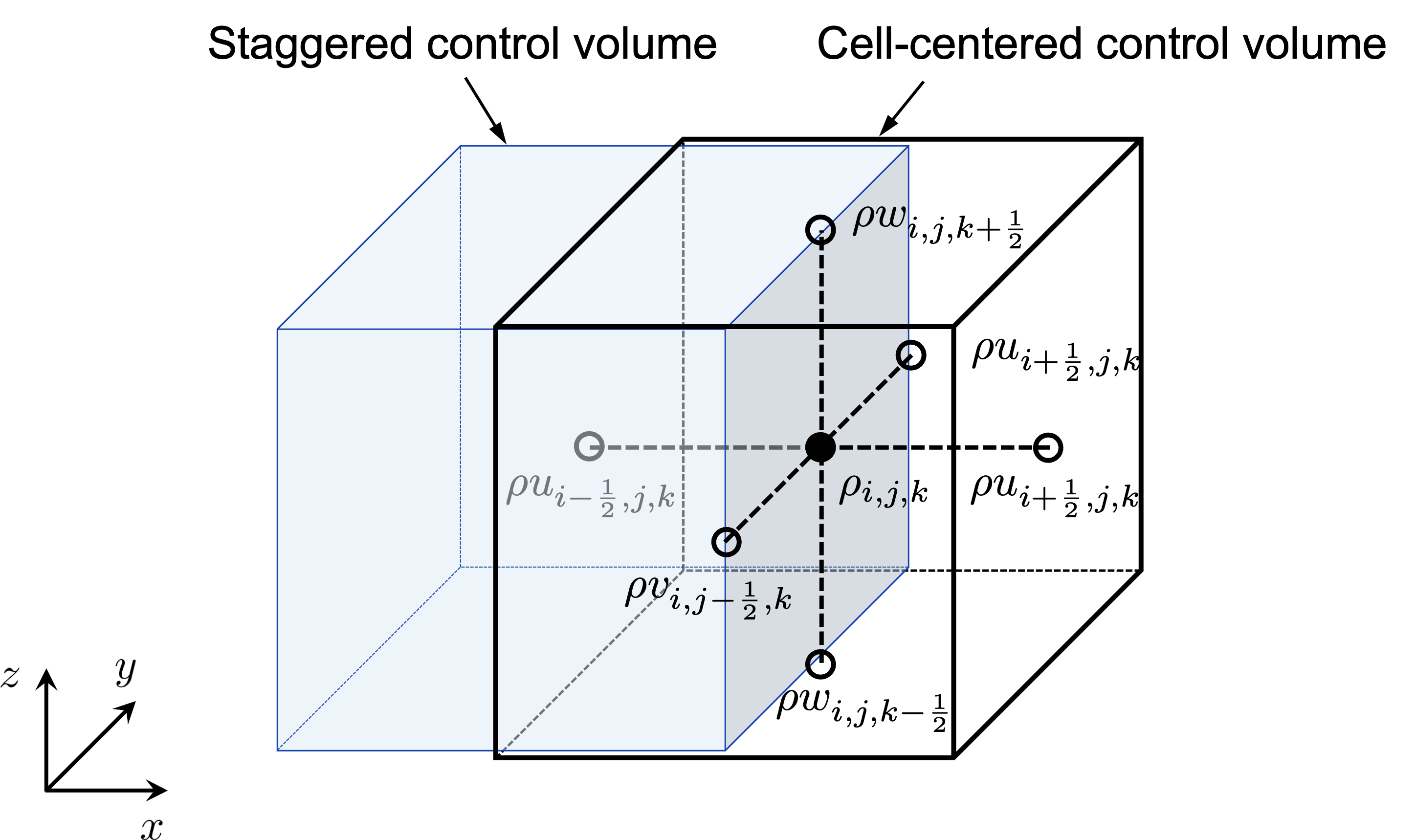}
    \end{center}
    \caption{Layout of the staggered grid with cell-centered and face-centered variables and their control volumes. The cube outlined with black lines is the control volume for the cell-centered variables; the blue shaded region is the control volume for the momentum on the low $x$ face of the cell-centered control volume.}\label{fig:staggered-grid}
\end{figure}

Figure~\ref{fig:staggered-grid} illustrates the layout of the staggered grid with the cell-centered and face-centered variables. The density, potential temperature and pressure are located at cell centers, while the velocity/momentum variables are located at face centers. The viscous stress components are computed at the edge centers. In this work, we consider a uniform Cartesian grid composed of rectangular cells with grid spacings $\dx$, $\dy$, and $\dz$ in each coordinate direction.


The sets of compressible and anelastic equations can be written in the compact form:
\begin{equation}
\frac{\partial \Ub}{\partial t} = - \nabla \cdot \Fb + \Sb,
\end{equation}
where $\Ub$ denotes the solution vector, $\Fb = (\Fb^x,\; \Fb^y,\; \Fb^z)$ is the flux with $\Fb^x$, $\Fb^y$, and $\Fb^z$ representing the fluxes through faces normal to the $x$-, $y$-, and $z$-directions, respectively, and $\Sb$ denotes the source term.

In a finite volume method, the fluxes used to update cell-centered variables are defined on the cell faces which form the boundary of the \textit{control volume} coincident with the cell itself. To update the momentum variables following the same paradigm, we define a control volume for each face-centered variable and the corresponding fluxes on the boundary of that control volume, as shown in Figure \ref{fig:staggered-grid}. The fluxes are then differenced to create an update to the solution vector in a discretized form of
\begin{equation}
\pdv{\Ub}{t} = \frac{1}{V}\int_{\partial \Omega} -\Fb \cdot \nb \; dA + \frac{1}{V}\int_\Omega \Sb \; dV, \label{eq:FV-integral}
\end{equation}
where $\Omega$ is the control volume region, $\partial \Omega$ is the boundary of that region, and $V=\abs{\Omega}$ is the volume of that region. 

Our EB solution methods can be viewed as modifications of the standard finite volume approach in which control volumes are cut by the embedded boundary, and the governing equations are integrated only over the irregularly shaped fluid region. Therefore, for cut cells, $\Omega$ in Eq.~\eqref{eq:FV-integral} represents the portion of the full cell that lies within the fluid domain. To characterize this irregular region, we define the geometric volume fraction $\alpha$ as the ratio of the fluid volume to the corresponding full rectangular cell volume ($\dx\dy\dz$), so that $\alpha\in[0,\;1]$. Similarly, an area fraction $\beta\in[0,\;1]$ is defined for each face of the control volume, representing the fraction of the face that is inside the fluid.


We write the fully discrete update of the solution variable from time level $n$ to $n+1$ with time step $\Delta t$ as
\begin{equation}
\Ub^{n+1} = \Ub^{n} + \Delta t \qty(\delta \Ub), \label{eq:FV-discrete}
\end{equation}
where $\delta\Ub$ denotes the time update of the solution variables, i.e., the right-hand side of the semi-discrete form of Eq.~\eqref{eq:FV-integral}. For explicit methods, $\delta\Ub$ is evaluated at time level $n$, whereas for semi-implicit methods, it is evaluated over the time interval $[n,n+1]$. The update is computed from the fluxes through the control-volume boundaries.


In the following discussion, we describe the time update assuming a forward Euler scheme for each variable. For clarity, we note that the ERF code actually employs third-order Runge-Kutta (RK3) and second-order Runge-Kutta (RK2) schemes for the compressible and anelastic formulations, respectively; here, a forward Euler scheme is considered for simplicity. Furthermore, we consider only the flux contribution and neglect the source term, i.e., we assume $\Sb = 0$.

To represent the discrete formulation, we use the index $(i,j,k)$ to denote cell-centered quantities and half-integer indices to denote face-centered quantities, as illustrated in Figure~\ref{fig:staggered-grid}. For example, $(i+\tfrac{1}{2},j,k)$ refers to an $x$-face. The first superscript ($c$ or $u$) denotes the location of the variable being updated: $c$ indicates a cell-centered variable, while $u$ indicates a variable located at an $x$-face. The second superscript ($x$, $y$, or $z$) denotes the coordinate direction of the face on which quantities such as fluxes and area fractions are defined.


The update of a cell-centered quantity, such as the density, is computed as follows:
\begin{align}
\label{eq:update-cc}
    (\delta\rho)_\ivec =& -\frac{1}{\alpha^c_{\ivec}}
    \qty[ 
    \begin{aligned}
    &\frac{1}{\dx}\qty( \beta^{c,x}_{\iph} F^{c,x}_{\iph} - \beta^{c,x}_{\imh} F^{c,x}_{\imh}) \\
    +& \frac{1}{\dy}\qty( \beta^{c,y}_{\ijph} F^{c,y}_{\ijph} - \beta^{c,y}_{\ijmh} F^{c,y}_{\ijmh}) \\
    +& \frac{1}{\dz}\qty( \beta^{c,z}_{\ijkph} F^{c,z}_{\ijkph} - \beta^{c,z}_{\ijkmh} F^{c,z}_{\ijkmh}) \\
    +& \frac{A^{c,\EB}_{\ivec}}{\dx\dy\dz} F^{c,\EB}_{\ivec}
    \end{aligned} ],
\end{align}
where $F^{c,x}$, $F^{c,y}$, and $F^{c,z}$ are the fluxes for the scalar equation in $x$-, $y$-, and $z$-directions, respectively, $\alpha^c$ and $\beta^c$ are the volume and area fractions on the cell-centered grid, $A^{c,\EB}$ is the area of the EB facet, and $F^{c,\EB}_{\ivec}$ is the flux through the EB facet.



Similarly, the update of a face-centered quantity, such as the $x$-momentum $\rho u$, is evaluated within its staggered control volume as follows:
\begin{align}
\label{eq:update-fc}
    (\delta\rho u)_{\imh} =& -\frac{1}{\alpha^u_{\imh}}
    \qty[ 
    \begin{aligned}
    &\frac{1}{\dx}\qty( \beta^{u,x}_{i,j,k} F^{u,x}_{i,j,k} 
    - \beta^{u,x}_{i-1,j,k} F^{u,x}_{i-1,j,k}) \\
    +& \frac{1}{\dy}\qty( \beta^{u,y}_{i-\frac{1}{2},j+\frac{1}{2},k} F^{u,y}_{i-\frac{1}{2},j+\frac{1}{2},k} 
    - \beta^{u,y}_{i-\frac{1}{2},j-\frac{1}{2},k} F^{u,y}_{i-\frac{1}{2},j-\frac{1}{2},k}) \\
    +& \frac{1}{\dz}\qty( \beta^{u,z}_{i-\frac{1}{2},j,k+\frac{1}{2}} F^{u,z}_{i-\frac{1}{2},j,k+\frac{1}{2}} 
    - \beta^{u,z}_{i-\frac{1}{2},j,k-\frac{1}{2}} F^{u,z}_{i-\frac{1}{2},j,k-\frac{1}{2}}) \\
    +& \frac{A^{u,\EB}_{\imh}}{\dx\dy\dz} F^{u,\EB}_{\imh}
    \end{aligned} ],
\end{align}
where $F^{u,x}$, $F^{u,y}$, and $F^{u,z}$ are the fluxes for the $x$-momentum equation in $x$-, $y$-, and $z$-directions, respectively, $\alpha^u$ and $\beta^u$ are the volume and area fractions on the $x$-staggered grid, $A^{u,\EB}$ is the area of the EB facet, and $F^{u,\EB}_{\imh}$ is the flux through the EB facet. The update to quantities on $y$-faces and $z$-faces would be defined analogously. 


\commentout{
For the discussion here, we will consider a single-step integration scheme typical of higher-order Godunov-type discretizations.  Generalizations to a method of lines approach are discussed at the end of Section \ref{sec:WSRD}.  We let $\U_\ivec$ and $F^x_\iph$, $F^y_\ijph$, and $F^z_\ijkph$ represent the discretized solution $U$, and the normal flux $F$ on $x$-faces,  $y$-faces and  $z$-faces, respectively.
We define a standard finite volume scheme in the form,
\begin{eqnarray*}
U_\ivec^{n+1} &=& U_\ivec^n + \Delta t \; \delta U^c_\ivec 
\end{eqnarray*}
where
\begin{eqnarray} 
\label{eq:cons_up}
\delta U^c_\ivec = 
&-& 
    \frac{\left (A^x_{\iph} F^x_{\iph} - A^x_{\imh} F^x_{\imh} \right )}{\vol_\ivec} \\ 
&-& 
    \frac{\left (A^y_{\ijph} F^y_{\ijph} - A^y_{\ijmh} F^y_{\ijmh} \right )}{\vol_\ivec} \nonumber \\
&-&   
    \frac{\left (A^z_{\ijkph} F^z_{\ijkph} - A^z_{\ijkmh} F^z_{\ijkmh} \right )}{\vol_\ivec} \nonumber \\
&-&  
    \frac{A^f_\ivec F^f_\ivec}{\vol_\ivec} \; \nonumber .
\end{eqnarray} 
\AddA{Here, $U_\ivec^{n+1} $ represents the integral average over the cell $(\ivec$).  For cut cells, this average refers to the average only over the portion of the cell that is inside the fluid region.} 
The update is conservative by construction, and the time accuracy of the scheme is determined by the details of the flux construction.
}

\subsection{Geometric representation of embedded boundaries}\label{sec:EBnot}

To calculate the updates given in Eqs.~\eqref{eq:update-cc}--\eqref{eq:update-fc}, we need geometric data describing the EB surface relative to both cell-centered and face-centered control volumes. Accordingly, we need four separate datasets, and each dataset contains the area fractions, area centroids, volume fractions, volume centroids, as well as the area, centroid, and unit normal vector of each EB facet. 


The construction of geometric EB data for staggered grids is a two-step procedure. First, the dataset is constructed on the cell-centered grid, which serves as the base grid. Second, the base dataset is extended to the staggered grids in each direction. In this work, we consider stationary embedded boundaries, and therefore, the procedure described here is performed once at the beginning of the simulations as a preprocessing step.

The EB surface is reconstructed as piecewise linear surface facets within a cell-centered grid. Its geometrical shape is defined by an implicit function, such as signed distance function or height function. First, we compute the level-set field for the EB surface at the nodes of a cell-centered grid. This information classifies each cell as
\textit{covered}, \textit{regular}, or \textit{cut}, and simplifies the handling of multiple cuts within a single cell. Second, the level-set field is used to compute the edge-boundary intersections and, subsequently, the cut-face data, including area fractions and area centroids. Third, the cut-cell data are constructed, including volume fractions, volume centroids, and the area and unit normal vector of each EB facet. For mesh refinement, this construction is performed first at the finest resolution level and then coarsened to coarser levels.

\begin{figure}[h!]
    \begin{center}
        \includegraphics[width=0.7\textwidth]{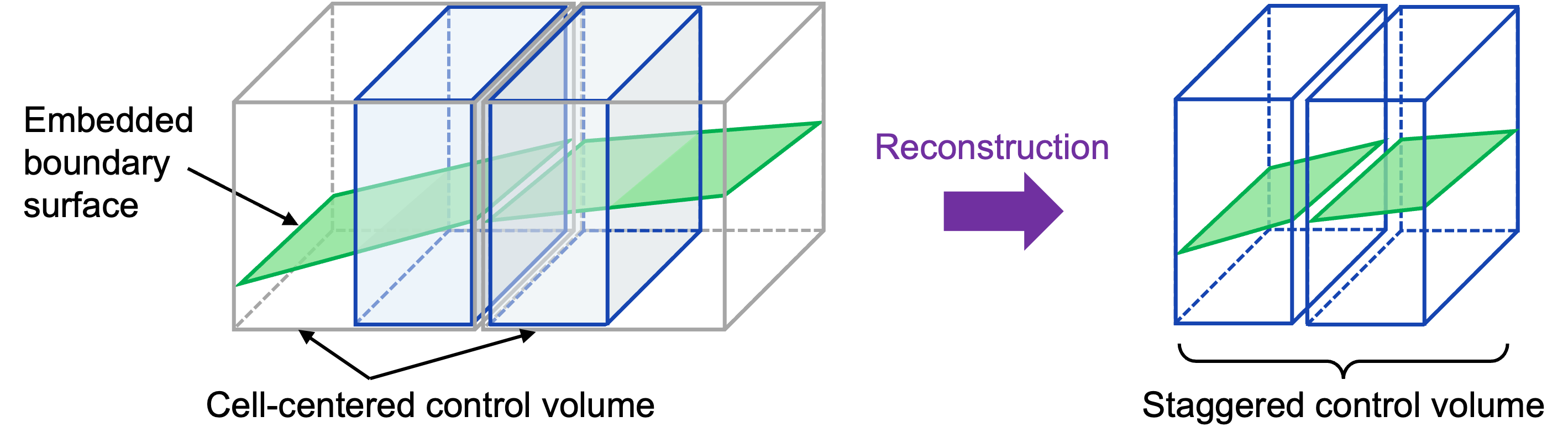}
    \end{center}
    \caption{Construction of cut-cell geometric data on a staggered grid.}\label{fig:eb_aux}
\end{figure}

The geometric properties of cut cells on staggered grids are constructed using precomputed data from the cell-centered grid. As illustrated in Figure~\ref{fig:eb_aux}, two half-cells from neighboring cell-centered control volumes are combined to form the cut-cell geometric data for a staggered control volume. If either half-cell corresponds to a cut cell, the resulting staggered cell is classified as a cut control volume. In this approach, the geometric data for half-cells in all three directions are constructed, where each half-cell is represented as a polyhedron bounded by planar facets. The cut-face area and cut-cell volume on the staggered grid are obtained by summation of the corresponding quantities from the two half-cells. Similarly, the cut-surface area of the embedded boundary is computed as the sum of the two contributions, while the unit normal vector and centroid are determined through area-weighted averaging. We reconstruct the dataset for staggered grids from the base cell-centered grid rather than constructing it independently to ensure consistency of volume and area fractions between the staggered and cell-centered grids. The same construction procedure is applied independently to the three staggered grids in the $x$-, $y$-, and $z$-directions.

With the geometric data generated, the cell-centered and face-centered variables are updated wherever the corresponding volume fractions are greater than zero, $\alpha^c > 0$ and $\alpha^u > 0$, respectively. Consequently, the momentum variables may be updated even when the corresponding face is completely covered and therefore not used in computing fluxes for cell-centered values, as long as the associated control volume contains a nonzero fluid fraction.


For implementation, the cell-centered dataset is generated using the EB module of AMReX. The reconstruction process for the staggered grids is implemented in ERF. All steps are GPU-accelerated. For additional details on the embedded boundary data structures, we refer to the AMReX documentation \cite{AMREX:Docs}.




\section{Numerical Method}\label{sec:numeth}

\subsection{Advection scheme}

For the advection terms in the compressible equation set, the flux is expressed as
\begin{align}
    \Fb_{\text{adv}} &= \begin{pmatrix}
    \rho\ub \\
    \rho\ub\otimes \ub \\
    \rho \ub \theta
    \end{pmatrix}.
\end{align}
For the anelastic model, $\rho$ is replaced by $\rho_0$ and the first component of the flux is removed. In the case of a no-slip or free-slip wall, the normal convective flux, $\Fb_{\text{adv}}\cdot\nb$, vanishes at the embedded boundaries, as the terms associated with the prescribed velocity become zero. The advective flux is constructed on the cell faces and the EB surface, while the pressure gradient term is added to the right-hand side vector as a source term in Eq.~\eqref{eq:FV-integral}.

High-order upwind discretizations for the advected variables are constructed using stencils that span multiple cells. In this work, we employ the third-order upwind scheme implemented in ERF, which can be written as a combination of a fourth-order centered scheme and a hyperviscosity term. Near the embedded boundaries, however, this construction is constrained by covered cells. If an upstream cell is covered, the required values of the variables are not available. To accommodate this limitation while maintaining high-order accuracy elsewhere, the advection scheme is adaptively reduced to a lower order, second-order centered scheme, when an upstream cell is covered. This selective reduction in order can be justified in the accuracy perspective, as the boundary layer along the embedded boundaries is typically diffusion-dominated rather than convection-dominated. Consequently, the additional numerical diffusion introduced by the low-order scheme is likely negligible.

\begin{figure}[h!]
    \begin{center}
        \includegraphics[width=0.45\textwidth]{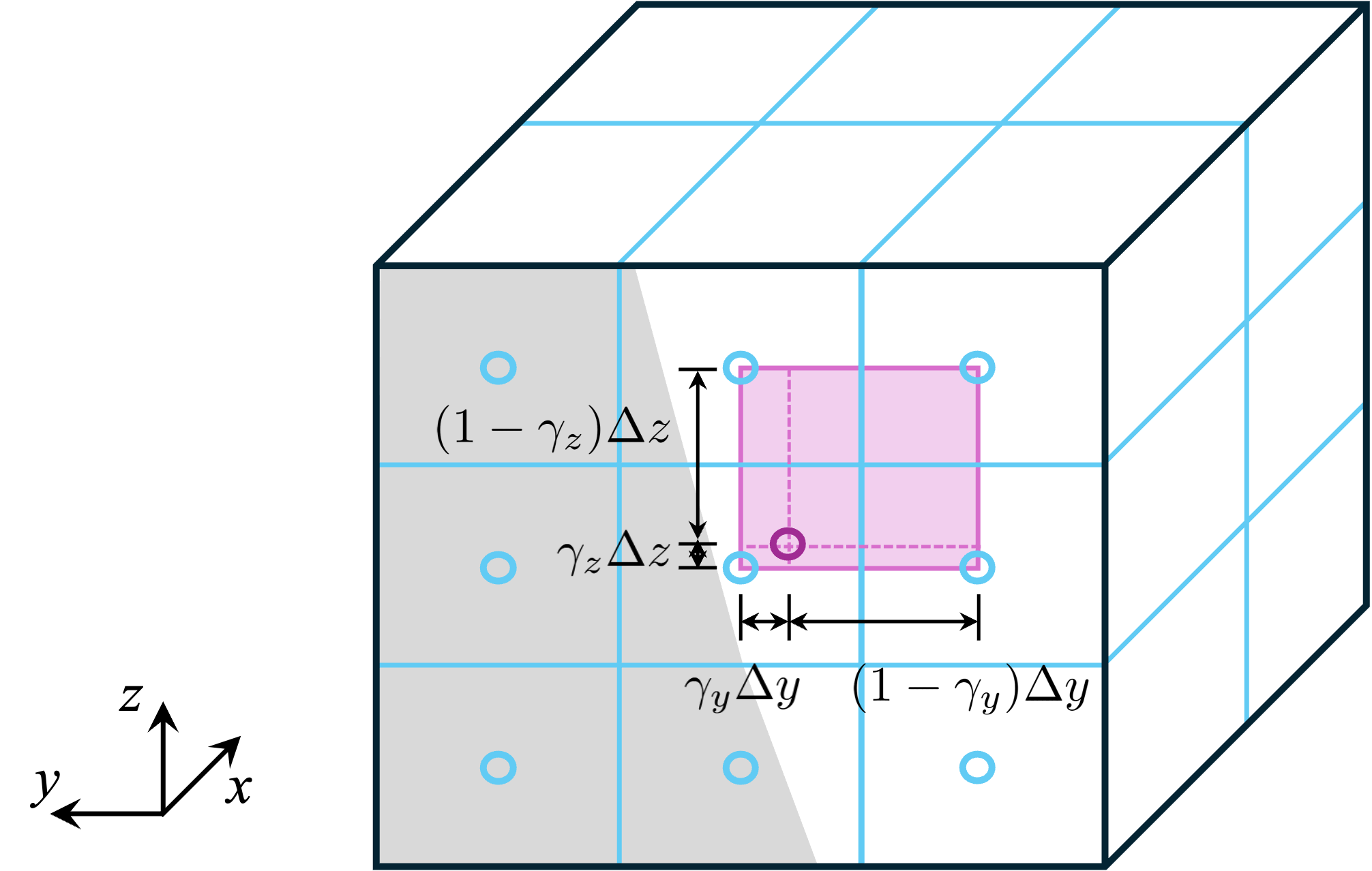}
    \end{center}
    \caption{Bilinear interpolation of the advective flux on the cut face.}\label{fig:flux-interpolation}
\end{figure}

In order to achieve consistency and consequent high-order accuracy using the discrete formulation in Eqs.~\eqref{eq:update-cc}--\eqref{eq:update-fc}, the advective flux needs to be evaluated at the centroid of the cut face. To achieve this, we reconstruct the flux using the directional interpolation method described in \cite{modiano2000higher}. In this approach, the flux at the cut face is obtained by bilinear interpolation of face-centered provisional fluxes from four surrounding cells, as illustrated in Figure \ref{fig:flux-interpolation}. The three neighboring cells are selected based on the direction toward which the cut-face centroid is offset. Consequently, the advective flux is calculated at the cell faces as
\begin{align}
\begin{split}
F^x_{\imh} =& \; \qty(1-\gamma_y)\qty(1-\gamma_z) \; \widehat{F}^x_{\imh} \; 
+\; \gamma_y\qty(1-\gamma_z) \; \widehat{F}^x_{{i-\sfrac{1}{2},j+1,k}} \; \\
&+\; \gamma_z\qty(1-\gamma_y) \; \widehat{F}^x_{{i-\sfrac{1}{2},j,k+1}} \; 
+\; \gamma_y\gamma_z \; \widehat{F}^x_{{i-\sfrac{1}{2},j+1,k+1}}
\end{split}
\end{align}
where $\gamma_y$ and $\gamma_z$ are the offset parameters of the cut-cell centroid in $y$- and $z$-directions ($\gamma_y,\gamma_z\in[-0.5,0.5]$), respectively, and $\widehat{F}^x$ denotes the provisional flux at the $x$-face.

\subsection{Diffusion scheme}

\begin{figure}[h!]
    \captionsetup[subfigure]{justification=centering}
    \centering
    \begin{subfigure}[b]{0.4\textwidth}
        \includegraphics[width=\textwidth]{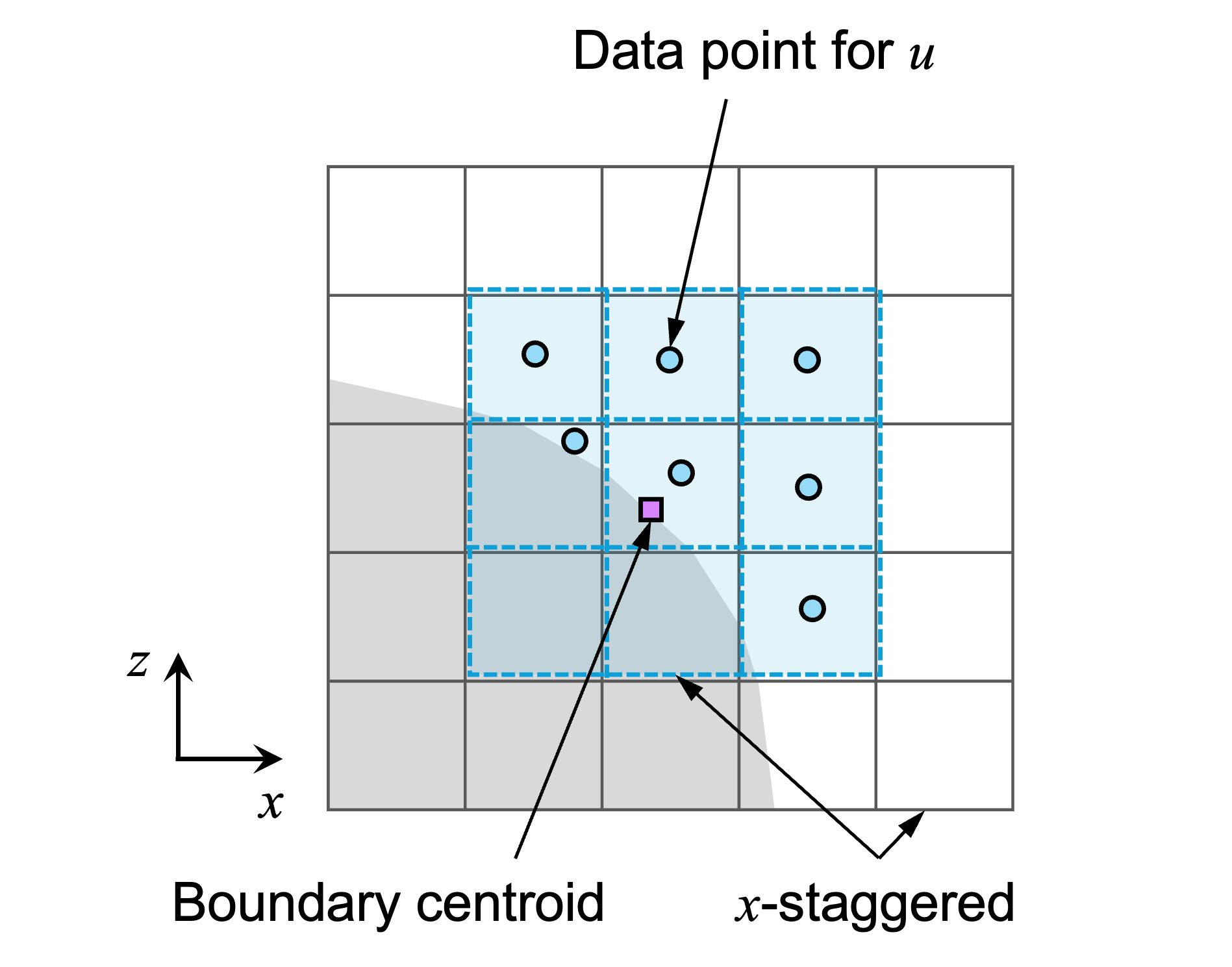}
        \caption{Gradient of $u$ on the $x$-staggered grid}
    \end{subfigure}
    \hspace{2em}
    \begin{subfigure}[b]{0.4\textwidth}
        \includegraphics[width=\textwidth]{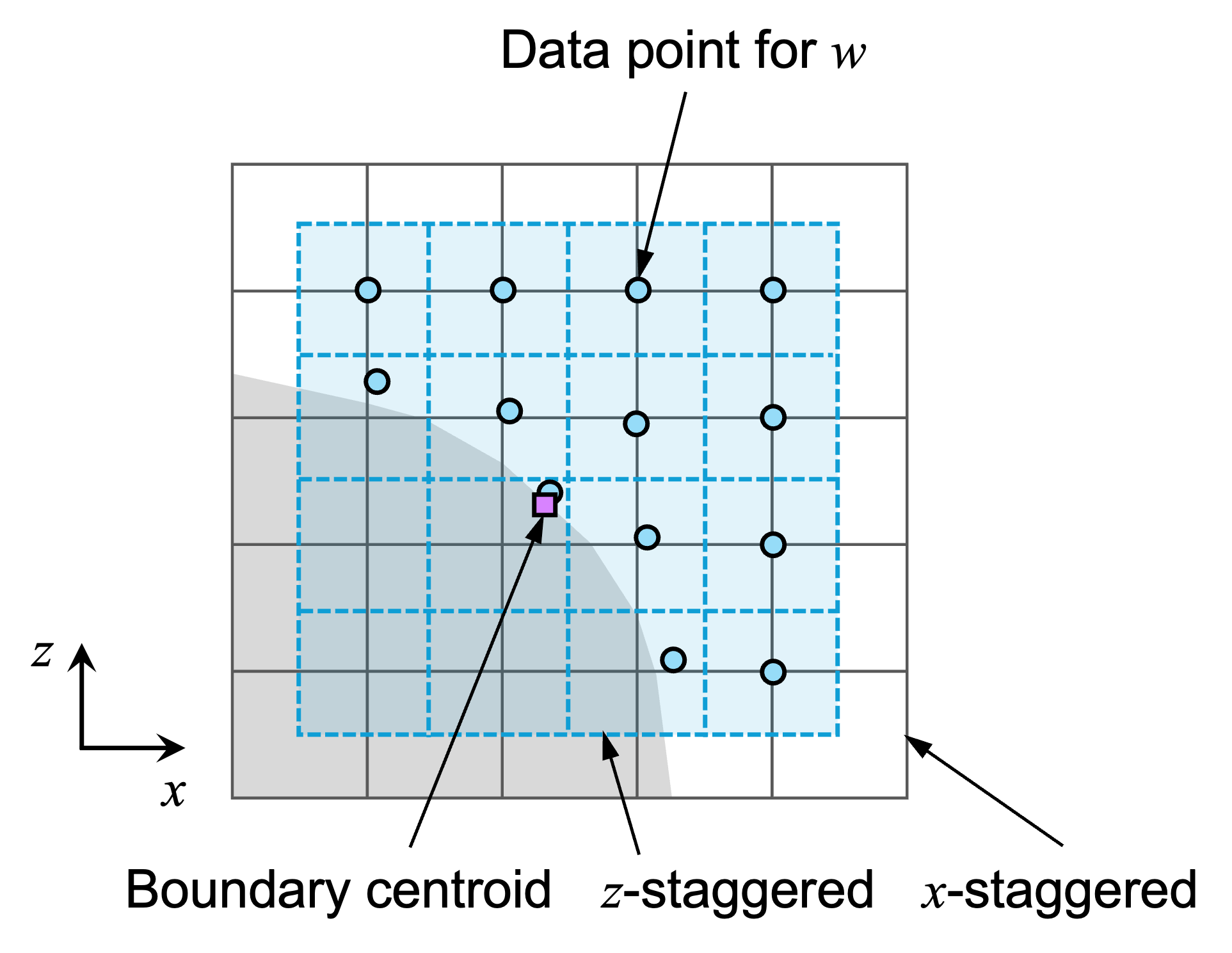}
        \caption{Gradient of $w$ on the $x$-staggered grid}
    \end{subfigure}
    \caption{Neighborhood stencils for calculating the gradient of $u$ and $w$ on the $x$-staggered grid using least-squares fitting. Blue circles represent data points used in the fitting, and the purple square marks the EB facet centroid where the gradients are evaluated.}\label{fig:LSF}
\end{figure}

For diffusive terms, the flux is given by
\begin{align}
    \Fb_{\text{diff}} &= \begin{pmatrix}
    0 \\
    \taub \\
    \rho \alpha_{\theta} \nabla\theta
    \end{pmatrix}.
\end{align}
The diffusive flux requires the calculation of the velocity and potential temperature gradients on the cut face. Several methods have been proposed to evaluate gradients on the cut surface, including a 3-point stencil \cite{johansen1998cartesian} and a least squares approximation \cite{natarajan2022moving}. In this work, we compute the gradients by fitting a second-order approximation to the state variables using a least-squares method applied to a local neighborhood. Figure \ref{fig:LSF} illustrates the neighborhood stencils used for the least-squares fitting of $u$ and $w$ on the $x$-staggered grid. Specifically, a neighborhood is defined as the set of regular and cut cells located within a $3\times 3\times 3$ region (in index space) when evaluating approximate gradients for variables on the same grid. For gradients of variables stored on different grid types, the stencil size is increased from 3 to 4 to preserve the symmetry of the stencil. The comprehensive formulation for the least-squares error minimization is provided in the appendix of \cite{natarajan2022moving}.

\subsection{Adaptive mesh refinement}

In this work, we employ a static block-structured mesh refinement strategy to improve computational efficiency. In this approach, overlapping grid boxes are hierarchically created to provide higher resolution in selected regions. Coarse-level cells are subdivided into fine subcells according to a uniform refinement ratio of 2 in this work. To preserve volume and area fractions in cut cells, these geometric quantities are constructed recursively from the finer level to the coarser level by merging subcells.

During time integration, the coarse and fine grids can be coupled in two ways. Dirichlet boundary conditions obtained from the coarse-grid solution are supplied at the fine-grid boundaries, and when using two-way coupling the fine-grid variables are transferred to the coarse grid through a restriction (averaging-down) operation at the end of each time-update stage. In the restriction operation, cell-centered and face-centered variables are treated differently. Cell-centered variables are volume-weighted using volume fractions to conserve mass over each coarse-cell region, whereas face-centered variables are area-weighted using area fractions to preserve the integral of mass fluxes over each coarse cell face. These conservative averaging procedures ensure local mass conservation during the restriction operation.


\section{Weighted state redistribution scheme} \label{sec:WSRD}

This section reviews the WSRD scheme for addressing the small-cell problem, which has been developed and applied to flow equations on collocated grids~\cite{giuliani2022weighted}. In this work, we extend the WSRD approach to staggered grids. The WSRD operators for all solution variables are constructed using the geometric datasets described in the previous section and applied independently to each variable. The incorporation of WSRD into the time-integration procedure is also discussed in this section.

\subsection{Weighted state redistribution on a staggered mesh}\label{sec:WSRD-algo}


In the absence of cut cells, where all cells have the same volume $V = \Delta x \Delta y \Delta z$, the Courant-Friedrichs-Lewy (CFL) condition for an explicit update constrains the time-step size $\Delta t$ based on the ratio of cell volume to face areas as well as the maximum wave speed, so the time-step size scales linearly with the mesh spacing. The small cell problem arises when an EB representation of geometry is used. The cut cells can have arbitrarily small volumes, i.e., vanishing volume fractions $\alpha$, which can make the update in Eqs.~\eqref{eq:update-cc}--\eqref{eq:update-fc} unstable \cite{almgren1997cartesian}. Reducing $\dt$ to account for these small cells would be computationally prohibitive.

The WSRD scheme addresses the small-cell problem by constructing logical merging neighborhoods, in which small cut cells are merged with neighboring cells with weights based on cut-cell volumes. To apply the WSRD scheme on a staggered grid, we apply the WSRD operator to each solution variable independently. The single-step time advance for the solution vector, including both cell-centered and face-centered variables, can be expressed as
\begin{align} \label{eq:basic_fv_srd}
\Ub^{n+1} = \mathcal{R}\;\qty\Big[\Ub^n + \dt\; (\delta\Ub^n)],
\end{align}
where $\mathcal{R}=\qty\big[R^c,R^u,R^v,R^w,R^c]$ for the compressible system in Eqs.~\eqref{eq:cons}--\eqref{eq:theta-comp} and $\mathcal{R}=\qty\big[R^u,R^v,R^w,R^c]$ for the anelastic system in Eqs.~\eqref{eq:mom-anelastic}--\eqref{eq:theta-anelastic}, and $R^{(\bullet)}$ represents the WSRD operator acting on variables located at either the cell center or the face center.

In the following, we describe the action of the WSRD operator for a cell-centered quantity. The same procedure can be readily applied to face-centered variables.

To create merging neighborhoods, we first identify small cells with a volume below a specified target threshold, $V_{\text{target}}$. These cells are then logically merged with neighboring cells until the neighborhood volume, i.e., the sum of the volumes of all cells in the neighborhood, reaches $V_{\text{target}}$. Here, we set $V_{\text{target}} = \frac{1}{2} \Delta x \Delta y \Delta z$. This choice is informed by \cite{berger2021state}, although in practice the appropriate threshold value may depend on factors such as the details of the numerical scheme and boundary conditions.



For merging, a small cell is first merged with its (regular or cut) neighbor in the direction of the boundary normal. If the addition of the neighbor's volume is not sufficient to reach the target threshold, additional cells are added to the merging neighborhood. In our implementation, the second-choice neighbor is selected based on the next largest component of the normal, creating an ``L-shaped'' neighborhood. The cell in the same plane as the first two other neighbors is automatically merged to form a $2 \times 2\times 2$ neighborhood. 






We define $M_{i,j,k}$ as the set of cells contained in the neighborhood associated with cell $(i,j,k)$, and $M^-_{i,j,k} = M_{i,j,k} - \{(i,j,k)\}$, i.e., the neighborhood of cell $(i,j,k)$ excluding the cell itself. We further define $W_{i,j,k}$ as the set of indices $(r,s,t)$ such that cell $(i,j,k)$ belongs to $M_(r,s,t)$, and $W_{i,j,k}^- = W_{i,j,k} - \{(i,j,k)\}$. Finally, $N_{i,j,k}$ denotes the number of neighborhoods that cell $(i,j,k)$ belongs to , i.e., the size of $W_{i,j,k}$.

For WSRD, at every regular or cut cell $(i,j,k)$ we define
\begin{equation}\label{eqn:kappa}
{\kappa}_{i,j,k} =\begin{cases}
			\;\qty(V_{\text{target}} - V_{i,j,k}) \;\; /  \displaystyle\sum_{(r,s,t) \in M^-_{i,j,k} } \, V_{r,s,t} \, & \text{if $V_{i,j,k} < V_{\text{target}}$}\\
            \; 0 & \text{if $V_{i,j,k} \geq V_{\text{target}}$}
		 \end{cases}
\end{equation}
and
\begin{equation}
\label{eqn:omega}
{\omega}_{i,j,k} =  1 \, - \,  \frac{1}{N_{i,j,k}} \, \sum_{(r,s,t) \in W^-_{i,j,k} } \, \kappa_{r,s,t} \, ,
\end{equation}
where $V_{i,j,k}=\alpha_{i,j,k}\dx\dy\dz$, and $0 \le \omega_{i,j,k}, \kappa_{i,j,k} \le 1$. We observe that $\kappa_{i,j,k}$ depends on the volume fractions of the cells in the neighborhood of cell $(i,j,k),$ while $\omega_{i,j,k}$ depends on the values of $\kappa$ in cells whose neighborhoods contain cell $(i,j,k)$. 



 

We define the weighted volume $\widehat V_{i,j,k}$ of the neighborhood associated with cell $({i,j,k})$ as
 \begin{equation}
 \label{eqn:nbhd_vol}
{\widehat V}_{i,j,k} =  \omega_{i,j,k} V_{i,j,k} +  \kappa_{i,j,k} \sum_{(r,s,t) \in M^-_{i,j,k}} \,  \frac{V_{r,s,t}}{N_{r,s,t}},
\end{equation}
and the centroid of the neighborhood by 
\begin{align}
\begin{aligned}
\qty({\widehat x}_{i,j,k},\, {\widehat y}_{i,j,k},\, {\widehat z}_{i,j,k}) = \frac{1}{\widehat V_{i,j,k}} \Biggl[
&\omega_{i,j,k} V_{i,j,k} \qty(x_{i,j,k},\, y_{i,j,k},\, z_{i,j,k}) \\
&+ \; \kappa_{i,j,k}  \sum_{(r,s,t) \in M^-_{i,j,k} }  \frac{V_{r,s,t}}{N_{r,s,t}} \qty(x_{r,s,t},\, y_{r,s,t},\, z_{r,s,t}) \Biggr],
\end{aligned}
\end{align}
where $\qty(x_{i,j,k},\; y_{i,j,k},\; z_{i,j,k})$ denotes the original centroid of cell $(i,j,k)$. In an efficient implementation, all of the steps described so far in this subsection can be performed when the EB information is first defined at the start of a simulation, or, in the case of AMR, after any step that modifies the grid hierarchy.




As noted in Section~\ref{sec:FV}, the solution procedure begins by computing a conservative update, $\delta U_{i,j,k}$, using Eqs.~\eqref{eq:update-cc}--\eqref{eq:update-fc} for cell-centered and face-centered variables, respectively. In WSRD, once $\delta U_{i,j,k}$ is computed, a provisional update is defined as
\begin{equation}
    \widehat{U}_{i,j,k} = U^n_{\ivec}+ \Delta t \; \qty(\delta U_{i,j,k}).
\end{equation}
 
Next, for each neighborhood, the weighted solution average is computed as
\begin{equation}
\label{eqn:qhat_omega}
\widehat{Q}_{i,j,k} =  \frac{1}{{\widehat V}_{i,j,k}} \, \left( \omega_{i,j,k} V_{i,j,k} \widehat{U}_{i,j,k} \, + \kappa_{i,j,k} \,
\sum_{(r,s,t) \in M^-_{i,j,k}}   
\frac{V_{r,s,t}}{N_{r,s,t}}  \widehat{U}_{r,s,t} \right).
\end{equation}
For regular cells or cut cells with $V_{\ivec} > V_{\text{target}}$, the neighborhood contains only the cell itself, and consequently, $\widehat{Q}_{i,j,k} = \widehat{U}_{i,j,k}$.

The weighted solution $\widehat{Q}_{i,j,k}$ is reconstructed linearly to achieve second order accuracy where possible. For linear reconstruction, we compute the gradient of $\widehat{Q}_{i,j,k}$, $(\widehat{\sigma}_x, \widehat{\sigma}_x, \widehat{\sigma}_x)$, in all neighborhoods containing at least two cells using a least squares fitting. For neighborhoods that contain only a single cell, the gradient is simply set to zero. The values of $\widehat{Q}$ are fitted over a $3\times3\times3$ stencil centered on $(i,j,k)$, accounting only for regular or cut cells within the stencil and assuming that each $\widehat{Q}$ is defined at the centroid of its neighborhood, $(\widehat{x}, \widehat{y}, \widehat{z})$. If the distance between centroids in any coordinate direction is below a given threshold (half the mesh spacing in that direction), the gradient stencil in that direction is increased from 3 to 5. If the increased stencil contains too few points or lacks sufficient coordinate separation, the gradient is set to zero.

To prevent generating new maxima or minima of $\widehat{Q}$, we limit the gradient components using a Barth-Jesperson-style limiter. The linear reconstruction for neighborhood $(i,j,k)$ is then
 \begin{equation}\label{eq:qrecon}
\widehat{q}_{i,j,k}(x,y,z) = \widehat{Q}_{i, j,k} + \widehat{\sigma}_{x,i,j,k}(x - \widehat{x}_{i,j,k}) + \widehat{\sigma}_{y,i,j,k}(y - \widehat{y}_{i,j,k}) + \widehat{\sigma}_{z,i,j,k}(z - \widehat{z}_{i,j,k}).
\end{equation}
The final solution is computed as
 \begin{equation} \label{eq:final_update_omega}
        U^{n+1}_{i,j,k} =   \omega_{i,j,k} \; \widehat{q}_{i,j,k}(x_{i,j,k},y_{i,j,k},z_{i,j,k}) + 
        \frac{1}{N_{i,j,k}}\sum_{(r,s,t)  \in W^-_{i,j,k}} \kappa_{r,s,t} \; 
        \widehat{q}_{r,s,t}(x_{i,j,k},y_{i,j,k},z_{i,j,k}).
\end{equation}
If a small cut cell $(i,j,k)$ belongs only to its own neighborhood, then $\omega_{i,j,k} = 1$ and ${U}^{n+1}_{i,j,k} = \widehat{q}_{i,j,k}(x_{i,j,k},\, y_{i,j,k},\, z_{i,j,k})$, regardless of the number of other cells in its neighborhood.






\subsection{Time integration}

The WSRD scheme is incorporated into the time integration of each solution variable. To integrate the compressible equations in time, we employ a fully explicit low-storage third-order Runge-Kutta scheme implemented in ERF \cite{lattanzi2025erf}. At each Runge-Kutta stage, the state vectors are reditributed immediately after their update with the right-hand side vectors.


For the temporal integration of the anelastic equations, we employ a second-order Runge–Kutta scheme combined with a fractional-step (projection) method. In this approach, at each Runge–Kutta stage, the momentum and potential temperature are first updated without enforcing the anelastic constraint. The velocity field is then projected onto a divergence-free space by solving a pressure Poisson equation at the end of the stage. This equation is solved using a geometric multigrid solver with a Gauss–Seidel smoother and a biconjugate gradient stabilized solver, both implemented within the AMReX framework~\cite{AMReX:IJHPCA}.

Algorithms~\ref{algo:ti-compressible} and \ref{algo:ti-anelastic} summarize the sequence of temporal integration with the WSRD scheme for a single time step of the compressible and anelastic equations, respectively.
\begin{algorithm}
\caption{Sequence for advancing a single time step for the compressible equations Eqs.~\eqref{eq:cons}--\eqref{eq:theta-comp} with the weighted state redistribution.}\label{algo:ti-compressible}
  \begin{algorithmic}
    \For{$s=1,stages$}
      \State{(1) Compute the diagnostic variable: the primitive variables $\ub$ and $\theta$, and the pressure $p$.}
      \State{(2) Compute the time update $\delta\Ub^s$.}
      \State{(3) Compute the provisional stage solution $\widehat{\Ub}^s$ using $\delta\Ub^s$.}
      \State{(4) Correct the solution with WSRD: $\Ub^s = \mathcal{R} (\widehat{\Ub}^s)$, where $\mathcal{R}=\qty\big[R^c,R^u,R^v,R^w,R^c]$.}
    \EndFor
  \end{algorithmic}
\end{algorithm}

\begin{algorithm}
\caption{Sequence for advancing a single time step for the anelastic equations Eqs.~\eqref{eq:mom-anelastic}--\eqref{eq:theta-anelastic} with the weighted state redistribution.}\label{algo:ti-anelastic}
  \begin{algorithmic}
    \For{$s=1,stages$}
      \State{(1) Compute the diagnostic variables: the primitive variables $\ub$ and $\theta$.}
      \State{(2) Compute the time update $\delta\Ub^s$.}
      \State{(3) Compute the provisional stage solution $\widehat{\Ub}^s$ using $\delta\Ub^s$.}
      \State{(4) Correct the solution with WSRD: $\Ub^s = \mathcal{R} (\widehat{\Ub}^s)$, where $\mathcal{R}=\qty\big[R^u,R^v,R^w,R^c]$.}
      \State{(5) Project the intermediate velocity components onto a divergence-free field and compute the pressure gradient $\nabla p$.}
    \EndFor
  \end{algorithmic}
\end{algorithm}

\section{Numerical results}\label{sec:num}

This section presents numerical test results to verify the accuracy of the new embedded boundary method for staggered mesh. The numerical methods described in the preceding sections are implemented in the ERF code in a fully GPU-compatible manner, including the preprocessing tasks for EB geometric data construction. All computational kernels are ported to GPUs through the native vendor-specific programming models within the AMReX code \cite{AMReX:IJHPCA}. In this work, the HIP~\cite{rocm_hip} backend is used and the simulations were carried out using GPUs on the Tuolumne high-performance computing system at Lawrence Livermore National Laboratory, where each compute node consists of 96 AMD EPYC cores and four AMD MI300A GPUs.


\subsection{Witch of Agnesi hill}\label{sec:2d_hill}

This is an effectively two-dimensional test case that simulates flow over an idealized hill, the Witch of Agnesi. The fully compressible equation set is used. The test is set up following the configuration used in \cite{lundquist2010immersed}. The domain is rectangular with a length of $L=595$ m and a height of $H=600$ m. The terrain height $h_t$ is defined as
\begin{equation}
    h_t(x) = \dfrac{h_p}{1+(x/a)^2},
\end{equation}
where $x$ is the horizontal distance from the mid-plane, $h_p$ and $a$ are the peak height and the half-width of the mountain, respectively, both set to 100 m. Periodic boundary conditions are imposed on the lateral boundaries, and the terrain surface is modeled as a no-slip wall. The dynamic viscosity is set to 60.0 kg/(m$\cdot$s). A constant pressure gradient of 0.005 Pa/m is applied. 

\begin{figure}[h!]
    \captionsetup[subfigure]{justification=centering}
    \centering
    \begin{subfigure}[b]{0.28\textwidth}
        \includegraphics[width=\textwidth]{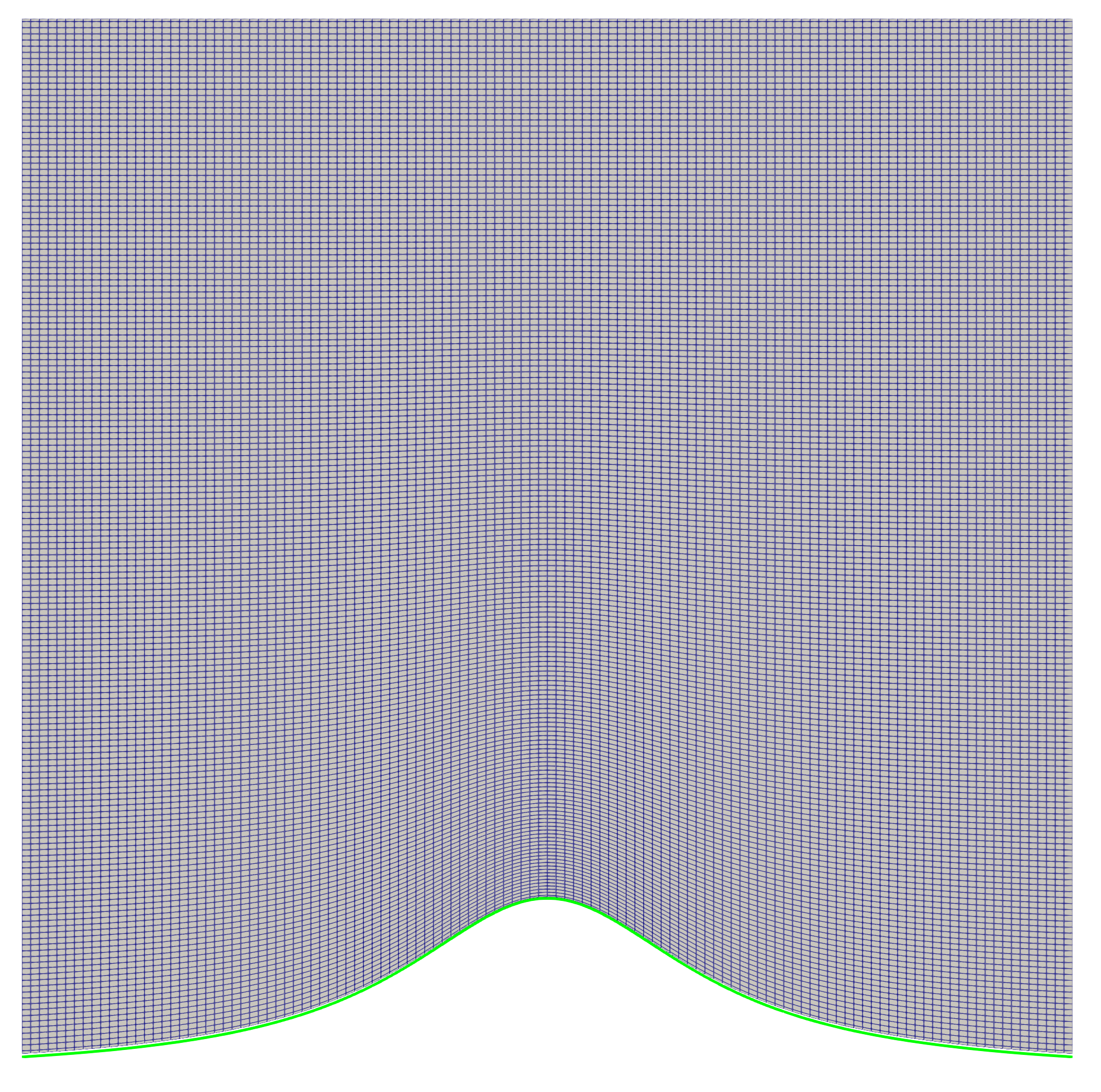}
        \caption{Fitted mesh}
    \end{subfigure}
    \begin{subfigure}[b]{0.28\textwidth}
        \includegraphics[width=\textwidth]{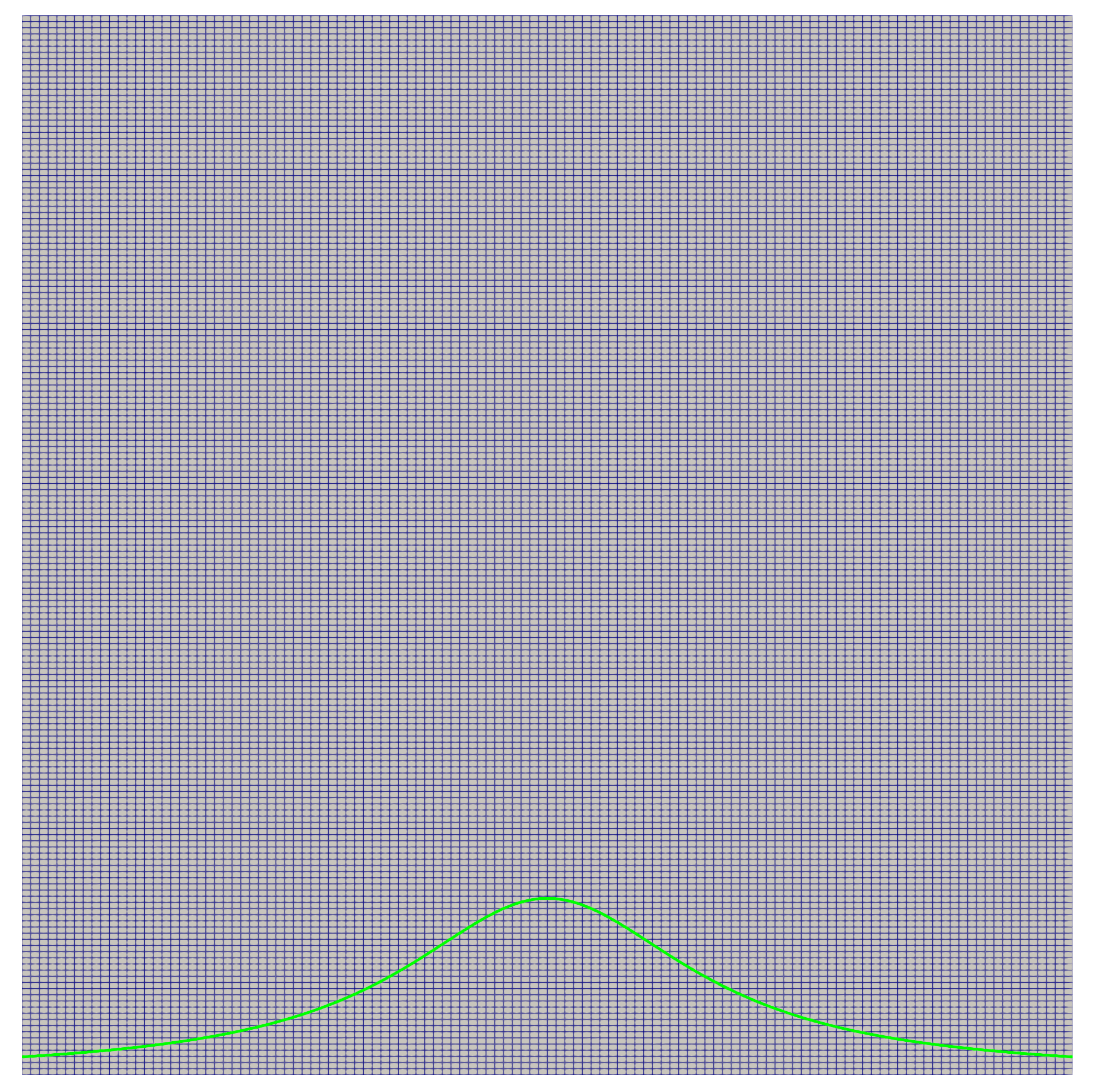}
        \caption{EB}
    \end{subfigure}
    \begin{subfigure}[b]{0.28\textwidth}
        \includegraphics[width=\textwidth]{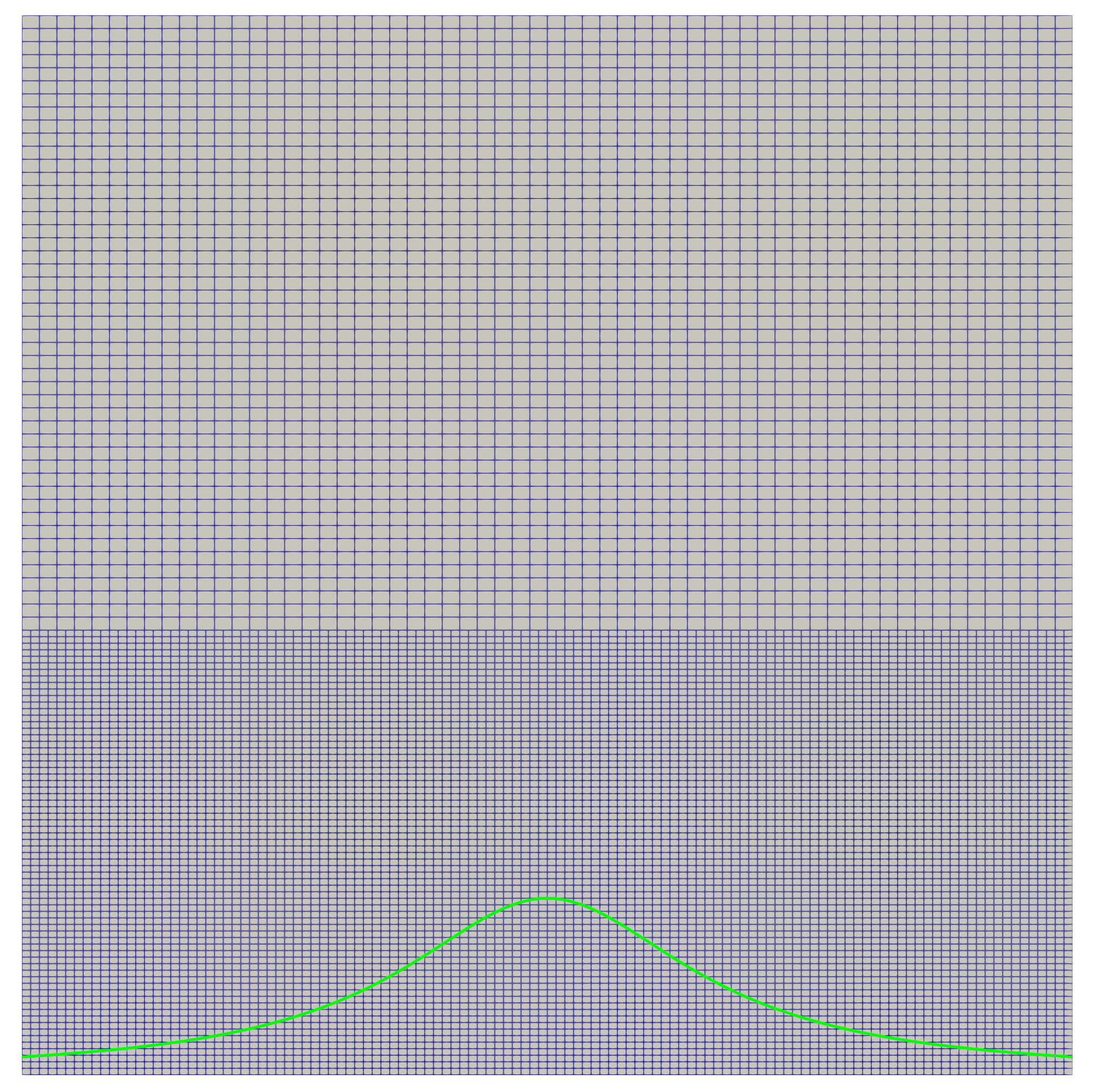}
        \caption{EB-AMR}
    \end{subfigure}
    \caption{Fitted, embedded boundary (EB), and EB with AMR grids for Witch of Agnesi test.}\label{fig:WoA-Mesh}
\end{figure}

We generate three grids: a fitted mesh, and embedded boundary (EB) mesh, and an EB with AMR, as shown in Figure~\ref{fig:WoA-Mesh}. The fitted and EB meshes consist of 120 and 172 cells in the horizontal and vertical directions, respectively. The EB-AMR grid has an equivalent resolution with the EB mesh in the bottom region up to a height of 237.2 m, and it is coarsened by a factor of 2 in the upper region. Advective fluxes are computed using a third-order upwind scheme. A Rayleigh damping layer of thickness 50 m is applied at the upper boundary of the domain, with a damping coefficient of 0.25. The time step size is adjusted dynamically to satisfy CFL$=0.5$.

\begin{figure}[h!]
    \centering
    \includegraphics[width=0.5\textwidth]{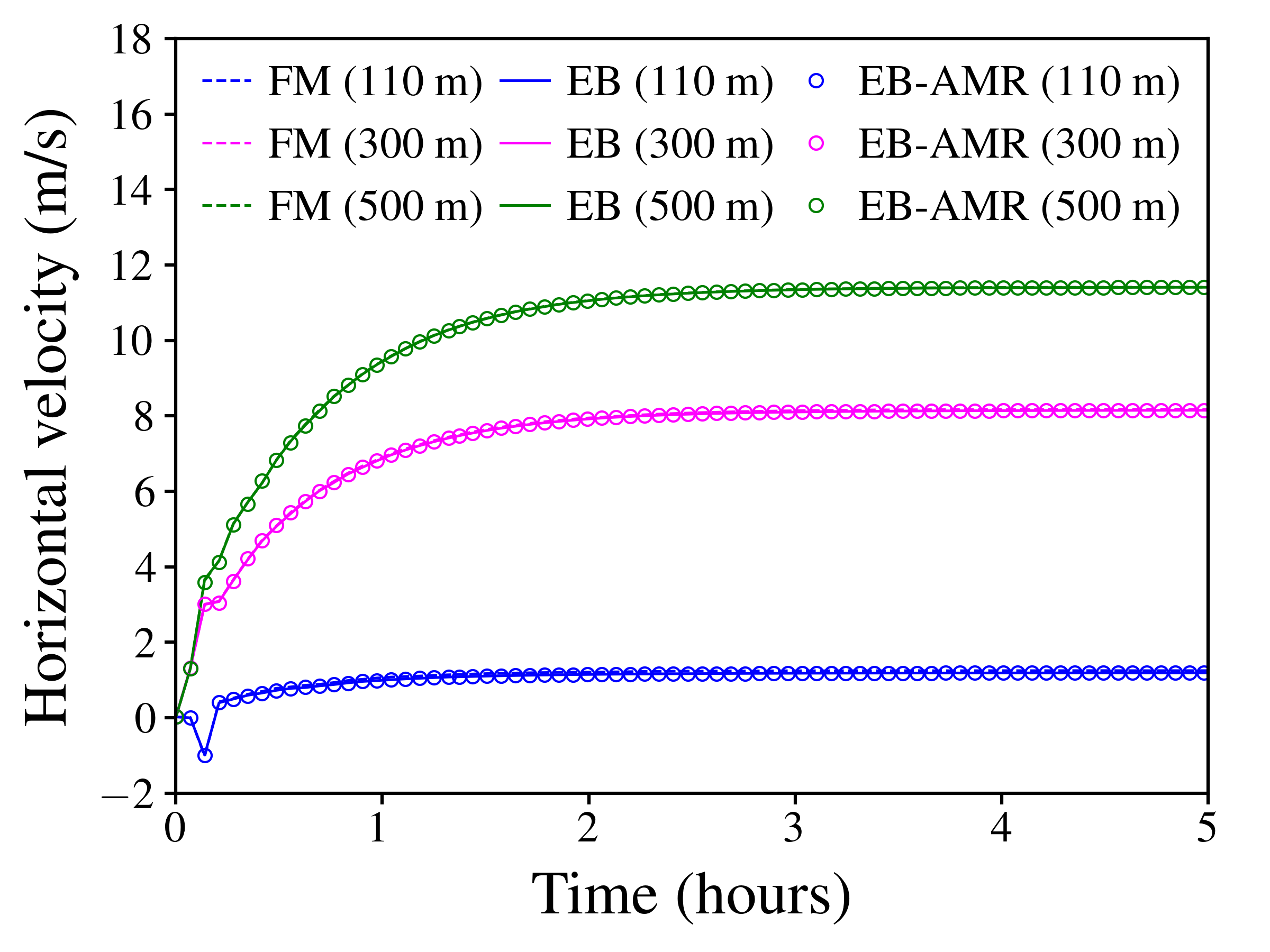}
    \caption{Horizontal velocity in time at three locations ($x=0.5L$ and $z=110$, 300, and 500 m) computed using fitted mesh (FM), embedded boundary (EB), and EB with AMR approaches.}\label{fig:WoA-u-time}
\end{figure}

\begin{figure}[h!]
    \captionsetup[subfigure]{justification=centering}
    \centering
    \begin{subfigure}[b]{0.3\textwidth}
        \includegraphics[width=\textwidth]{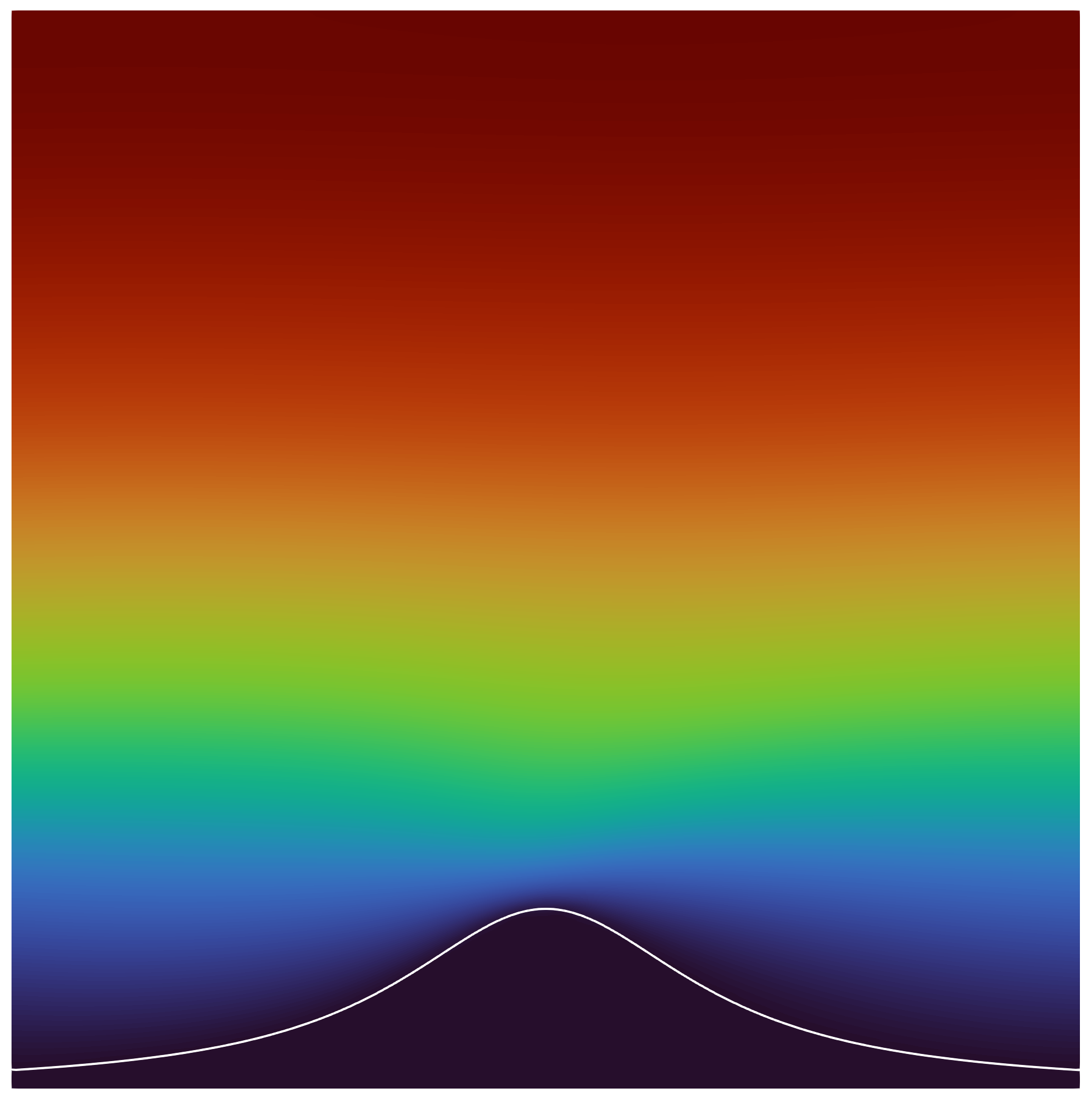}
        \caption{Horizontal velocity}
    \end{subfigure}
    \begin{subfigure}[b]{0.3\textwidth}
        \includegraphics[width=\textwidth]{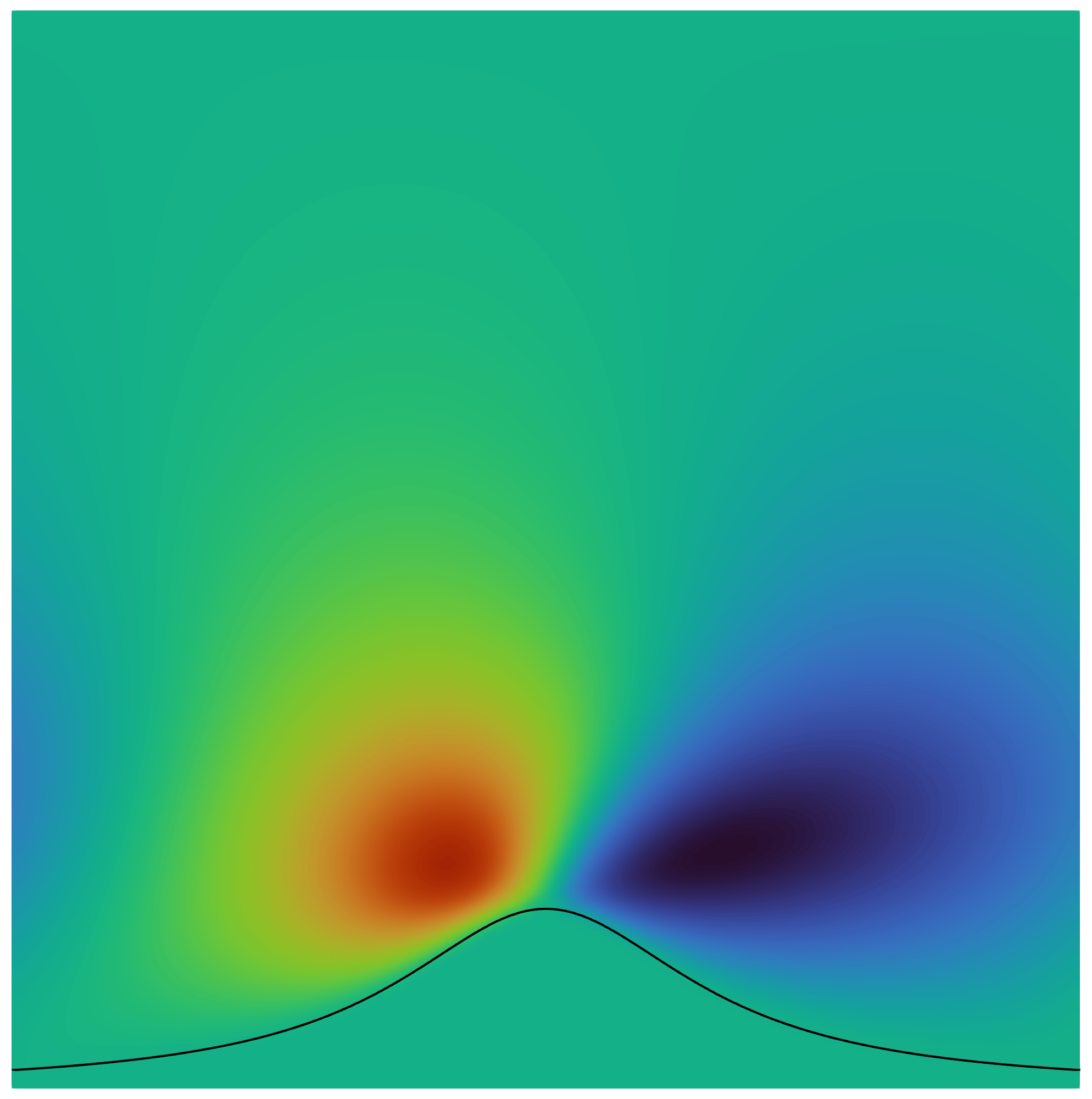}
        \caption{Vertical velocity}
    \end{subfigure}
    \\
    \begin{subfigure}[b]{0.24\textwidth}
        \includegraphics[width=\textwidth]{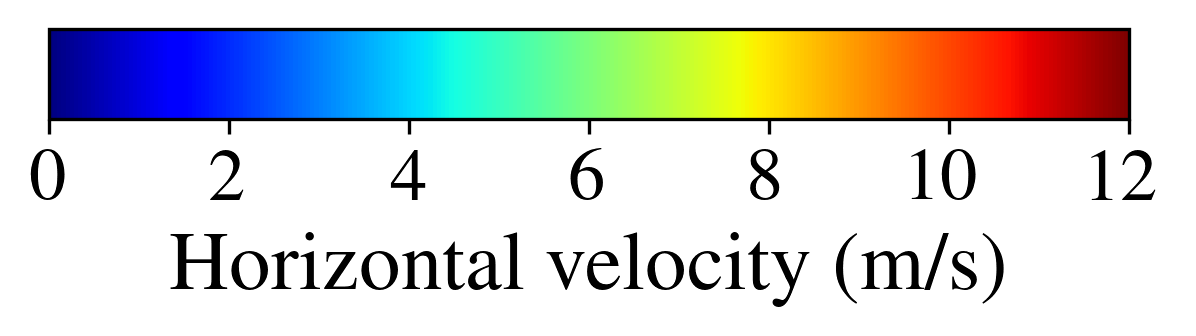}
    \end{subfigure} 
    \hspace{0.04\textwidth}
    \begin{subfigure}[b]{0.24\textwidth}
        \includegraphics[width=\textwidth]{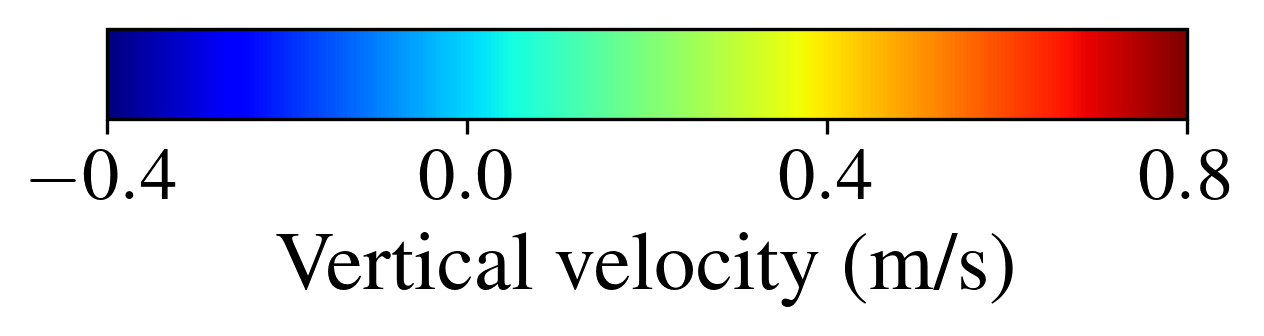}
    \end{subfigure}        
    \caption{Instantaneous horizontal and vertical velocity fields in the Witch of Agnesi test, computed using the embedded boundary approach.}\label{fig:WoA-velocity}
\end{figure}

Figure~\ref{fig:WoA-u-time} presents the time evolution of the horizontal velocity at three locations along the mid-plane ($x=0.5L$) at heights of 110, 300, and 500 m, computed using the three grid configurations. The flow develops from the zero initial state and reaches a steady state within approximately four hours. The results obtained from the three grids are in close agreement throughout the entire simulation, which demonstrates that the EB and AMR approaches produce solutions consistent with the terrain-fitted simulation. Instantaneous horizontal and vertical velocity fields at steady state are shown in Figure~\ref{fig:WoA-velocity}. The velocity fields remain smooth across cut cells near the embedded terrain, which indicates that the WSRD scheme provides a consistent and stable treatment of cut cells.

\begin{figure}[h!]
    \captionsetup[subfigure]{justification=centering}
    \centering
    \begin{subfigure}[b]{0.28\textwidth}
        \includegraphics[width=\textwidth]{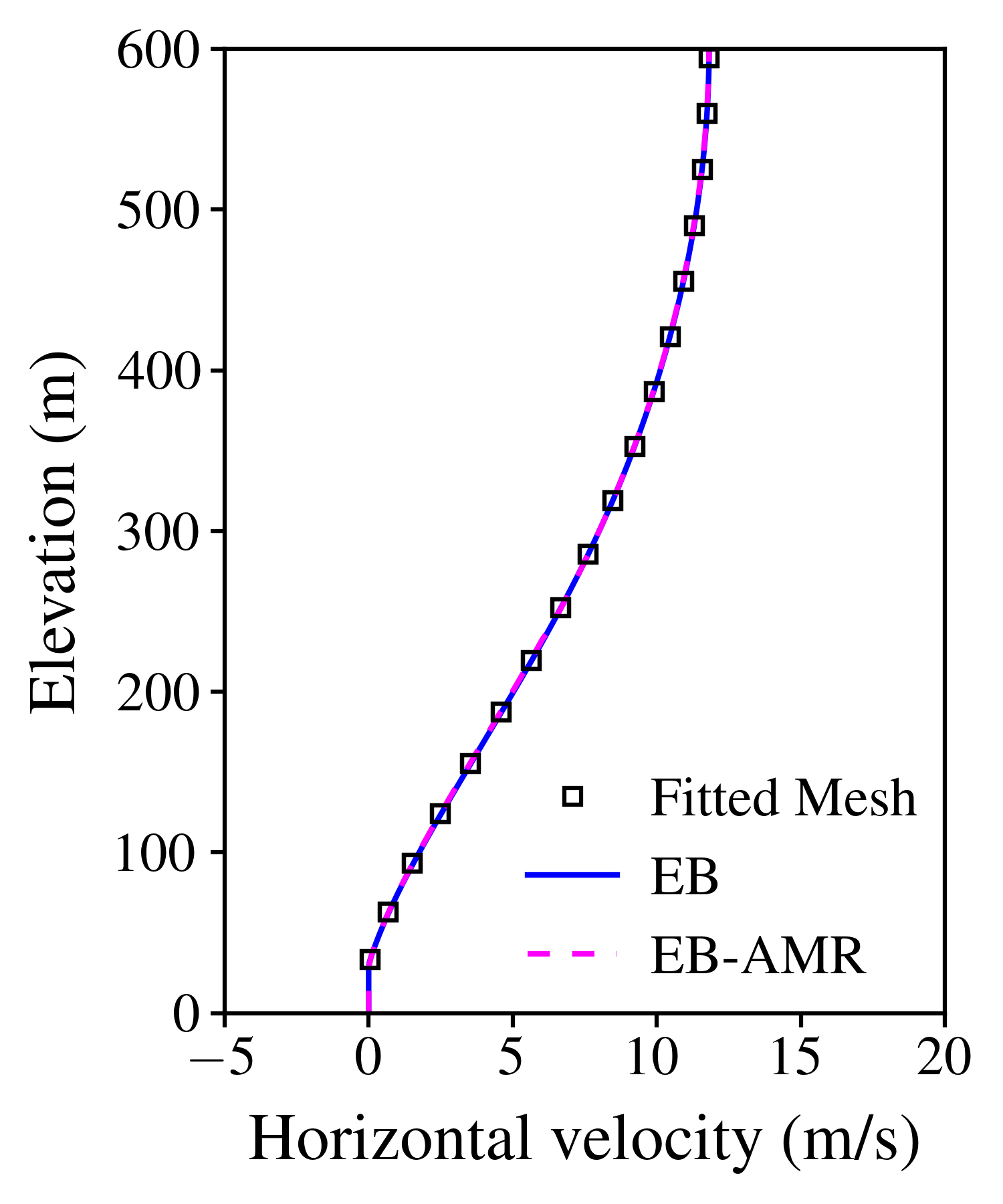}
        \caption{$x=0.25L$}
    \end{subfigure}
    \begin{subfigure}[b]{0.28\textwidth}
        \includegraphics[width=\textwidth]{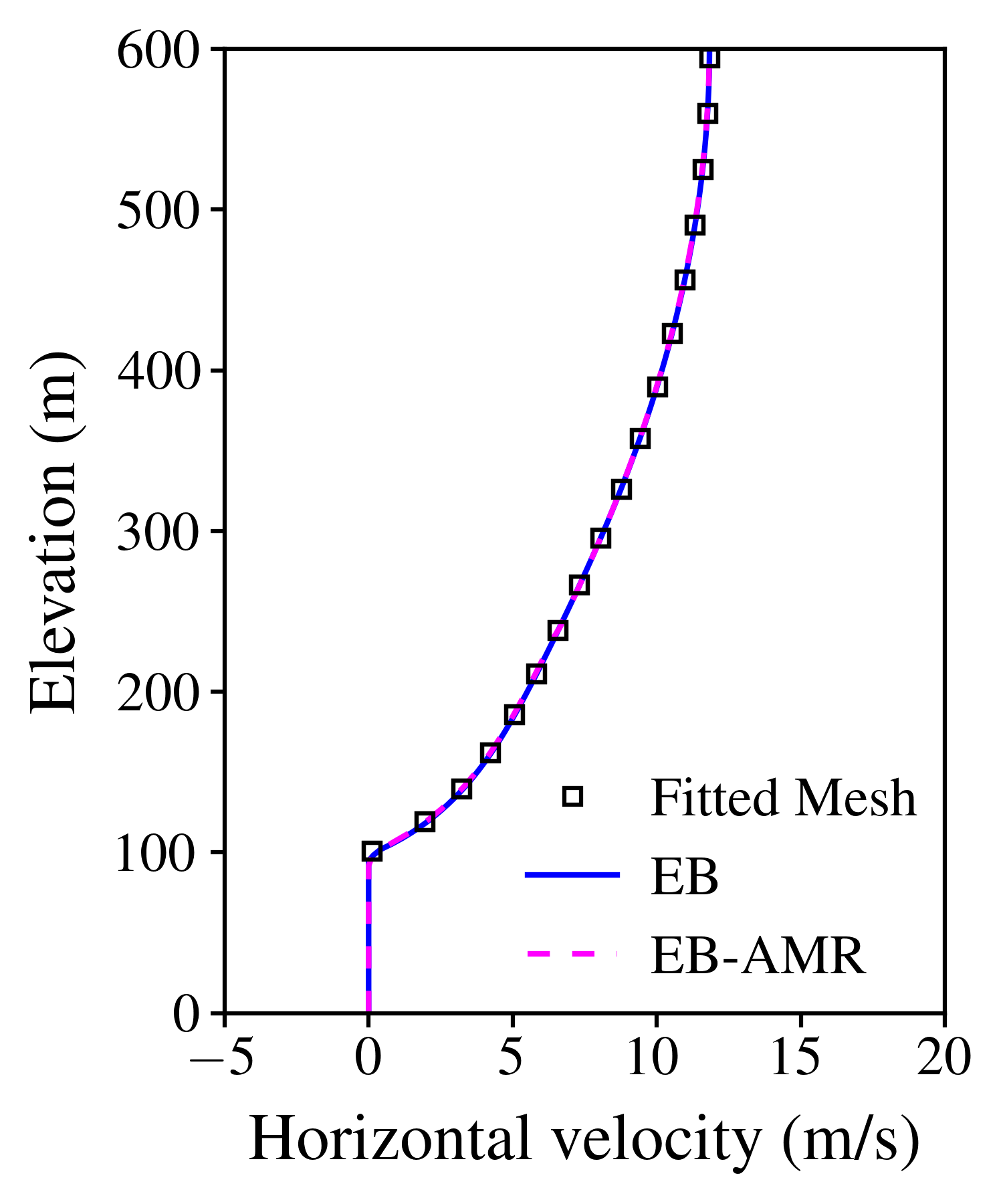}
        \caption{$x=0.5L$}
    \end{subfigure}
    \begin{subfigure}[b]{0.28\textwidth}
        \includegraphics[width=\textwidth]{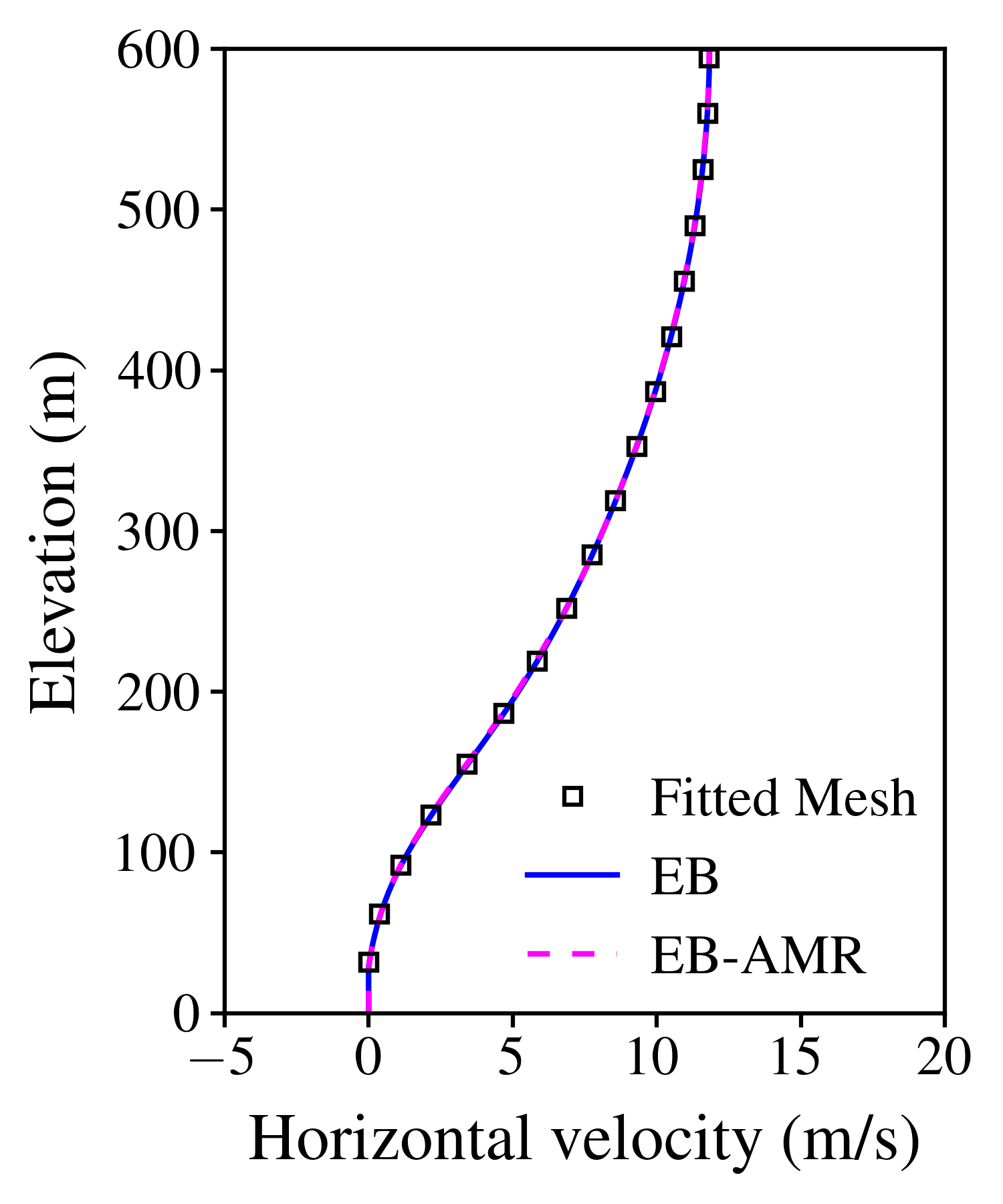}
        \caption{$x=0.75L$}
    \end{subfigure}
    \caption{Horizontal velocity profiles along vertical lines at $x=0.25L$, $0.5L$, and $0.75L$ after the solution reaches a steady state.}\label{fig:WoA-xvel-z}
\end{figure}

\begin{figure}[h!]
    \captionsetup[subfigure]{justification=centering}
    \centering
    \begin{subfigure}[b]{0.28\textwidth}
        \includegraphics[width=\textwidth]{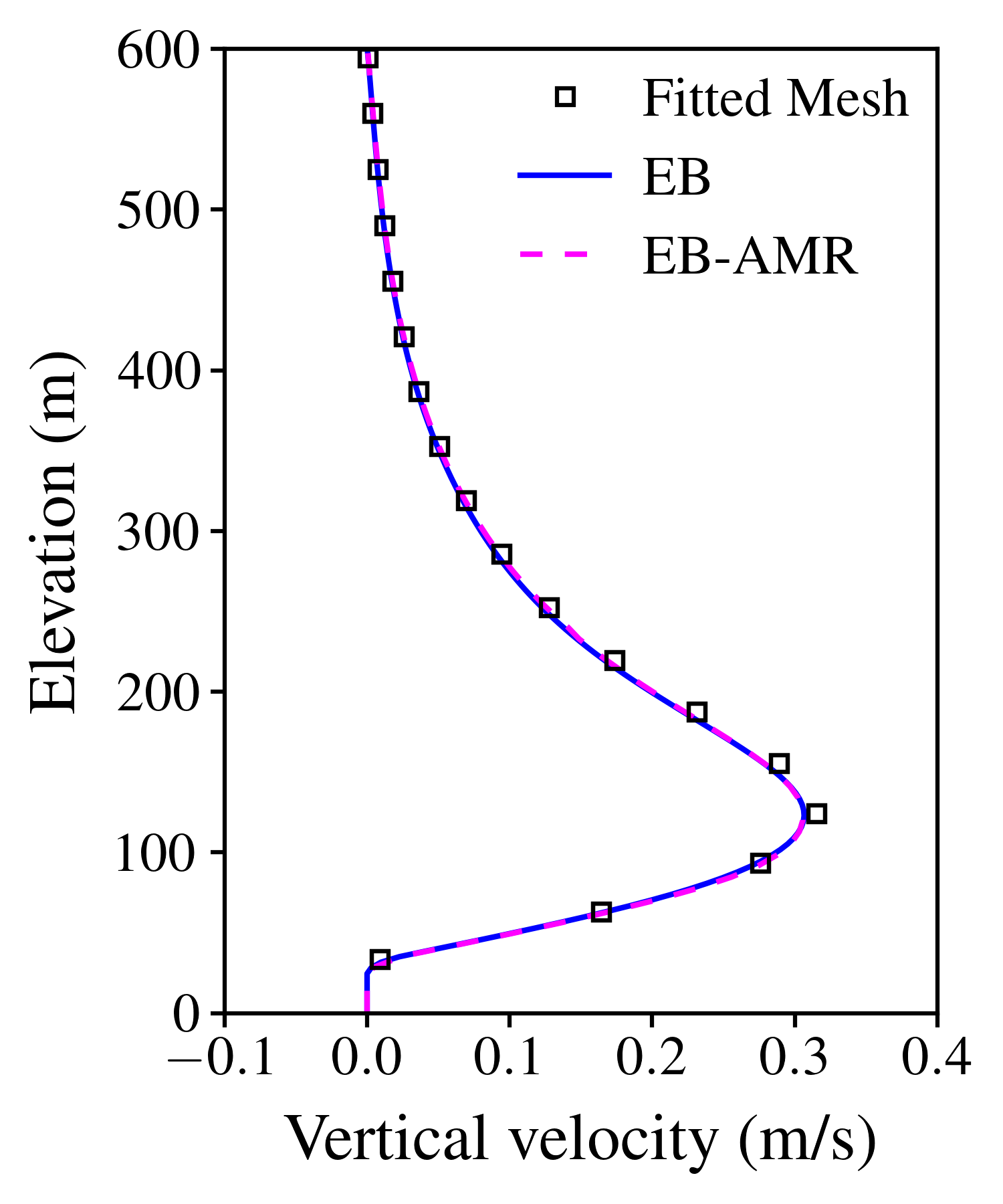}
        \caption{$x=0.25L$}
    \end{subfigure}
    \begin{subfigure}[b]{0.28\textwidth}
        \includegraphics[width=\textwidth]{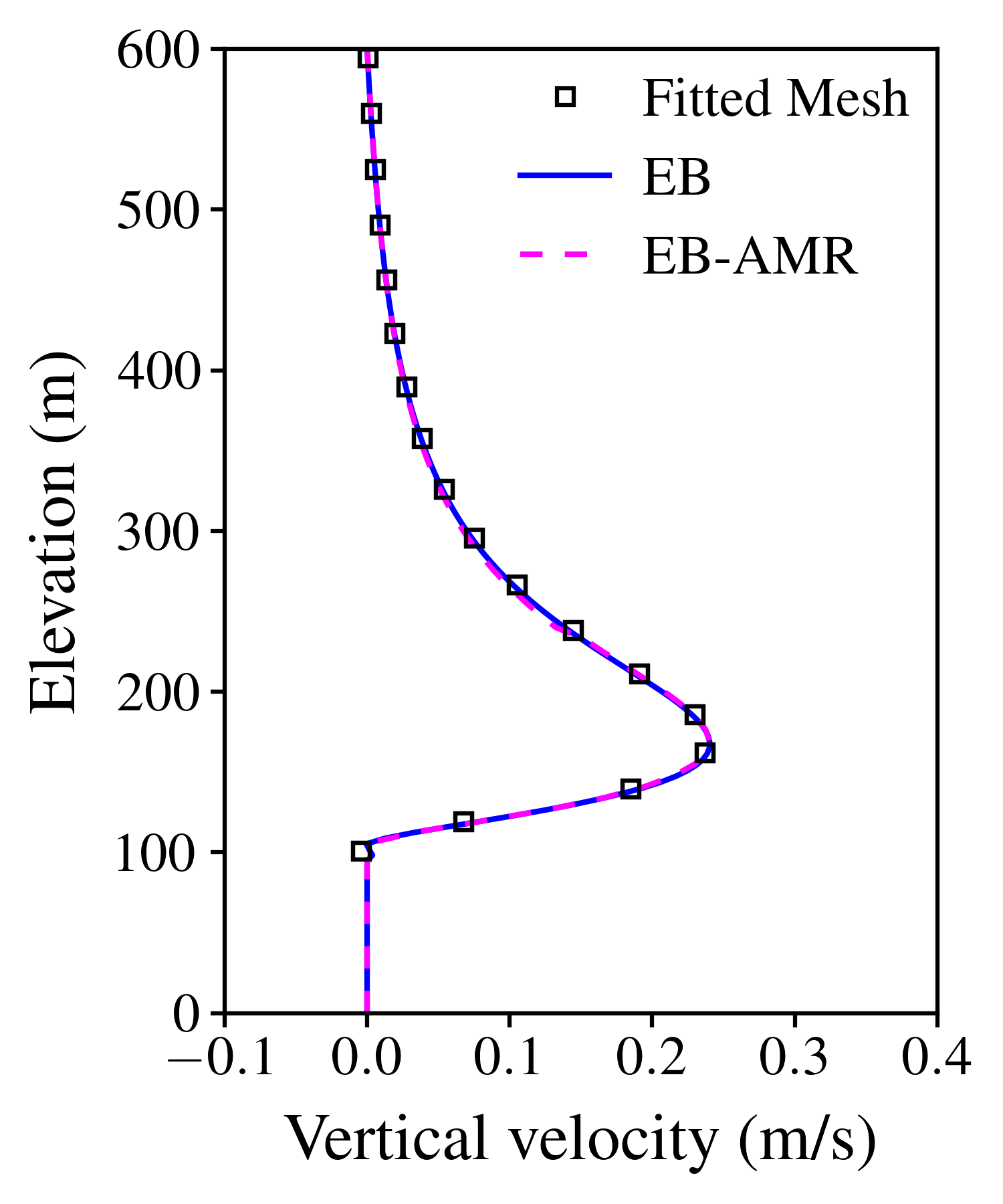}
        \caption{$x=0.5L$}
    \end{subfigure}
    \begin{subfigure}[b]{0.28\textwidth}
        \includegraphics[width=\textwidth]{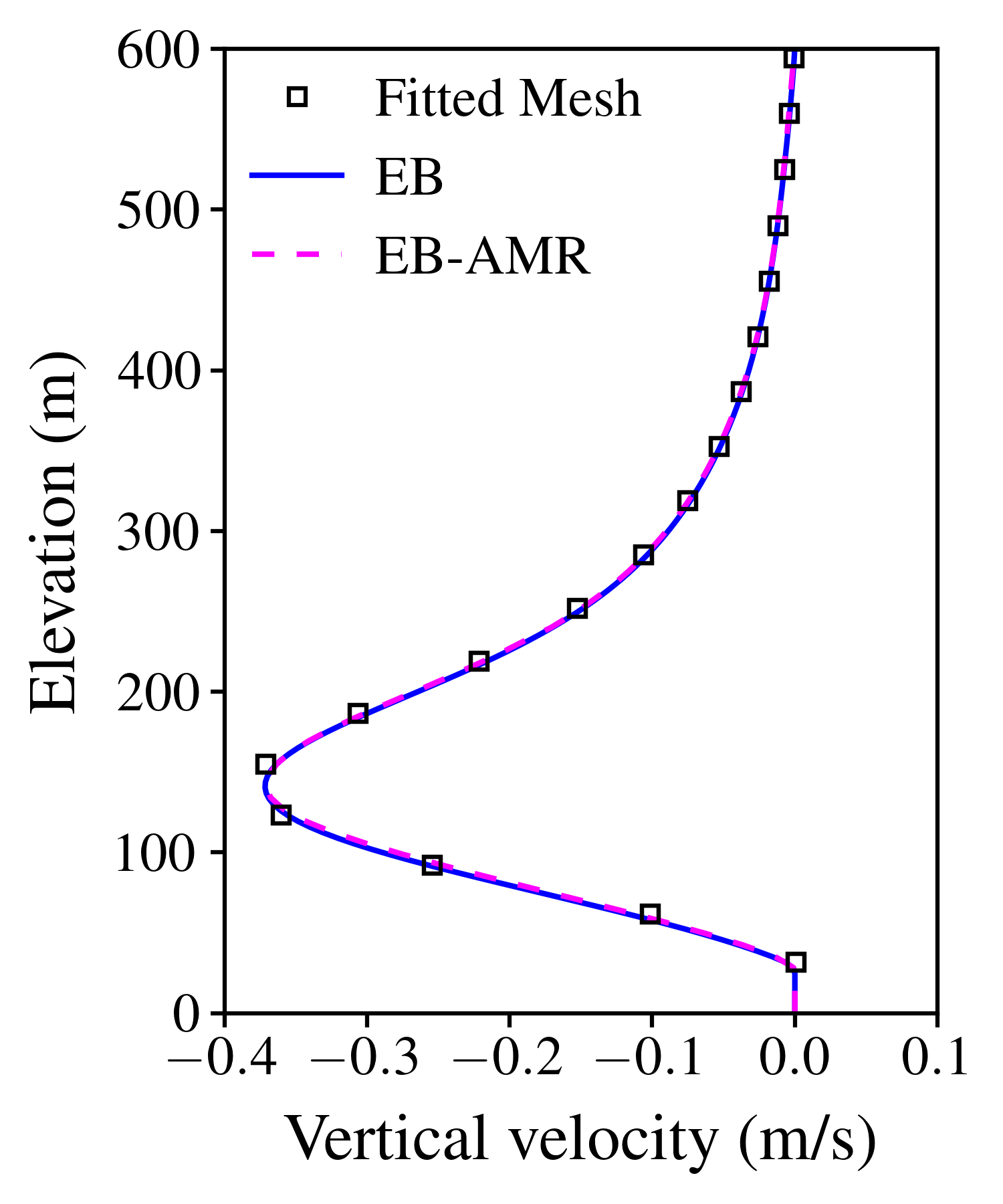}
        \caption{$x=0.75L$}
    \end{subfigure}
    \caption{Vertical velocity profiles along vertical lines at $x=0.25L$, $0.5L$, and $0.75L$ after the solution reaches a steady state.}\label{fig:WoA-zvel-z}
\end{figure}

Figures~\ref{fig:WoA-xvel-z} and \ref{fig:WoA-zvel-z} compare the steady-state horizontal and vertical velocity profiles obtained using the terrain-fitted, embedded boundary, and EB with AMR approaches along vertical lines at  $x=0.25L$, $0.5L$, and $0.75L$. The terrain-fitted and EB approaches show good agreement from the hill surface up to the domain top. The EB approach accurately captures the shear layer near the viscous no-slip boundary. This agreement confirms the accuracy of the EB method supported by stabilization provided by the WSRD scheme. The EB-AMR results almost overlap with the EB results, highlighting that the coarse and fine grids are seamlessly coupled across the interface.

\begin{figure}[h!]
    \captionsetup[subfigure]{justification=centering}
    \centering
    \begin{subfigure}[b]{0.47\textwidth}
        \includegraphics[width=\textwidth]{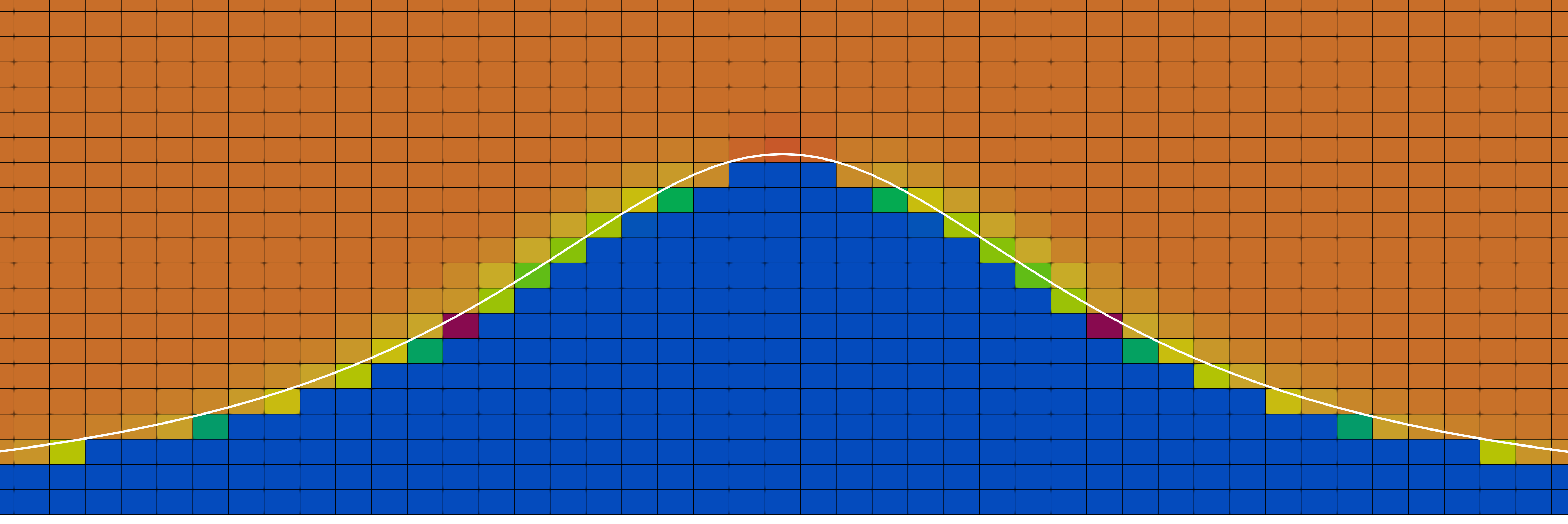}
        \caption{Without WSRD}
    \end{subfigure}
    \hspace{0.02\textwidth}
    \begin{subfigure}[b]{0.47\textwidth}
        \includegraphics[width=\textwidth]{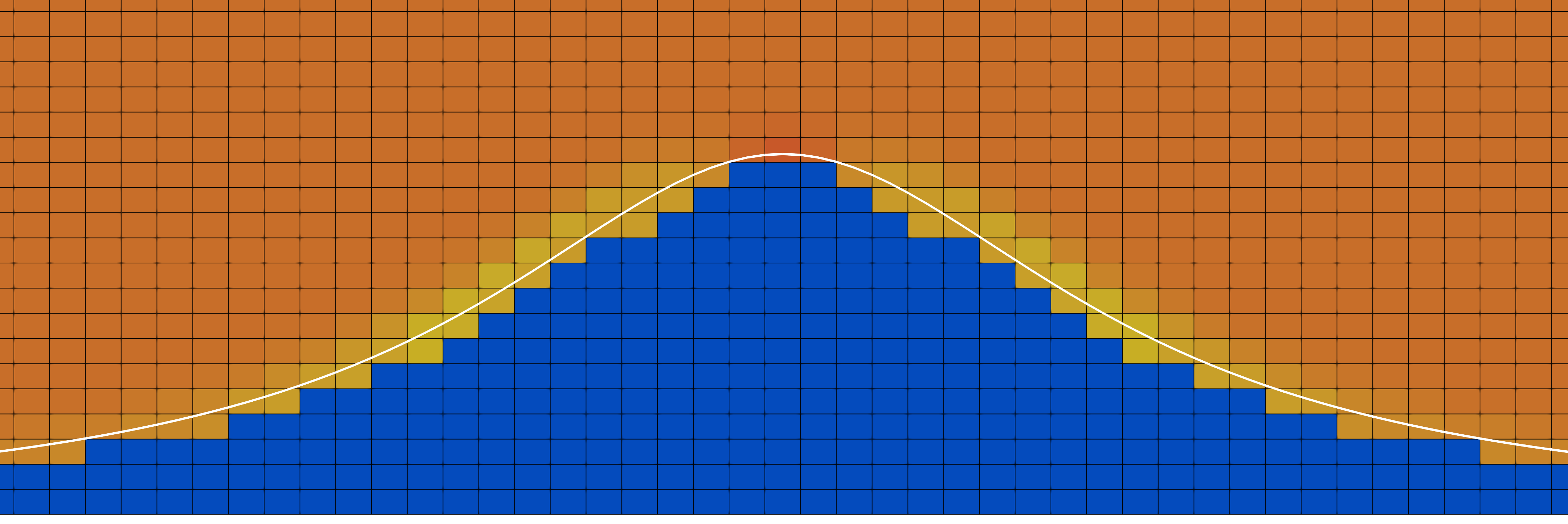}
        \caption{With WSRD}
    \end{subfigure} 
    \\
    \begin{subfigure}[b]{0.24\textwidth}
    \includegraphics[width=\textwidth]{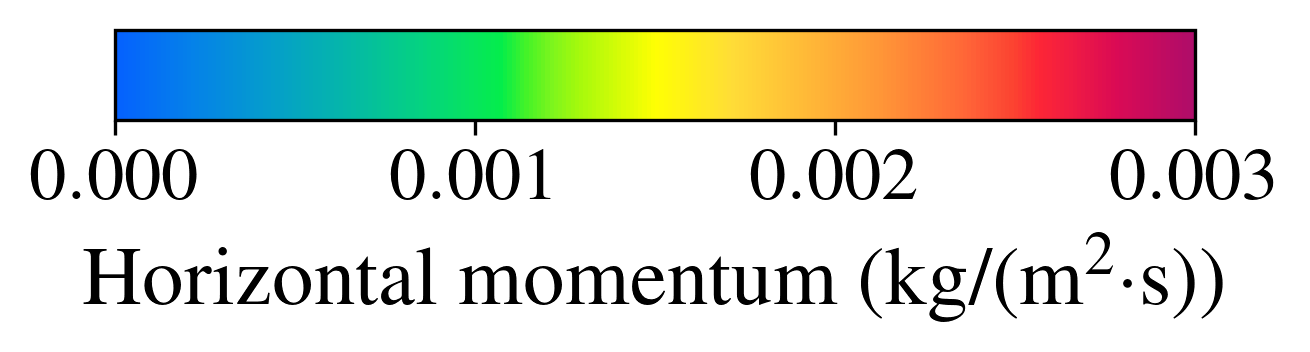}
    \end{subfigure}
    \caption{Distribution of $x$-momentum with and without applying the weighted state redistribution scheme. The terrain profile is indicated by the white line.}\label{fig:WoA-SRD}
\end{figure}

Figure~\ref{fig:WoA-SRD} highlights the effect of applying the WSRD scheme. The scheme gathers and redistributes the state variables across the neighborhood of cut cells based on their volume fractions, as detailed in Section~\ref{sec:WSRD}. As a result, it smooths the field of state variables while conserving the total mass. This process is invariably necessary for all state variables to achieve stable solutions. To examine the stabilization effect of WSRD, the EB simulation is run on a coarse grid with half the original resolution for five time steps, both with and without applying the WSRD scheme. Figure~\ref{fig:WoA-SRD} shows the distributions of the $x$-momentum on the $x$-staggered grid at the fifth time step. In fully covered cells below the terrain profile, the momentum variable is not updated from the zero initial value. Without WSRD, nonphysical values appear in cut cells with very small volume fractions, i.e., cells containing only a small portion of fluid, which leads to spurious oscillations. In contrast, applying WSRD yields a smooth momentum field and a stable solution.

\subsection{Flow over a hemisphere}\label{sec:3d_hill}

This test case represents three-dimensional potential flow over a hemisphere for which an exact solution exists. The flow is assumed to be inviscid and irrotational. We model this problem using the compressible equations without buoyancy forcing. The exact solutions for the radial and polar velocity components in the spherical coordinates are given by
\begin{align}
    u_r (r,\theta) &= - u_{\infty} \qty(1-\frac{a^3}{r^3})\cos\theta, \\
    u_{\theta} (r,\theta) &= u_{\infty} \qty(1+\frac{a^3}{2r^3})\sin\theta,
\end{align}
where $r$ is the radial distance from the center of the sphere, $\theta$ is the polar angle, $a$ is the radius of the sphere, and $u_{\infty}$ is the free stream velocity. Note that the azimuthal velocity component is zero.

\begin{figure}[h!]
    \centering
        \includegraphics[width=0.5\textwidth]{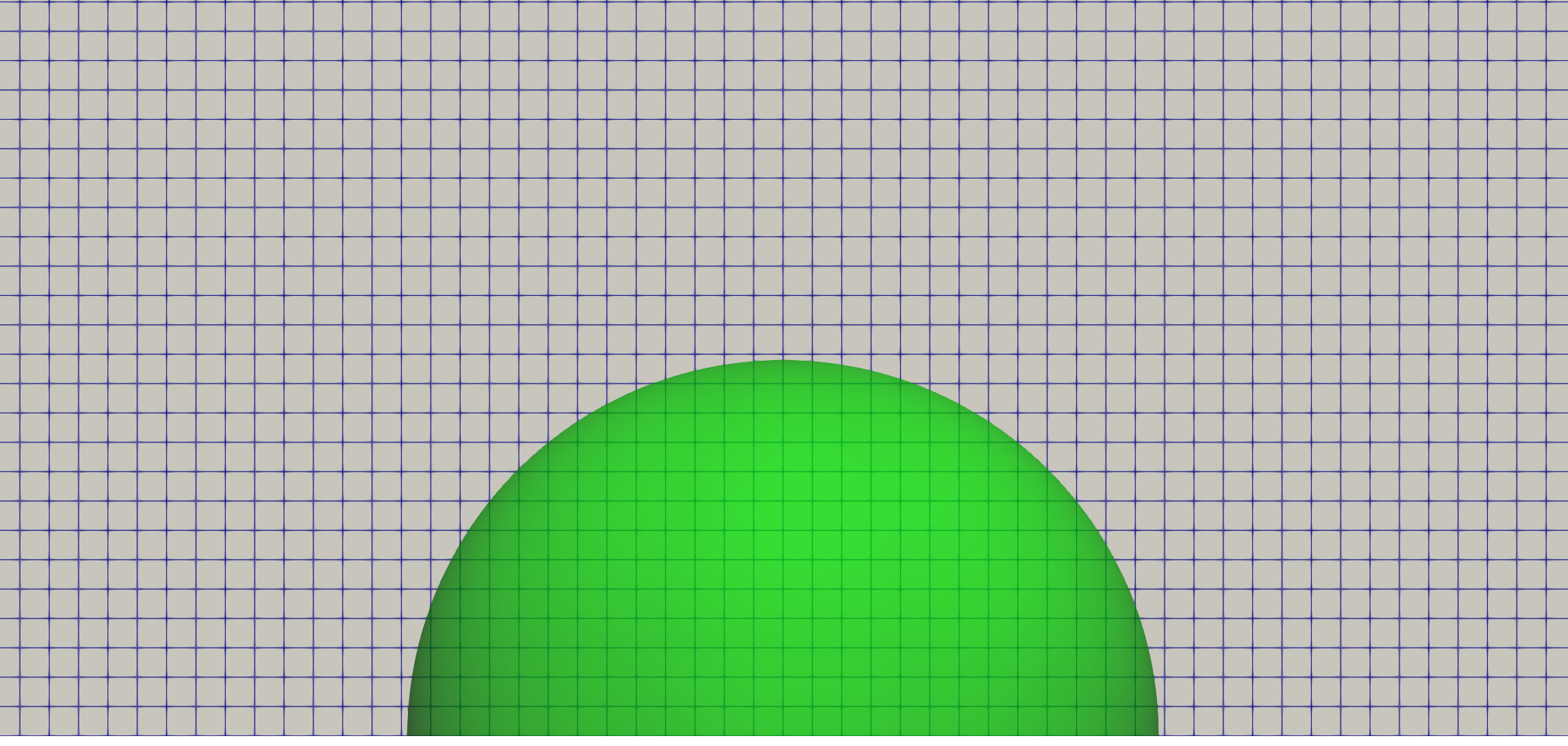}
    \caption{Computational grid around an embedded hemisphere.}\label{fig:HS-mesh}
\end{figure}

For the numerical test, the computational domain is a cubic box with an edge length of 10. The streamwise direction is aligned with the $x$-axis. A hemisphere of radius $a=0.5$ is placed at the center of the bottom surface and is modeled using an embedded boundary, as shown in Figure~\ref{fig:HS-mesh}. The inflow velocity is prescribed by the free stream velocity $u_{\infty}=10$, while the periodic boundary condition is imposed on the lateral boundaries in the $y$-direction. The bottom and spherical surfaces are treated as slip walls. The computational domain is discretized using a uniform grid consisting of 256 cells along each edge. The third-order upwind scheme is employed to calculate the advective flux. An artificial viscosity of $\mu=1$ is applied to suppress spurious oscillations throughout the domain, except in the vicinity of the hemisphere. To prevent wave reflections, a sponge layer of thickness 2 is applied at the top and lateral boundaries.

\begin{figure}[h!]
    \captionsetup[subfigure]{justification=centering}
    \centering
    \begin{subfigure}[b]{0.5\textwidth}
        \includegraphics[width=\textwidth]{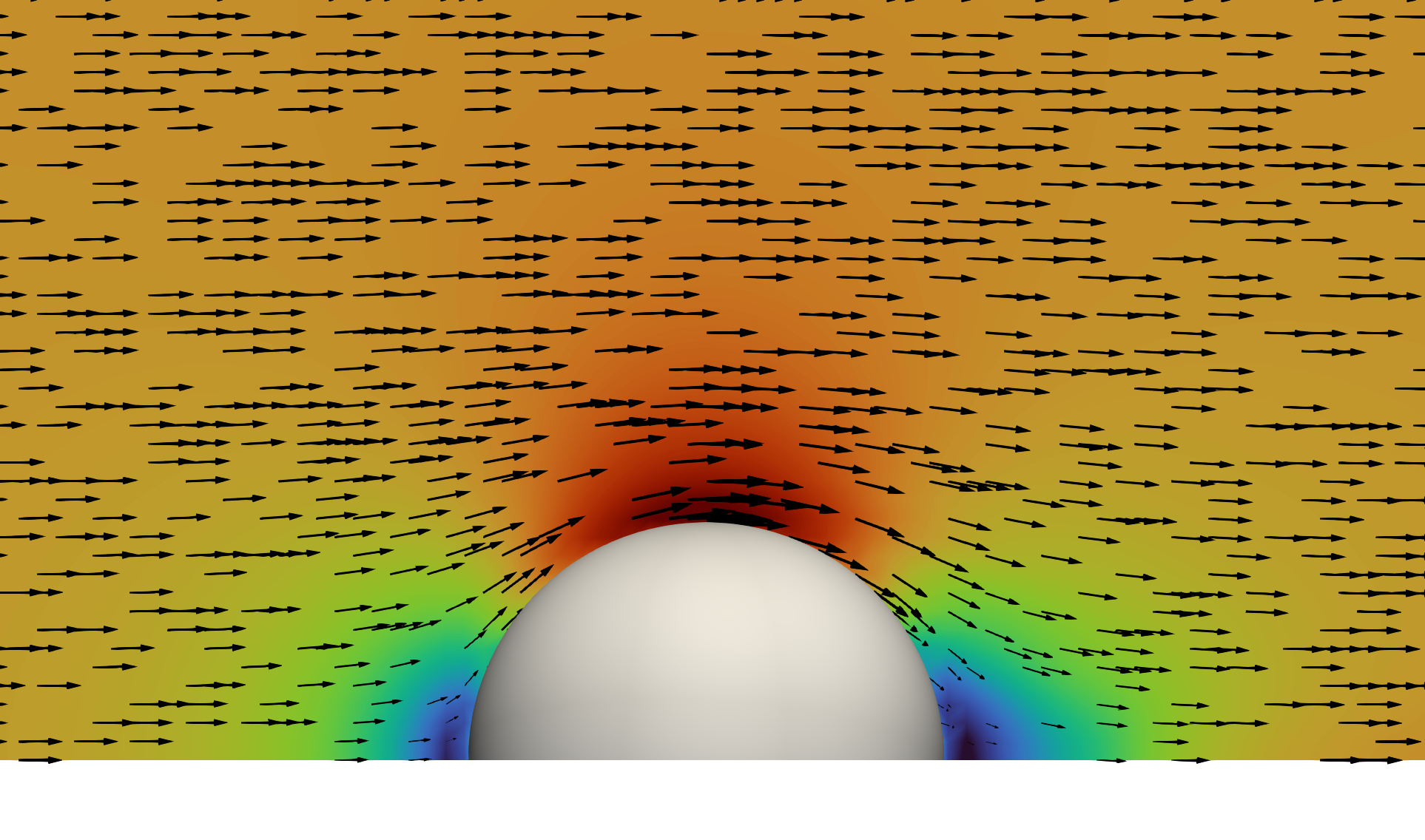}
    \end{subfigure}
    \\
    \begin{subfigure}[b]{0.24\textwidth}
        \includegraphics[width=\textwidth]{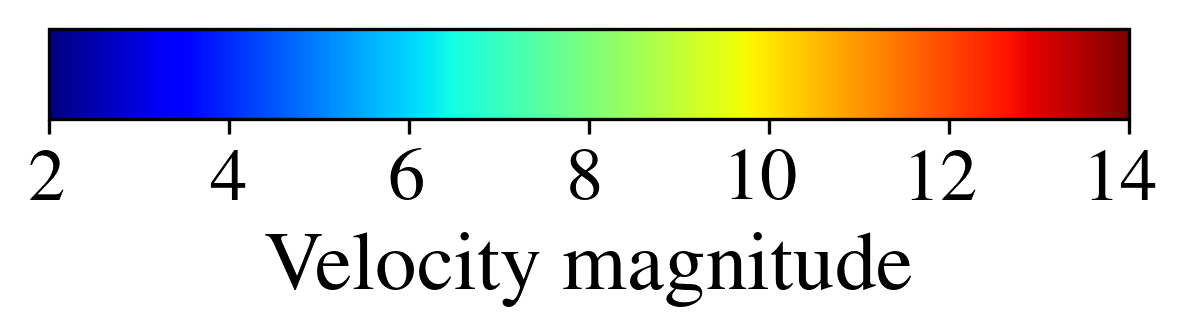}
    \end{subfigure} 
    \caption{Velocity field around an embedded hemisphere.}\label{fig:HS-vel-vector}
\end{figure}

\begin{figure}[h!]
    \captionsetup[subfigure]{justification=centering}
    \centering
    \begin{subfigure}[b]{0.35\textwidth}
        \includegraphics[width=\textwidth]{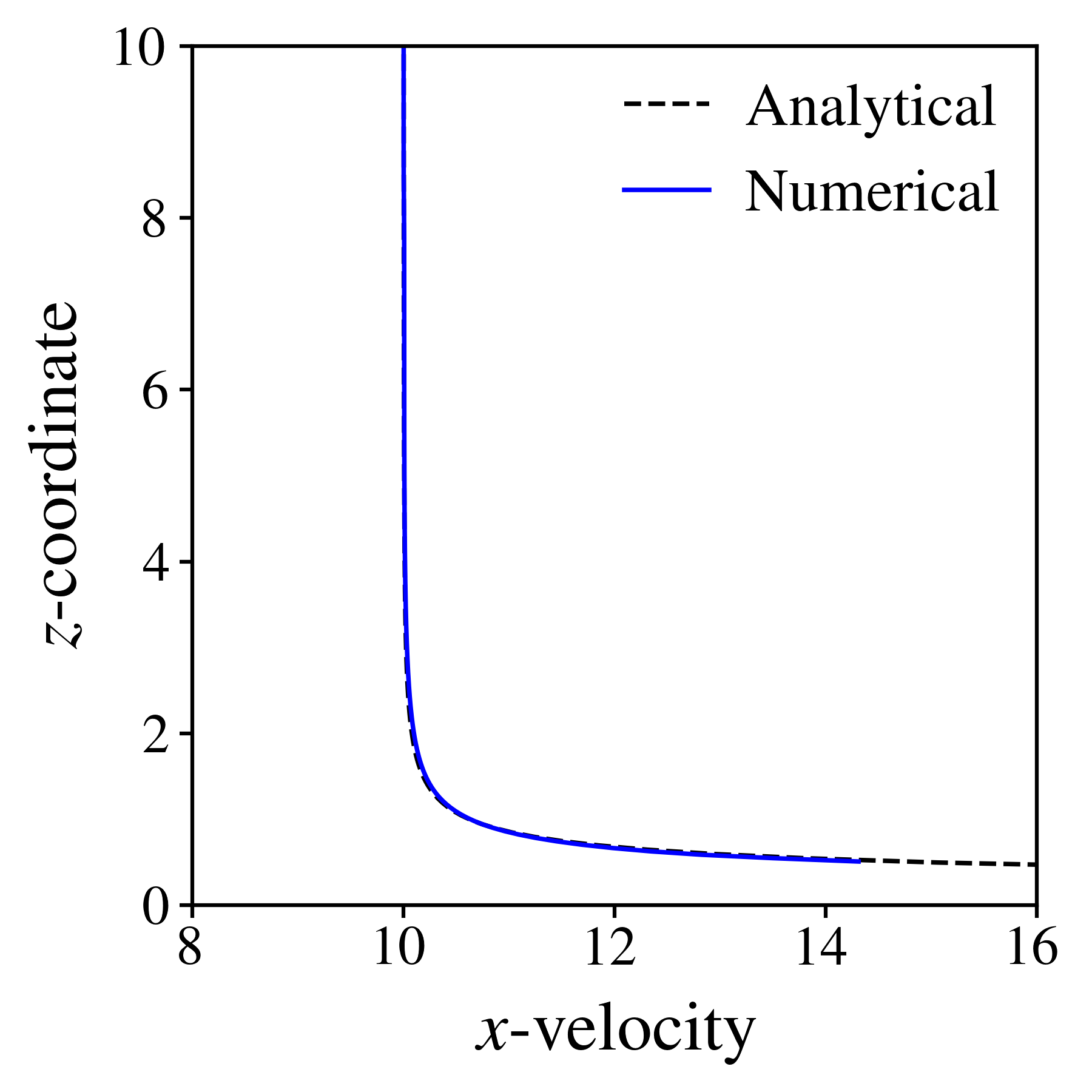}
        \caption{$x=5$}
    \end{subfigure}
    \begin{subfigure}[b]{0.35\textwidth}
        \includegraphics[width=\textwidth]{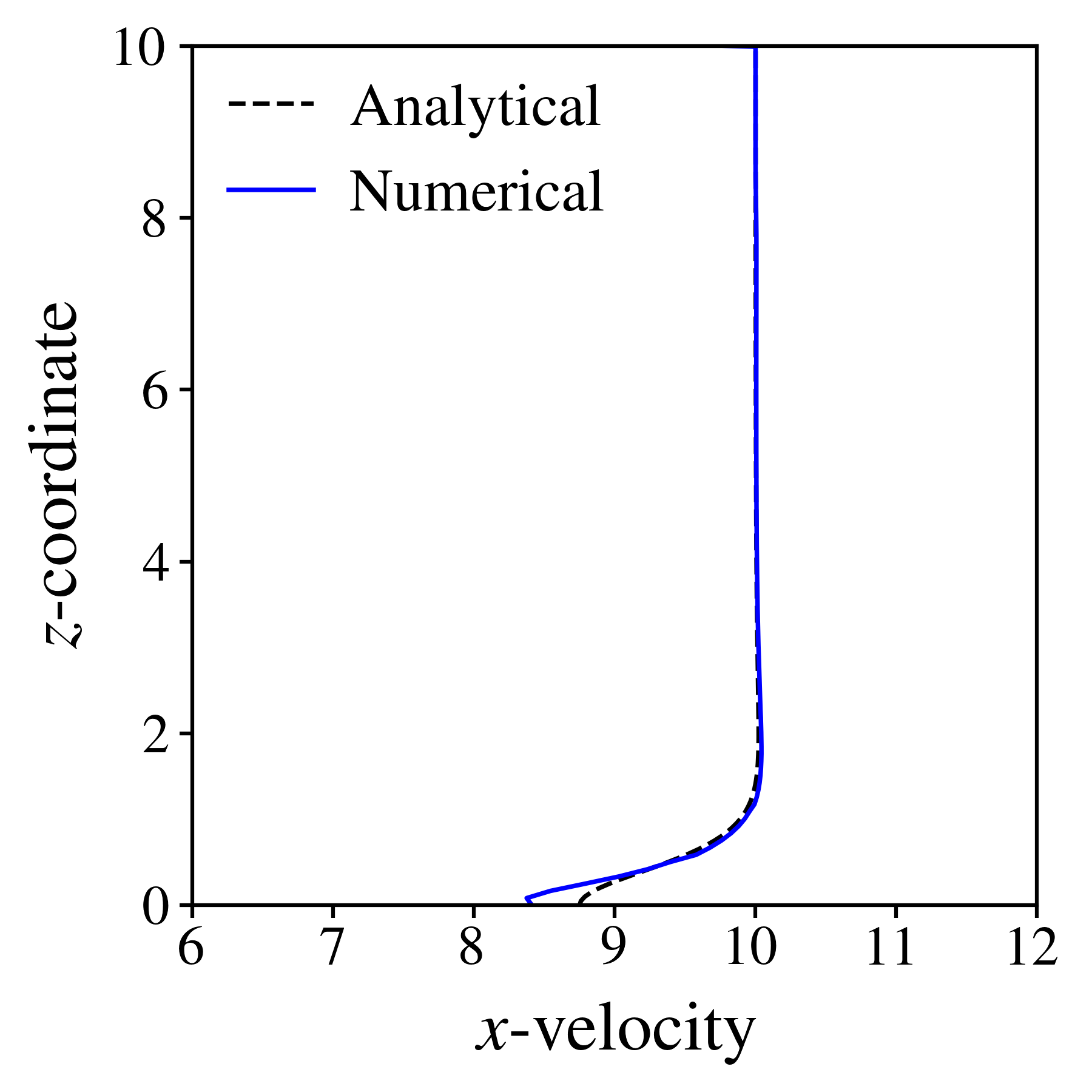}
        \caption{$x=6$}
    \end{subfigure} 
    \caption{Exact and numerical solutions for $x$-velocity along the $z$-axis at $x=5$ and $x=6$.}\label{fig:HS-xvel}
\end{figure}

Figure~\ref{fig:HS-vel-vector} presents the velocity field around an embedded hemisphere. The results clearly show the stagnation on the windward side, flow acceleration over the top, and flow separation on the leeward side. 
Figure~\ref{fig:HS-xvel} compares the numerical and exact solutions for the streamwise $x$-velocity along the vertical axis. The profiles of numerical results closely match the exact solutions, which demonstrates the accuracy of the EB method. A discrepancy between the exact and numerical results appears near the bottom boundary in the wake at $x=6$. This deviation can be attributed to the artificial viscosity, which is necessary to maintain stability in the inviscid flow simulation. 

\begin{figure}[h!]
    \captionsetup[subfigure]{justification=centering}
    \centering
    \begin{subfigure}[b]{0.38\textwidth}
        \includegraphics[width=\textwidth]{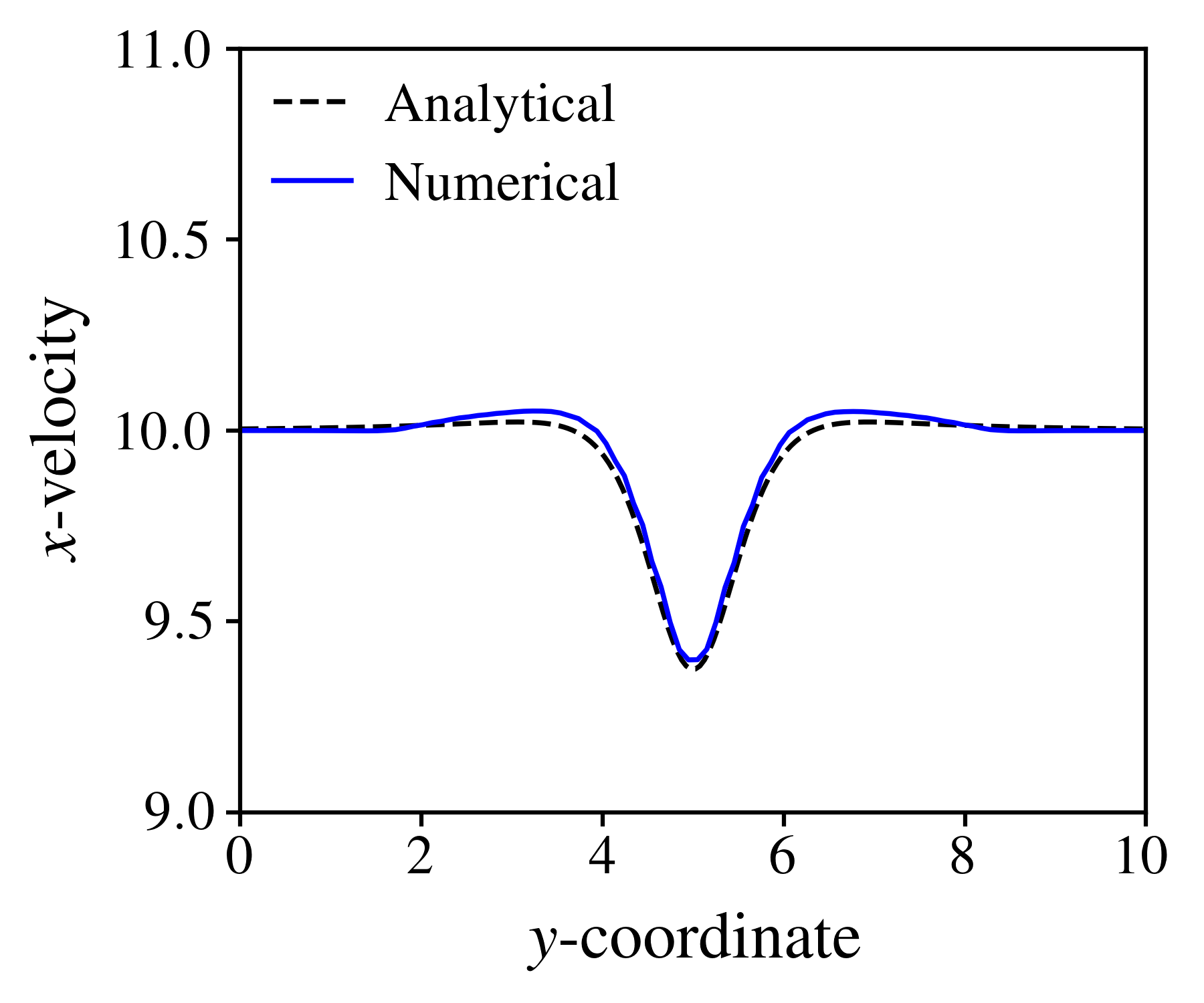}
        \caption{$x$-velocity}
    \end{subfigure}
    \begin{subfigure}[b]{0.38\textwidth}
        \includegraphics[width=\textwidth]{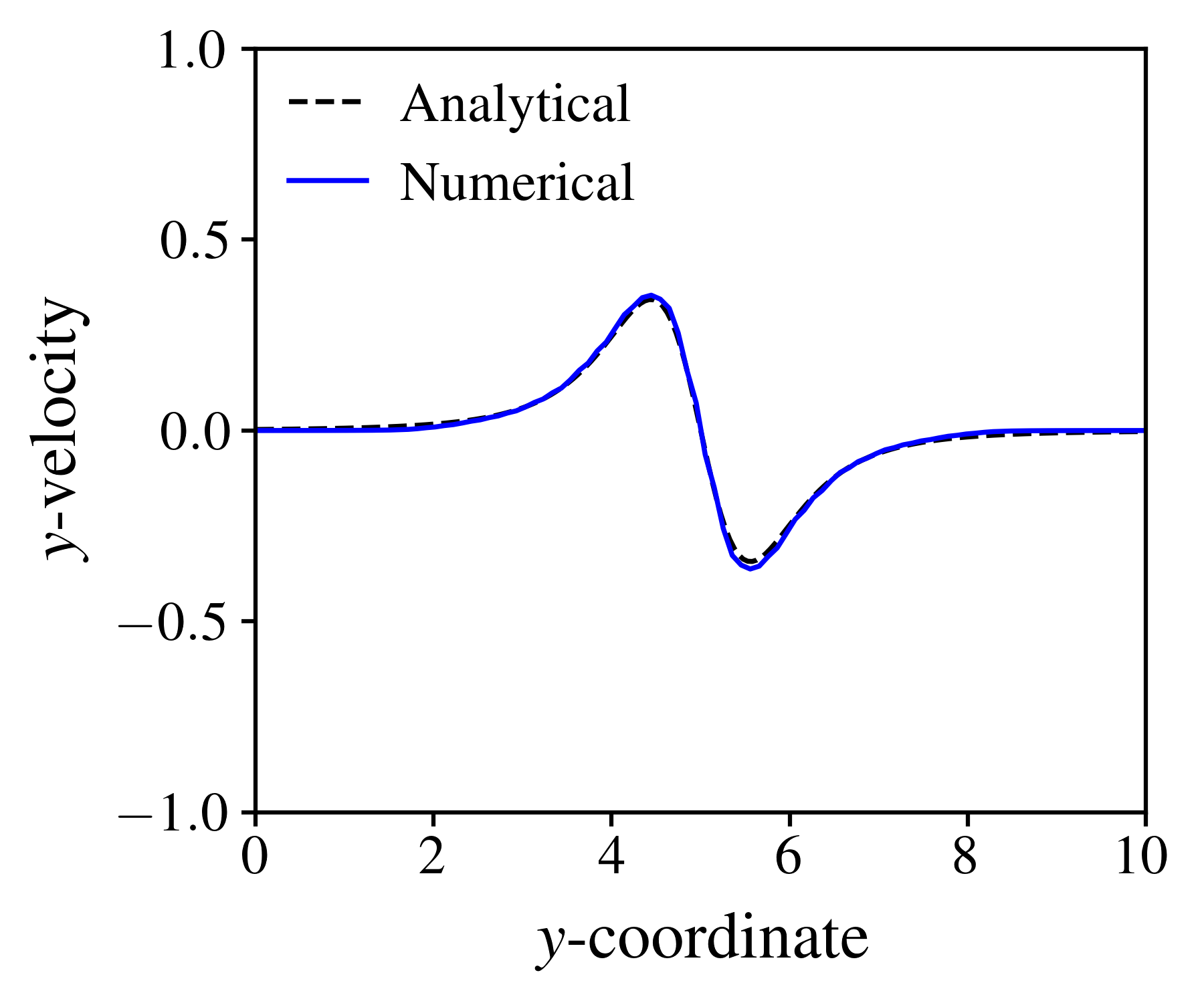}
        \caption{$y$-velocity}
    \end{subfigure} 
    \caption{Exact and numerical solutions for the streamwise $x$- and spanwise $y$-velocity components along the $y$-axis at $x=6$ and $z=0.5$.}\label{fig:HS-xvel-yaxis}
\end{figure}

Figure~\ref{fig:HS-xvel-yaxis} presents streamwise and crosswind velocity components along the horizontal axis at $x=6$ and $z=0.5$ in the wake. The numerical results are in good agreement with the exact solution, which confirms that the EB approach accurately represents the hemisphere geometry and preserves the wake symmetry.

Table~\ref{table:HS-time} compares the wall clock time for time step between the fitted-mesh and EB simulations. The computational cost was measured over 1000 time steps on four AMD MI300A GPUs and then averaged. The EB approach requires approximately 1.985 times more computational time per step than the fitted-mesh approach. This increased cost in the EB case may be attributed to the overhead of the WSRD, as well as the additional cost of computing advective fluxes and performing least-squares fitting for calculating diffusive fluxes. In practice, this additional cost is not expected to be prohibitive, as atmospheric flow simulations typically involve other computational processes that are substantially more expensive.


\begin{table}[h!]
\centering
\caption{Averaged wall clock time per a time step measured on 4 GPUs for flow over a hemisphere.}\label{table:HS-time}
\begin{tabular}{ccc}
\hline
 & Fitted Mesh & EB \\
\hline
Time per step (seconds) & 0.0341 & 0.0677 \\
\hline
\end{tabular}
\end{table}

\subsection{Flow past a wall-mounted square cylinder}\label{sec:SqCyl}

This test case investigates three-dimensional laminar flow around a truncated square cylinder \cite{dousset2010formation,saha2013unsteady} to validate the accuracy of the EB scheme. We consider the anelastic flow model without buoyancy forcing in this test. Reynolds number is set to $Re=$250. The numerical results obtained using a direct numerical simulation (DNS) approach \cite{saha2013unsteady} serves as a benchmark for comparison.

The computational domain is a three-dimensional rectangular channel with streamwise ($x$), transverse ($y$), and vertical ($z$) directions. A wall-mounted square cylinder of width $d=1$ and varying height $h$ is positioned on the bottom wall. The origin of the coordinate system is defined at the centroid of the cylinder base, located $6.1d$ downstream of the inlet. Three aspect ratios are investigated: $h/d=$3, 4, and 5. The length and width of the domain are $L=25.5d$ and $W=15.3d$, respectively. The domain height is $H=10.2$ for aspect ratio $h/d=3$ and 4, while $H=12.7$ is used for $h/d=5$ to maintain sufficient blockage ratio. The origin of the coordinate system is located at the center of the cylinder base, which is positioned $6d$ downstream from the inflow boundary.

\begin{figure}[h!]
    \centering
        \includegraphics[width=0.75\textwidth]{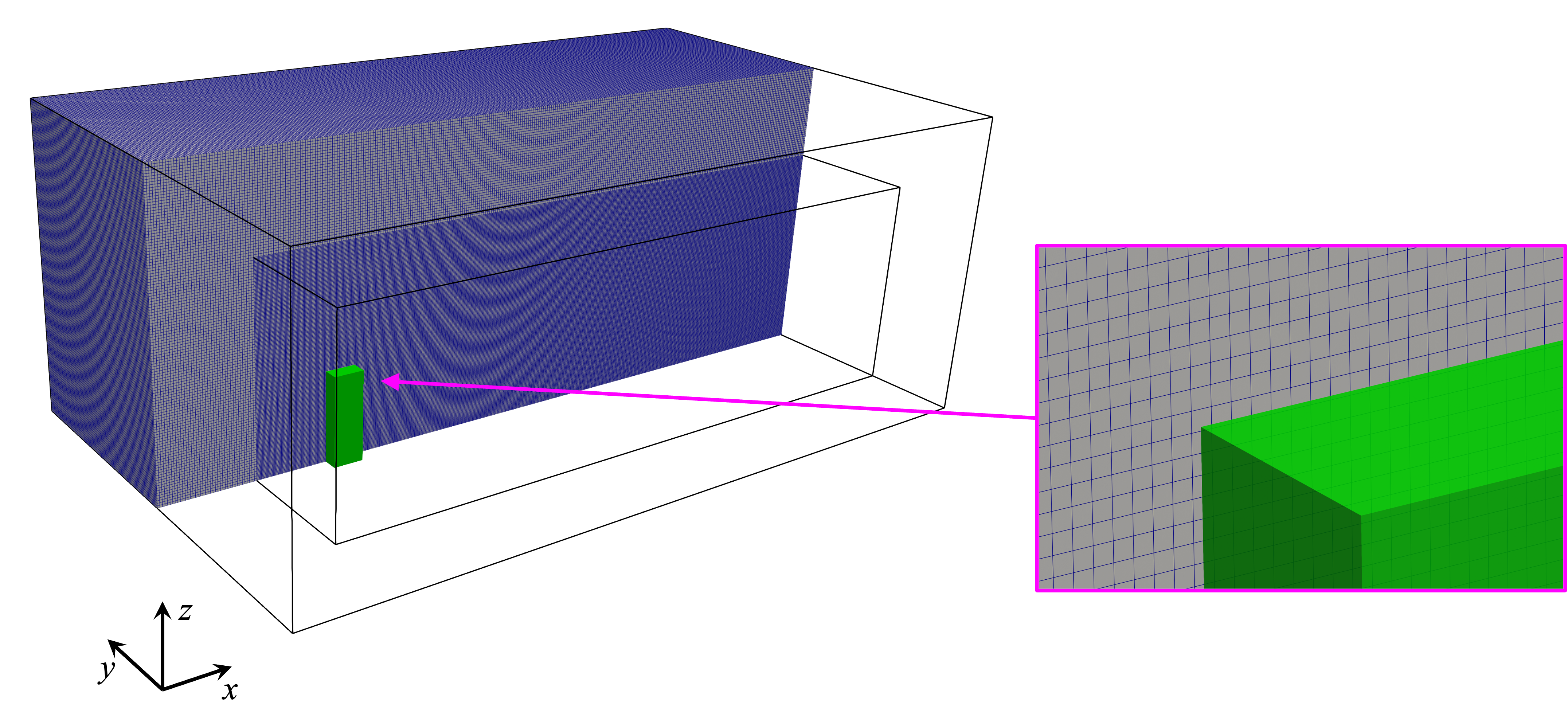}    
    \caption{Two-level refined grids and an embedded square cylinder with an aspect ratio of 
    $h/d=3$.}\label{fig:SqCyl-mesh}
\end{figure}

A uniform inflow velocity $u_{\infty}=1$ is imposed. No-slip boundary conditions are applied to the bottom wall and cylinder surface, while free-slip conditions are imposed on the lateral and top surfaces. The domain is discretized using a two-level adaptively refined mesh, as illustrated in Figure~\ref{fig:SqCyl-mesh}. To assess the EB scheme, the surface of the cylinder is misaligned with the underlying grid lines. All simulations are run at CFL$=$0.5.

\begin{figure}[h!]
    \captionsetup[subfigure]{justification=centering}
    \centering
    \begin{subfigure}[b]{0.32\textwidth}
        \includegraphics[width=\textwidth]{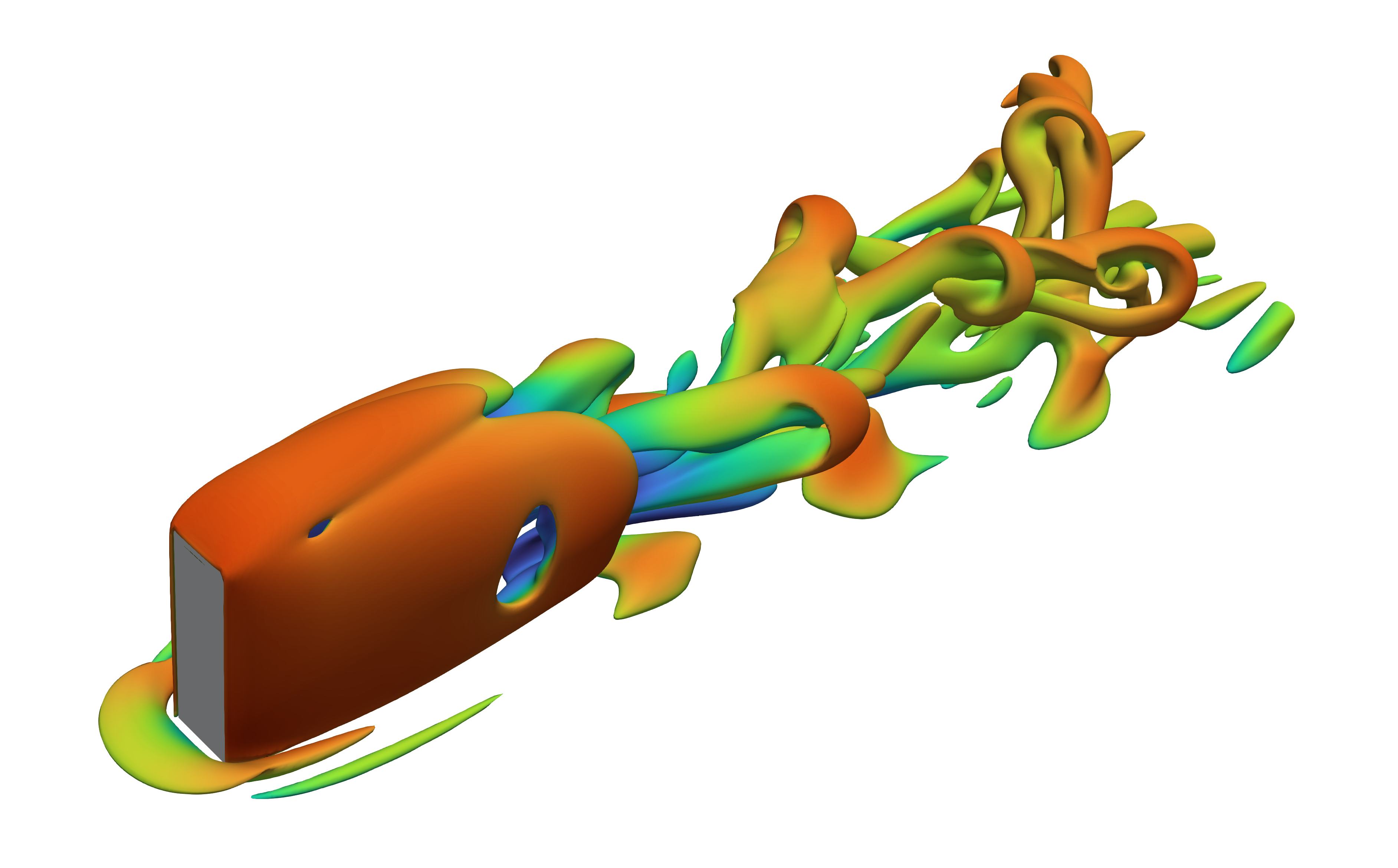}
        \caption{$h/d=3$}
    \end{subfigure}
    \begin{subfigure}[b]{0.32\textwidth}
        \includegraphics[width=\textwidth]{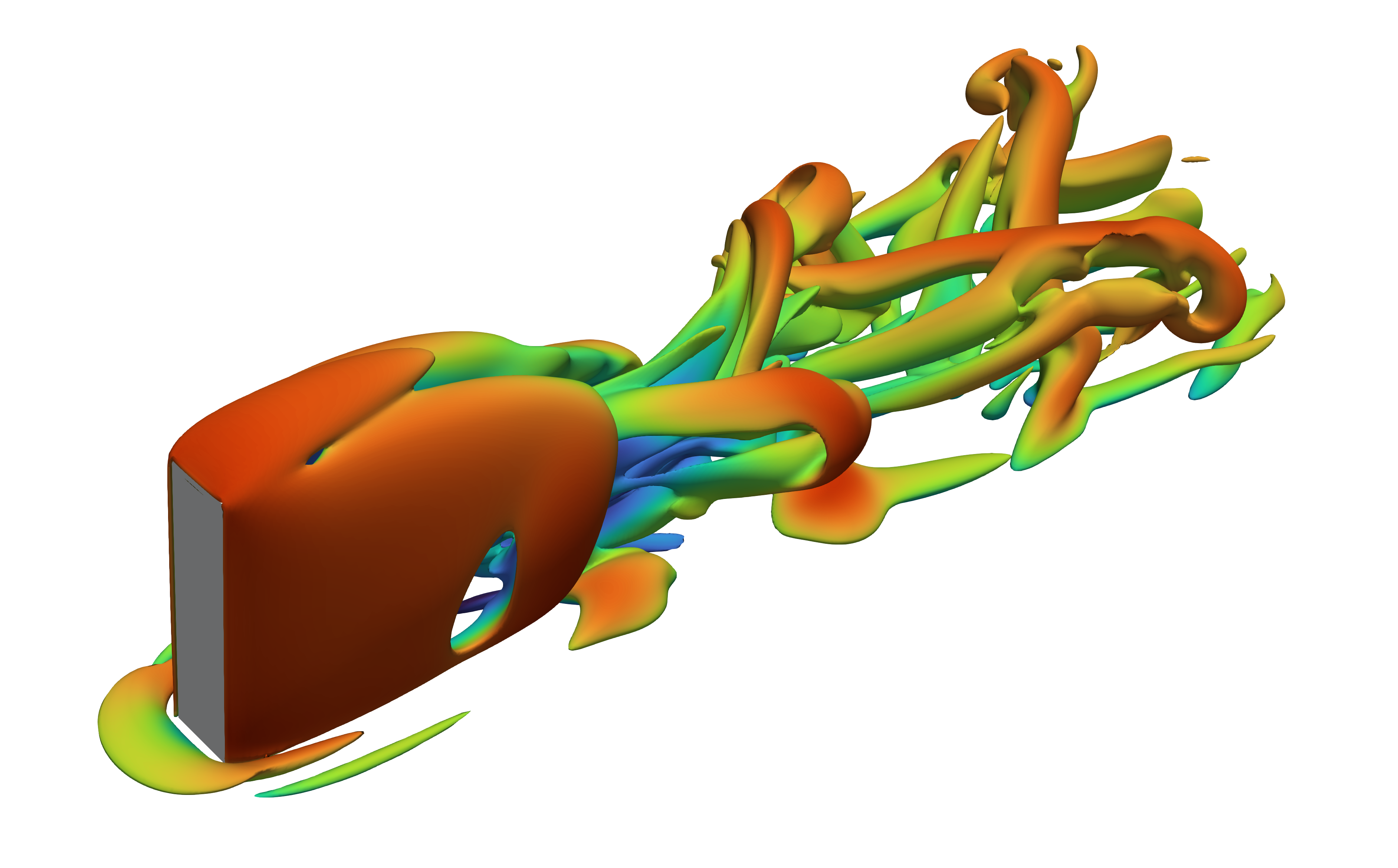}
        \caption{$h/d=4$}
    \end{subfigure}
    \begin{subfigure}[b]{0.32\textwidth}
        \includegraphics[width=\textwidth]{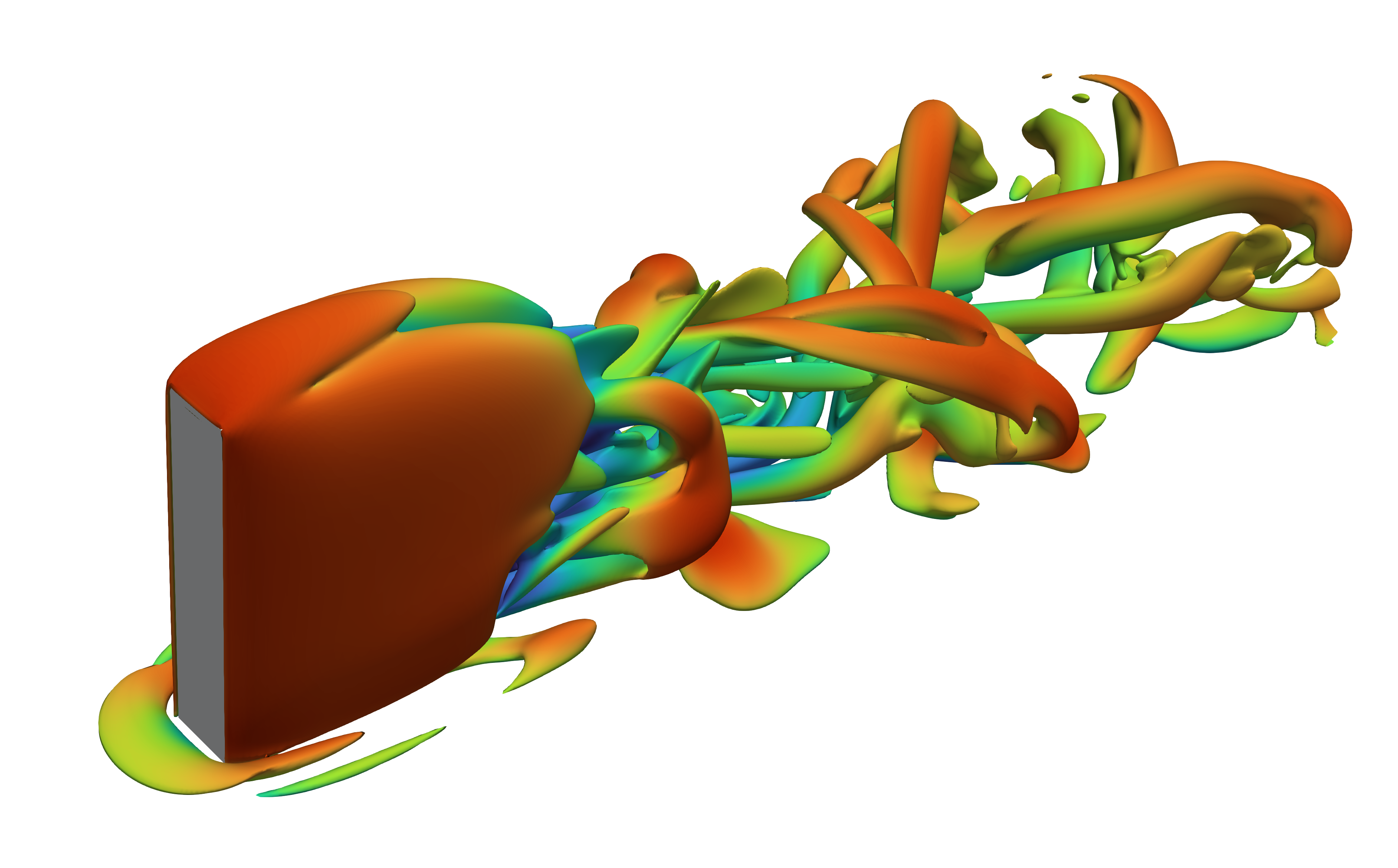}
        \caption{$h/d=5$}
    \end{subfigure}
    \\
    \begin{subfigure}[b]{0.24\textwidth}
        \includegraphics[width=\textwidth]{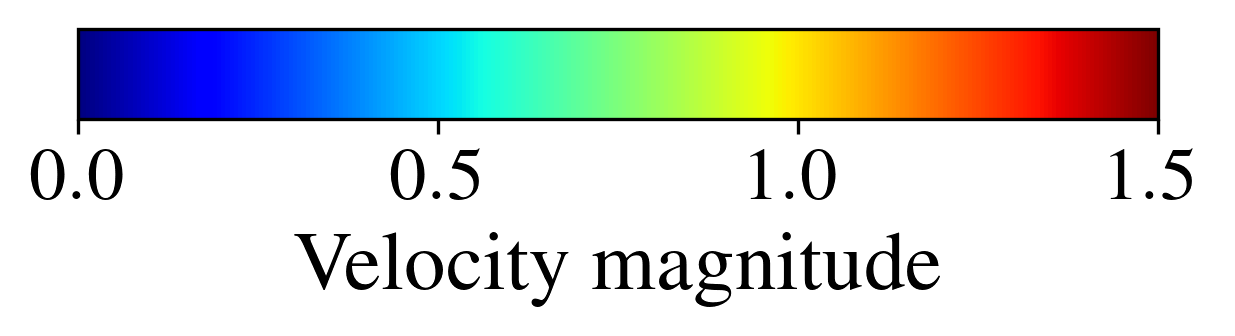}
    \end{subfigure}
    \caption{Vortical structures in the wake of square cylinders with different aspect ratios. Vortices are identified using $Q$-criterion iso-surfaces and colored by velocity magnitude.}\label{fig:SqCyl-Q}
\end{figure}

\begin{figure}[h!]
    \captionsetup[subfigure]{justification=centering}
    \centering
    \begin{subfigure}[b]{0.32\textwidth}
        \includegraphics[width=\textwidth]{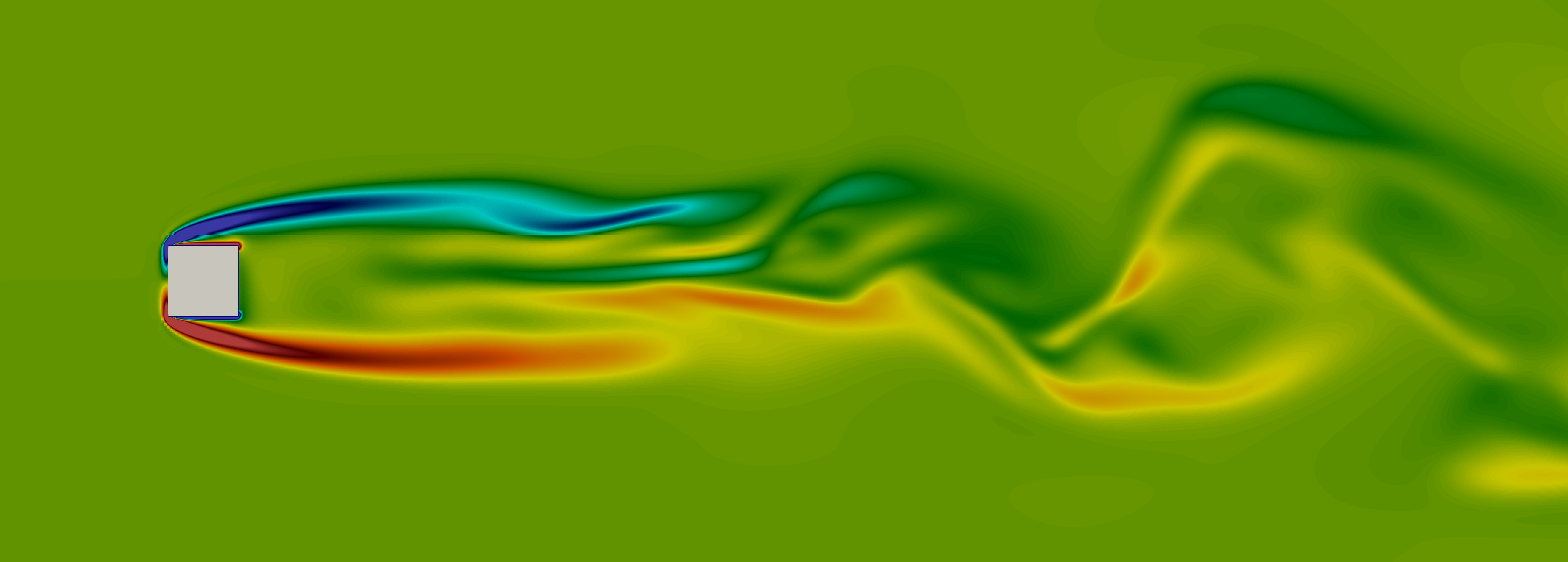}
        \caption{$z$-vorticity ($h/d=3$)}
    \end{subfigure}
    \begin{subfigure}[b]{0.32\textwidth}
        \includegraphics[width=\textwidth]{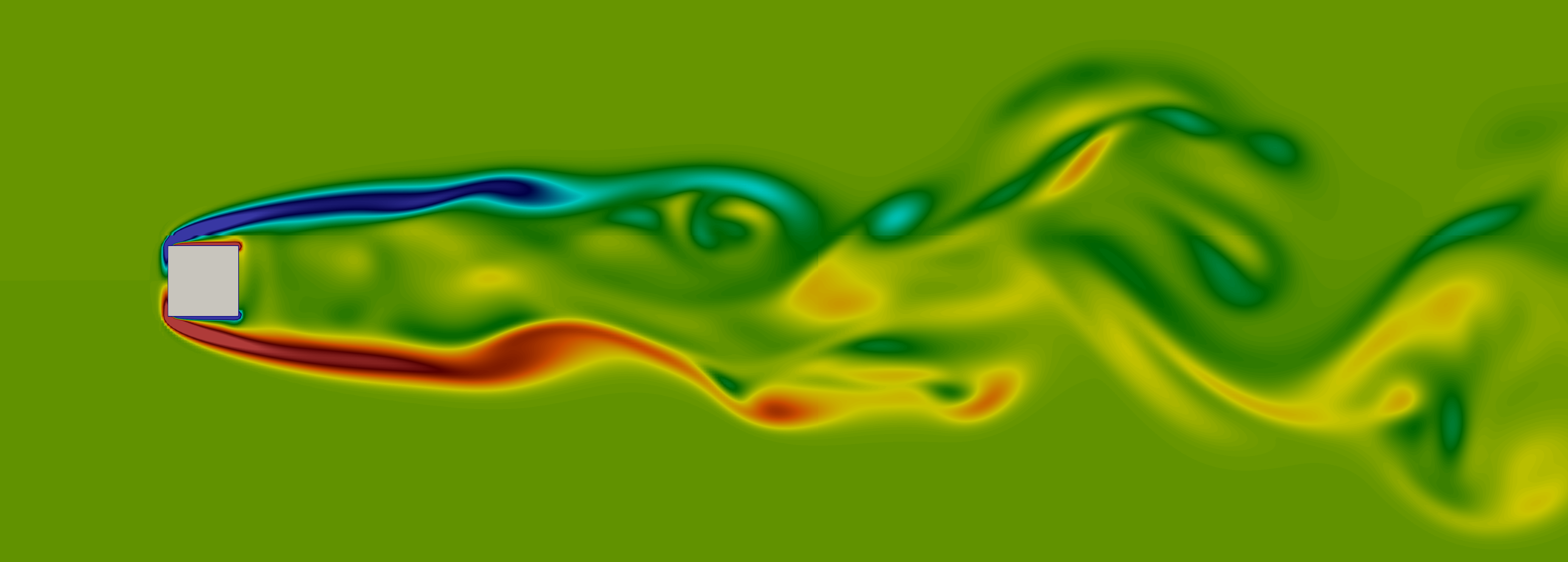}
        \caption{$z$-vorticity ($h/d=4$)}
    \end{subfigure}
    \begin{subfigure}[b]{0.32\textwidth}
        \includegraphics[width=\textwidth]{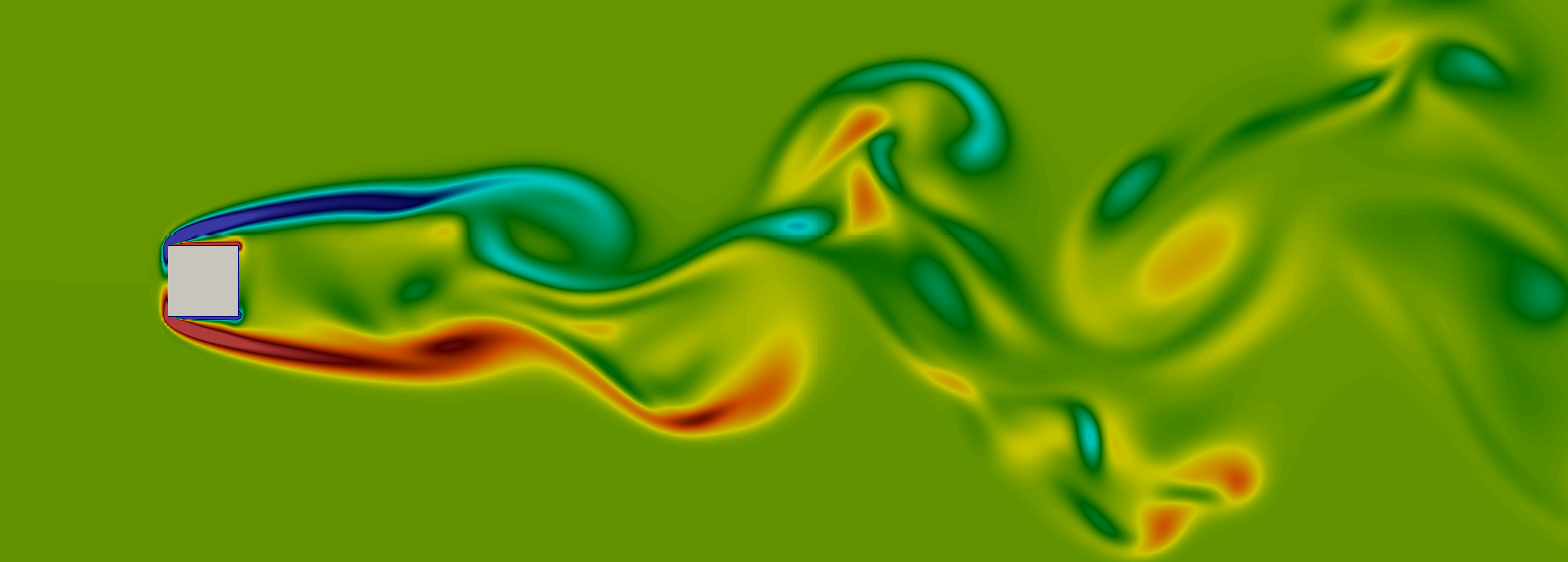}
        \caption{$z$-vorticity ($h/d=5$)}
    \end{subfigure}
    \\
    \begin{subfigure}[b]{0.32\textwidth}
        \includegraphics[width=\textwidth]{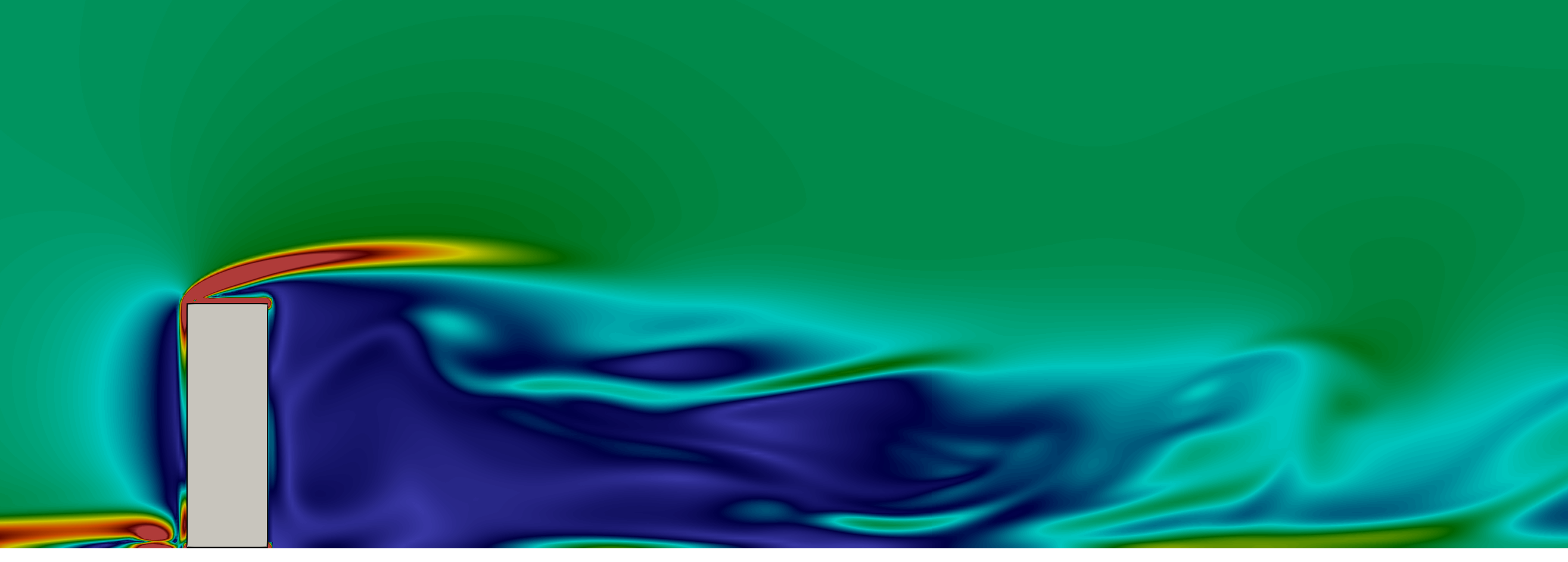}
        \caption{$y$-vorticity ($h/d=3$)}
    \end{subfigure}
    \begin{subfigure}[b]{0.32\textwidth}
        \includegraphics[width=\textwidth]{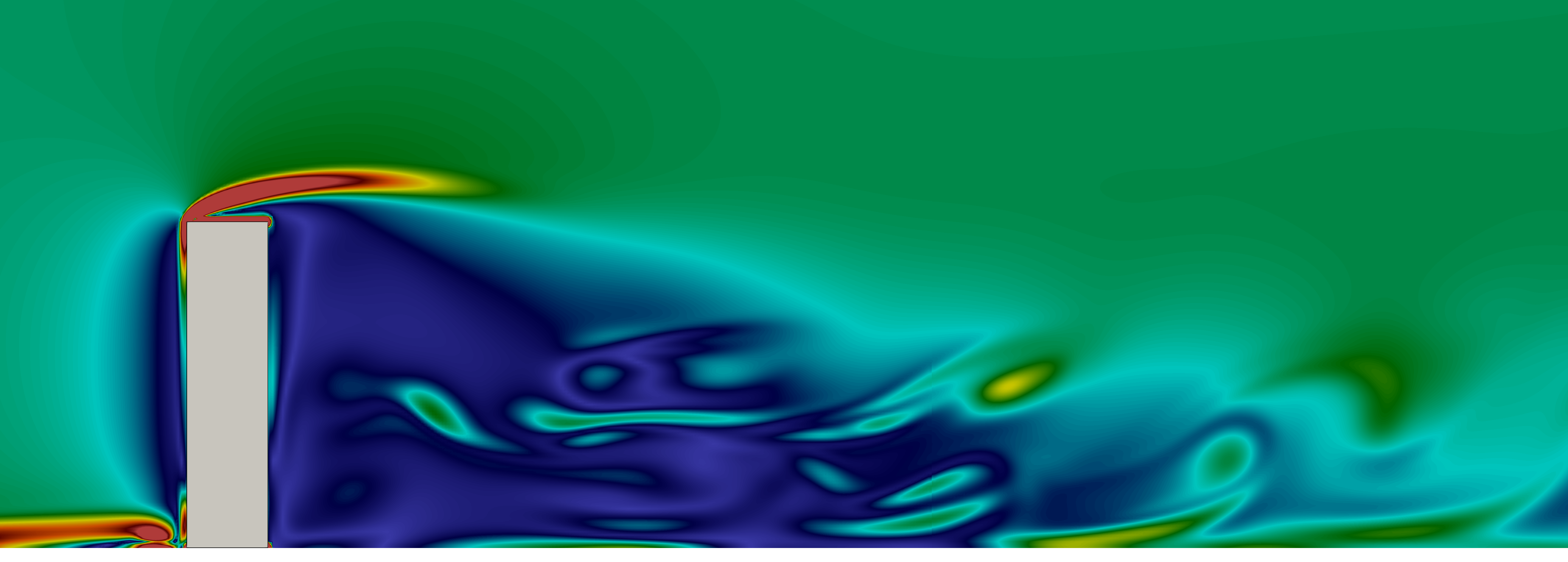}
        \caption{$y$-vorticity ($h/d=4$)}
    \end{subfigure}
    \begin{subfigure}[b]{0.32\textwidth}
        \includegraphics[width=\textwidth]{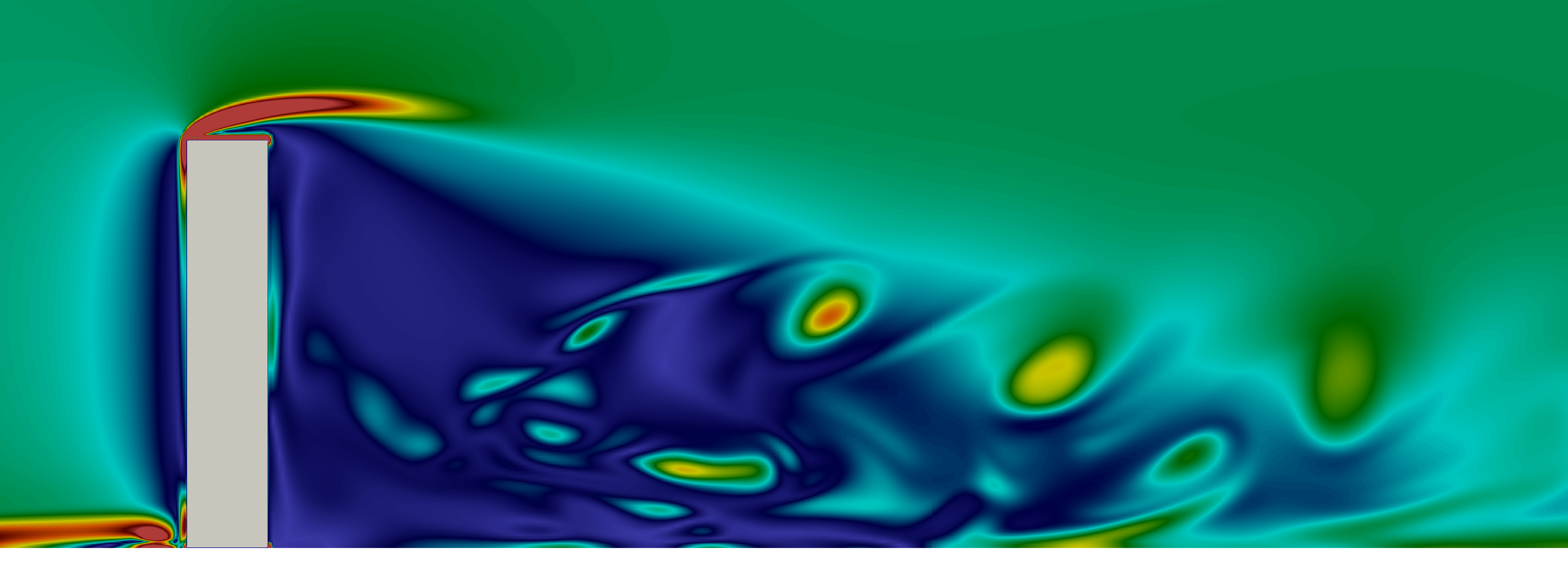}
        \caption{$y$-vorticity ($h/d=5$)}
    \end{subfigure}
    \\
    \begin{subfigure}[b]{0.24\textwidth}
        \includegraphics[width=\textwidth]{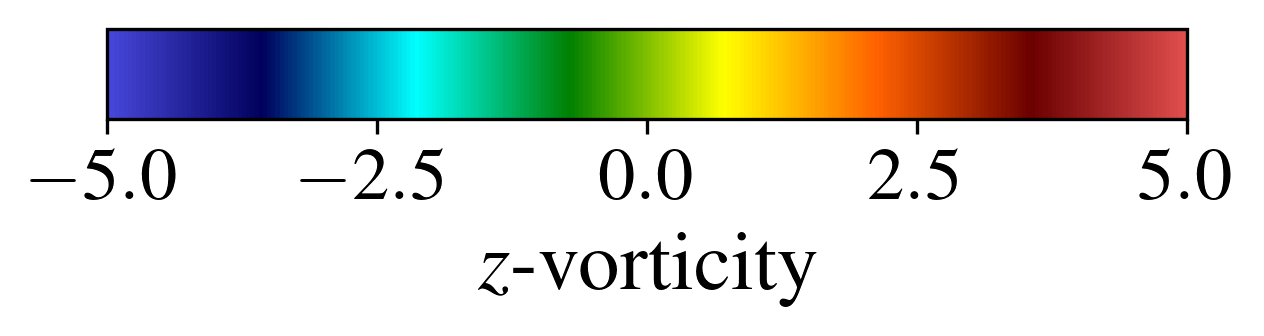}
    \end{subfigure}
    \hspace{0.5em}
    \begin{subfigure}[b]{0.24\textwidth}
        \includegraphics[width=\textwidth]{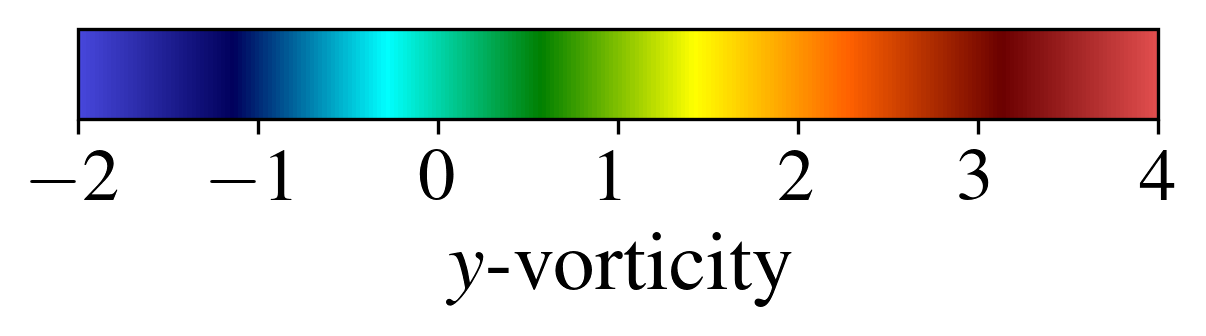}
    \end{subfigure}
    \caption{Instantaneous out-of-plane vorticity: $z$-vorticity at the mid-height ($z=0.5h$) and $y$-vorticity mid-plane ($y=0$) for flow past square cylinders of varying aspect ratios.}\label{fig:SqCyl-vorticity}
\end{figure}

Figure~\ref{fig:SqCyl-Q} shows instantaneous vortical structures in the wake of square cylinders with different aspect ratios. In all cases, a steady horseshoe vortex is observed at the upstream face near the bottom wall. This vortex forms due to flow separation at the upstream face and interaction with bottom shear layer, and it is expected to remain steady in this flow regime with low Reynolds number \cite{dousset2010formation}. Figure~\ref{fig:SqCyl-vorticity} presents instantaneous out-of-plane vorticity. The $z$-vorticity snapshots show that the flow is relatively straight and symmetric in the near wake immediately behind the cylinder and breaks into shedding vortices. Cylinders with lower aspect ratio produce a shorter near-wake region and a narrower vortex street.

The vortex shedding behavior is characterized by the Strouhal number. The Strouhal number, $St$, is defined as $f d/u_{\infty}$, where $f$ is the frequency of vortex shedding. The frequency is determined as the dominant peak in the power spectrum of the transverse velocity component measured at the point $(17,0,2)$. Table~\ref{table:SqCyl-St} summarizes the Strouhal number for different aspect ratios. The Strouhal number increases gradually with aspect ratio. The computed values of $St$ compare well with the values reported in the literature, which implies that the present method accurately captures the frequency of vortex shedding.

\begin{table}[h!]
\centering
\caption{Strouhal number, $St$, in flow over a square cylinder with varying aspect ratio for $Re=250$.}\label{table:SqCyl-St}
\begin{tabular}{ccc}
\hline
Aspect ratio $h/d$ & Present work & DNS \cite{saha2013unsteady} \\
\hline
3 & 0.119 & 0.114\\
4 & 0.122 & 0.124\\
5 & 0.125 & 0.130\\
\hline
\end{tabular}
\end{table}

\begin{figure}[h!]
    \captionsetup[subfigure]{justification=centering}
    \centering
    \begin{subfigure}[b]{0.32\textwidth}
        \includegraphics[width=\textwidth]{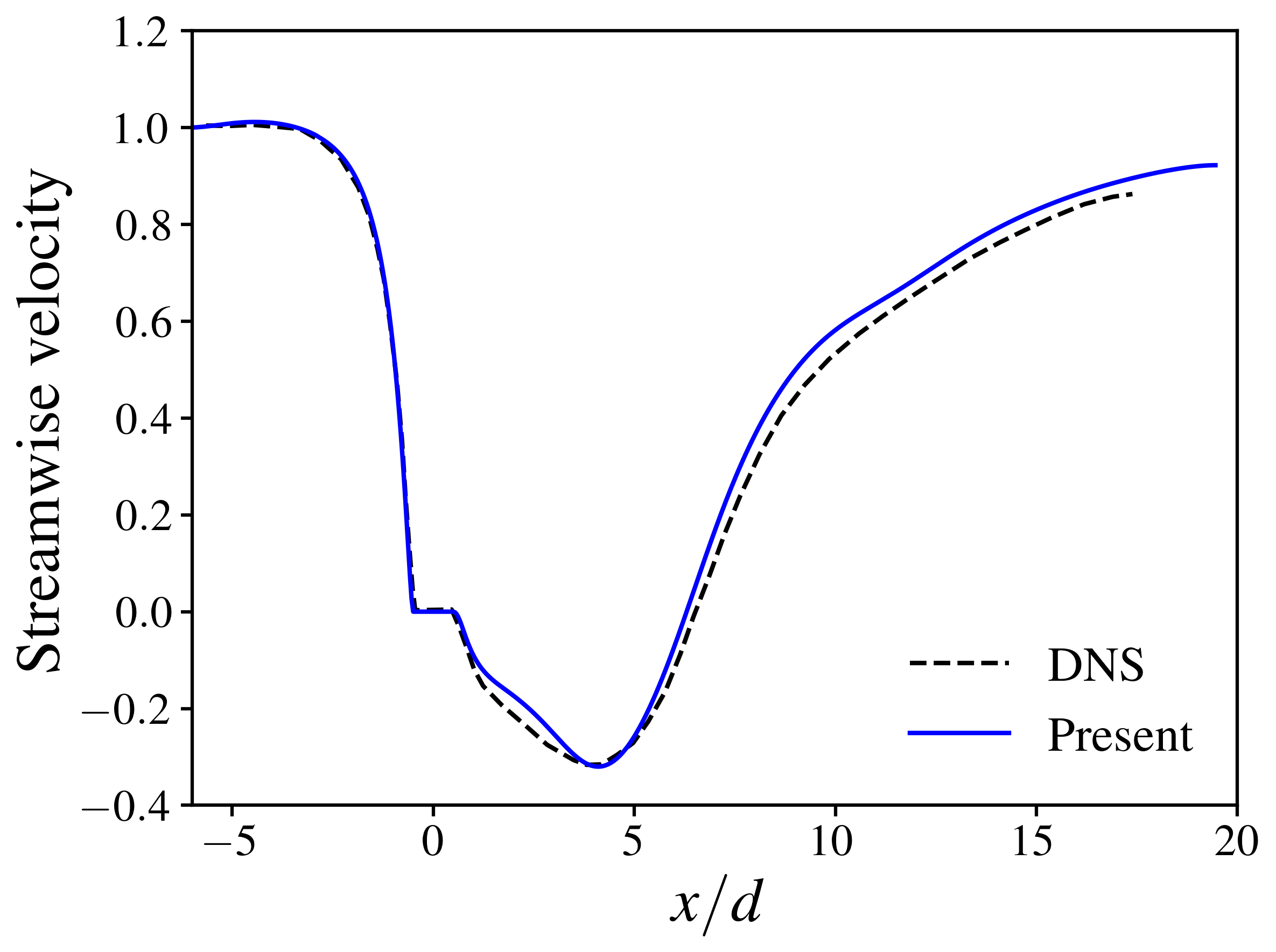}
        \caption{$h/d=3$}
    \end{subfigure}
    \begin{subfigure}[b]{0.32\textwidth}
        \includegraphics[width=\textwidth]{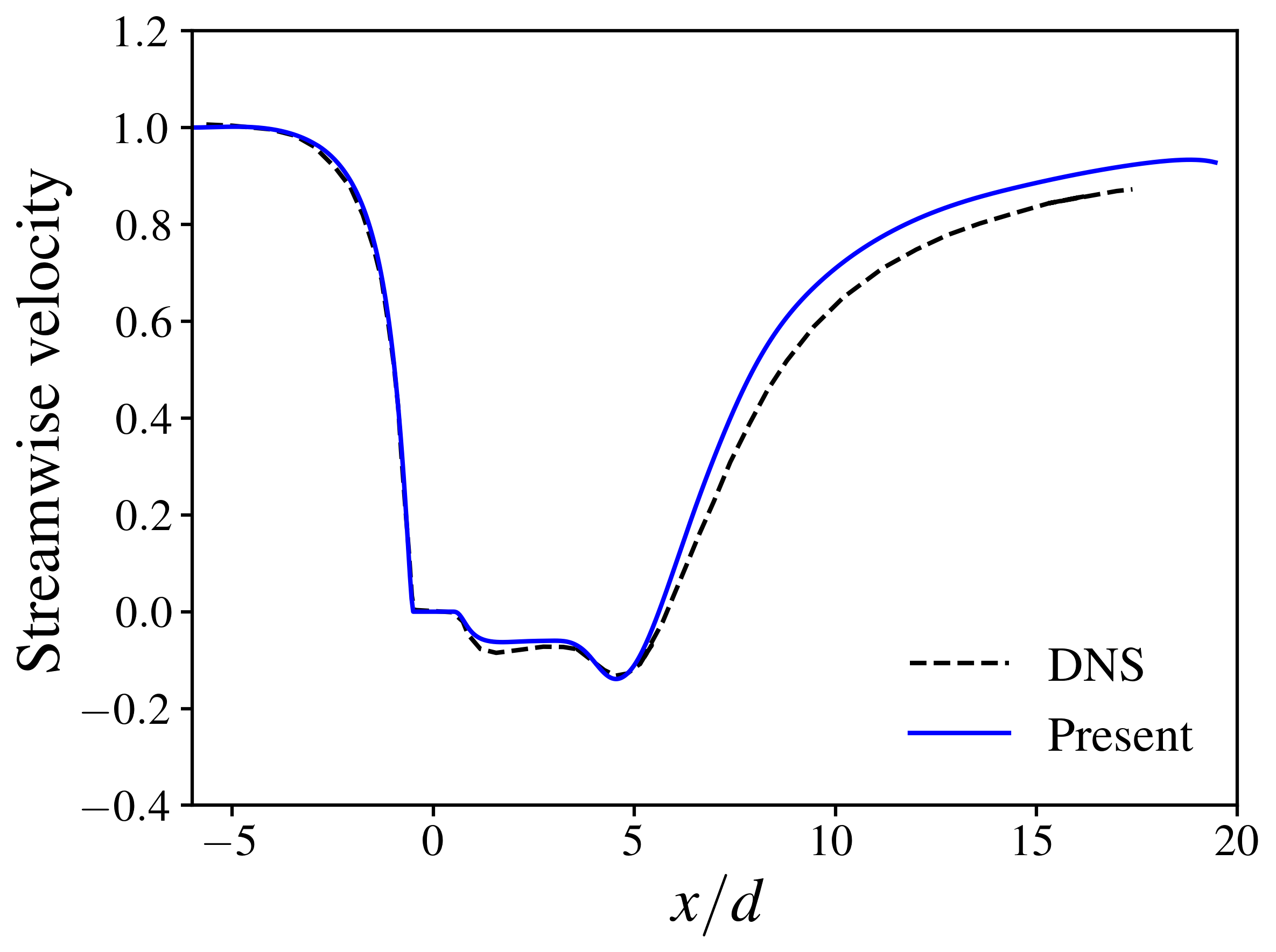}
        \caption{$h/d=4$}
    \end{subfigure}
    \begin{subfigure}[b]{0.32\textwidth}
        \includegraphics[width=\textwidth]{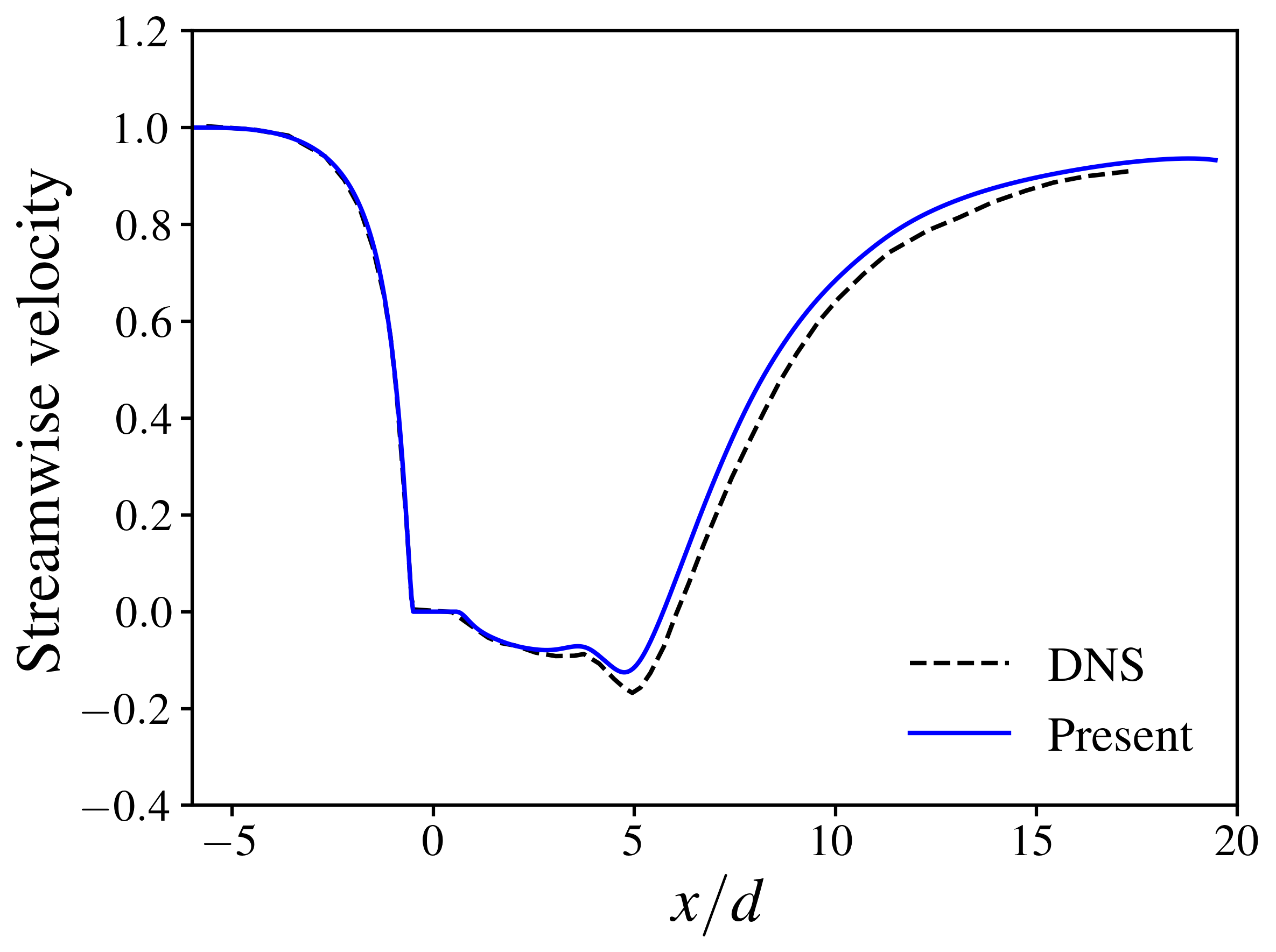}
        \caption{$h/d=5$}
    \end{subfigure}
    \caption{Time-averaged streamwise velocity along the centerline at the mid-height of the cylinder with different aspect ratios $h/d$. Reference DNS results are obtained from \cite{saha2013unsteady}.}\label{fig:SqCyl-centerline}
\end{figure}

Figure~\ref{fig:SqCyl-centerline} compares the time-averaged streamwise velocity along the centerline with the benchmark DNS results. The profiles of the present method show that the no-slip boundary condition is accurately imposed at the location of the cylinder surface, $x=-0.5$ and 0.5, and the gradients are also captured.

\begin{figure}[h!]
    \captionsetup[subfigure]{justification=centering}
    \centering
    \begin{subfigure}[b]{0.32\textwidth}
        \includegraphics[width=\textwidth]{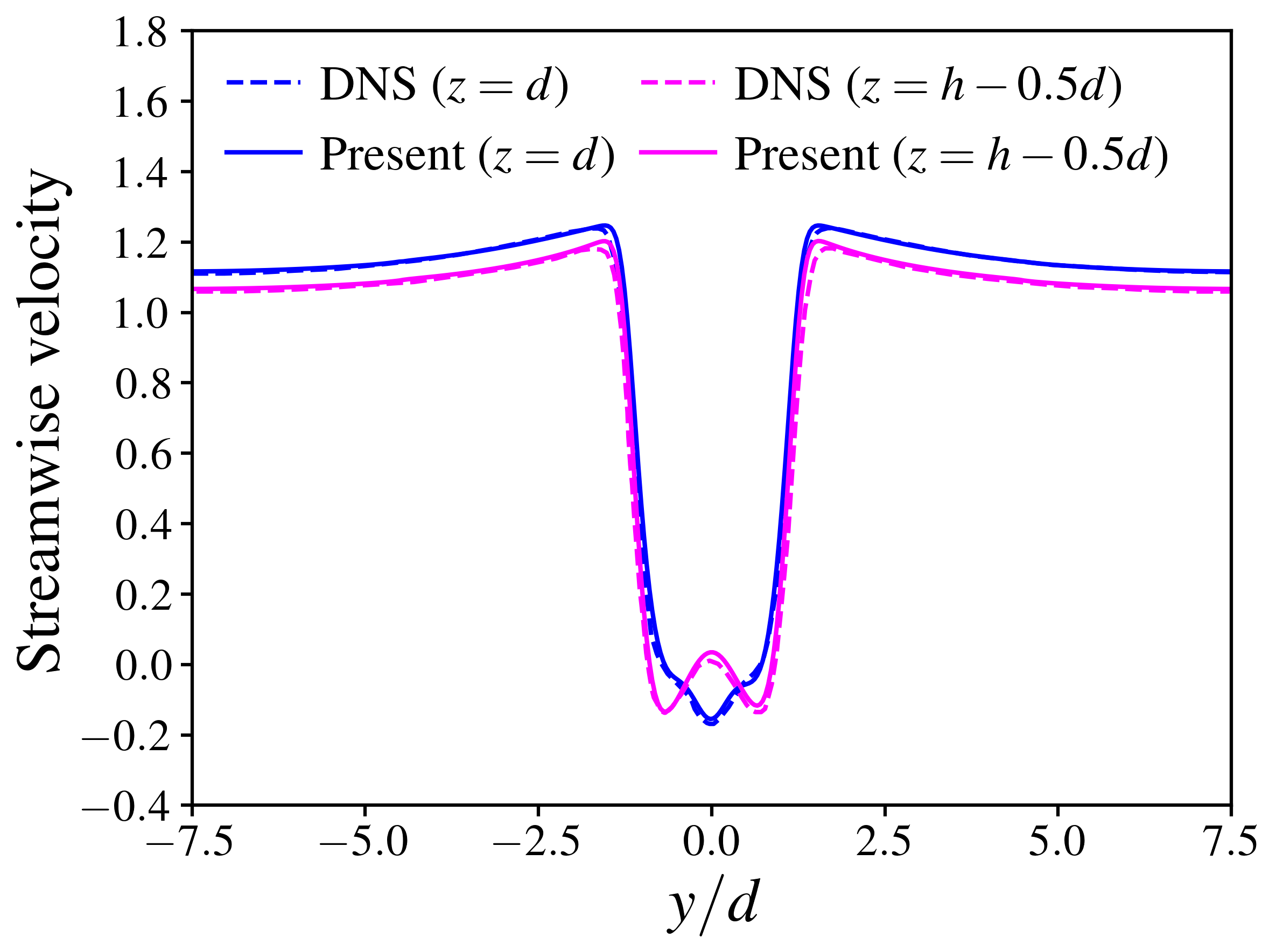}
        \caption{$h/d=3$}
    \end{subfigure}
    \begin{subfigure}[b]{0.32\textwidth}
        \includegraphics[width=\textwidth]{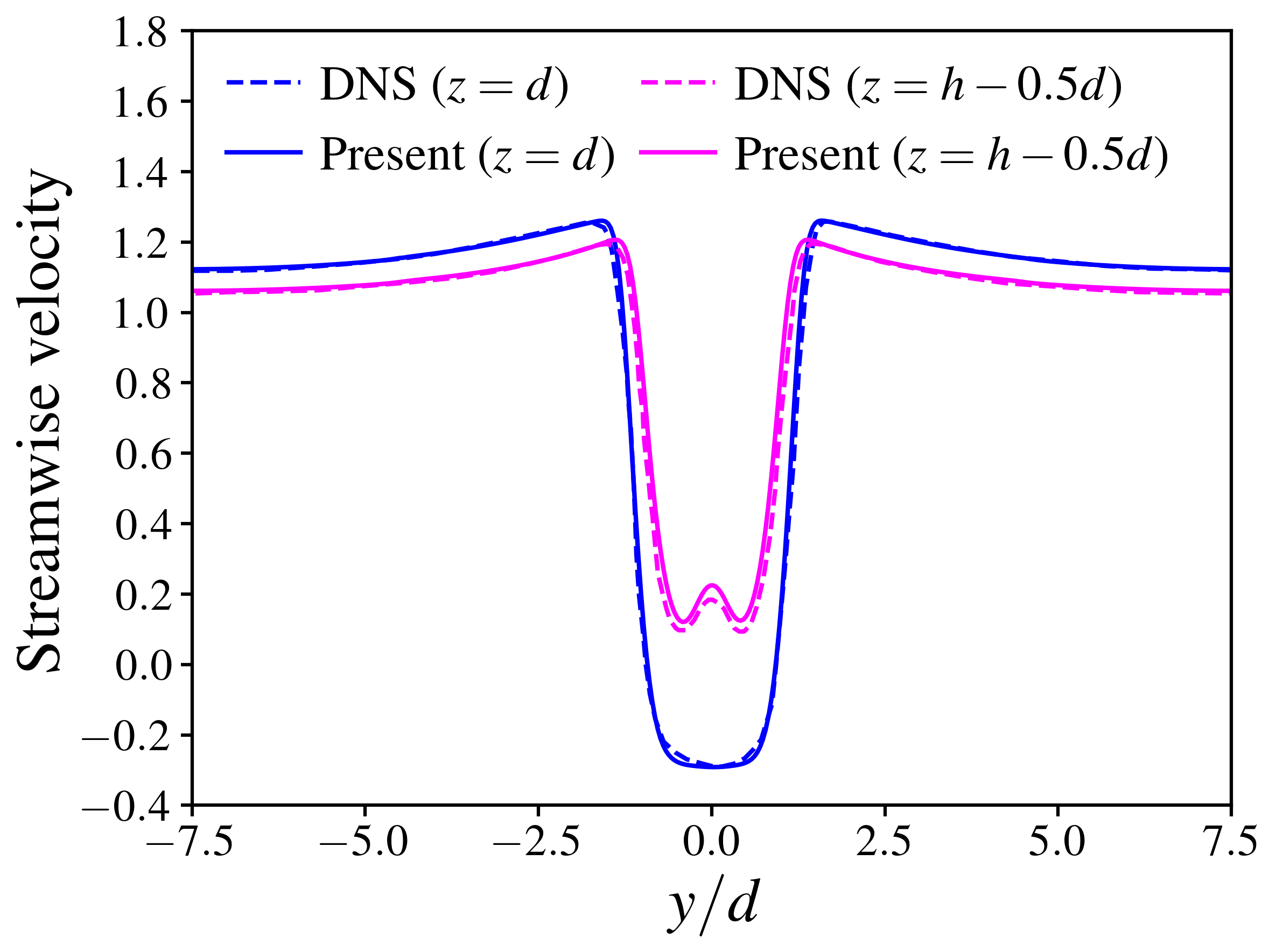}
        \caption{$h/d=4$}
    \end{subfigure}
    \begin{subfigure}[b]{0.32\textwidth}
        \includegraphics[width=\textwidth]{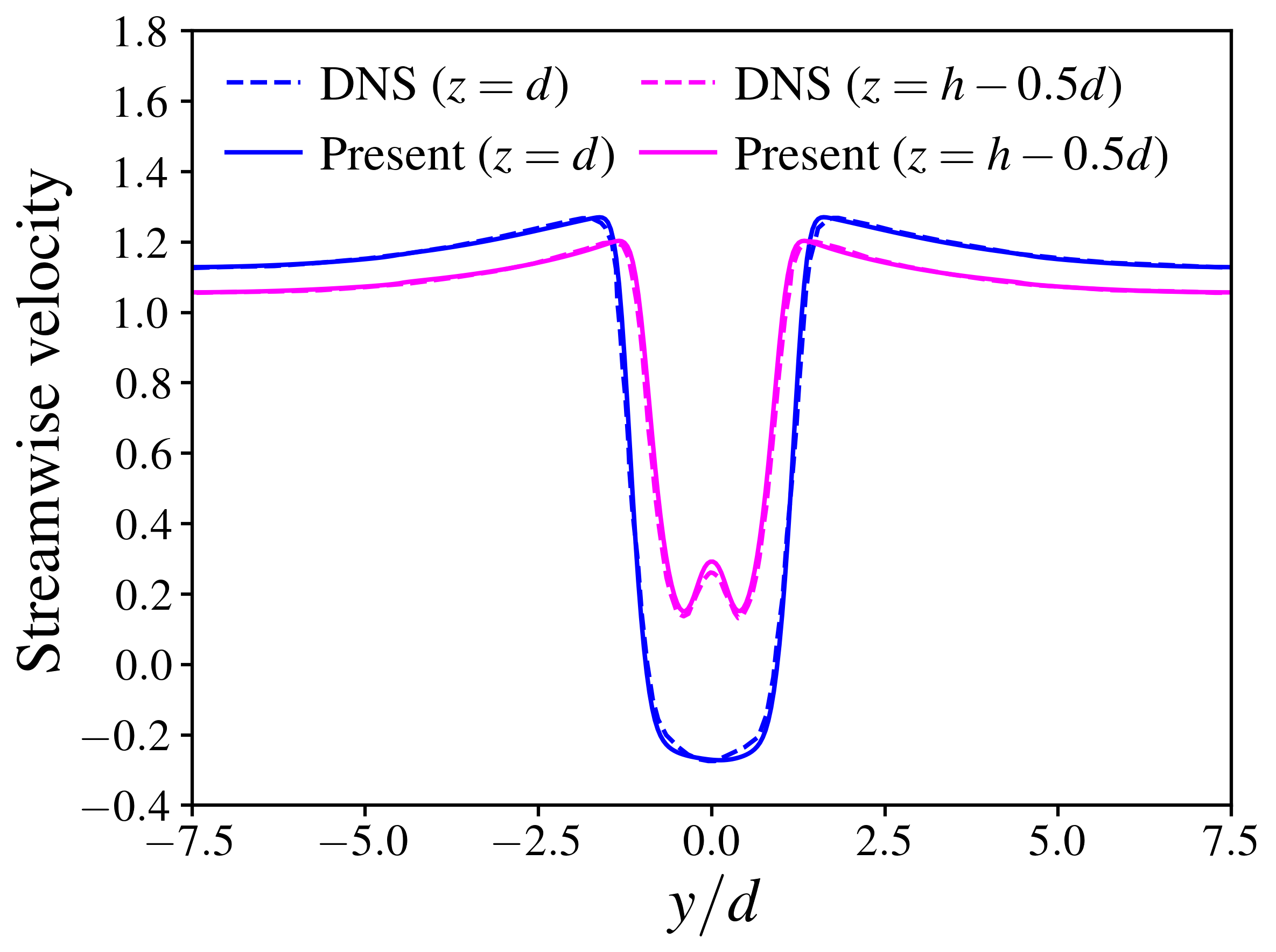}
        \caption{$h/d=5$}
    \end{subfigure}
    \\
    \begin{subfigure}[b]{0.32\textwidth}
        \includegraphics[width=\textwidth]{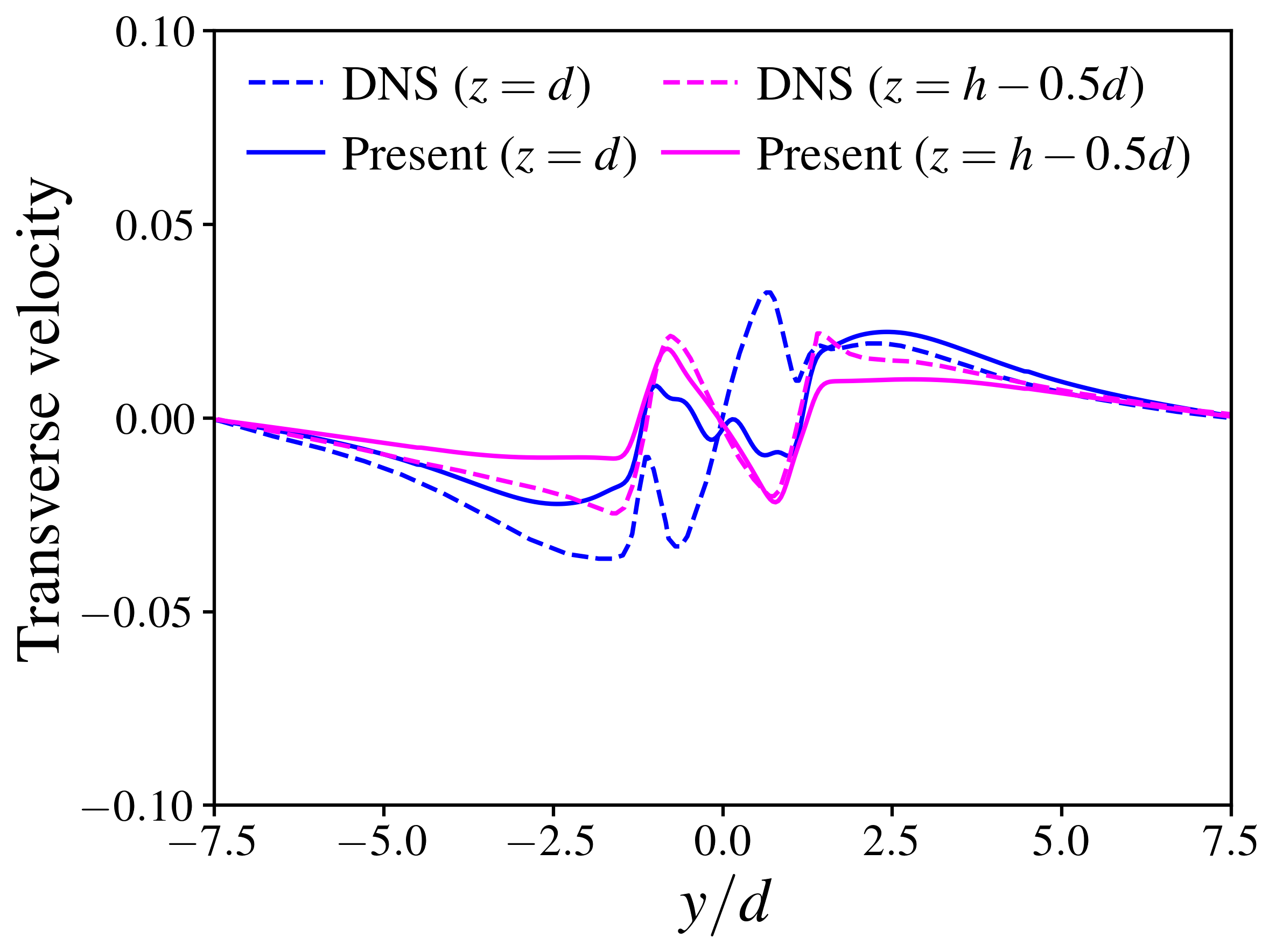}
        \caption{$h/d=3$}
    \end{subfigure}
    \begin{subfigure}[b]{0.32\textwidth}
        \includegraphics[width=\textwidth]{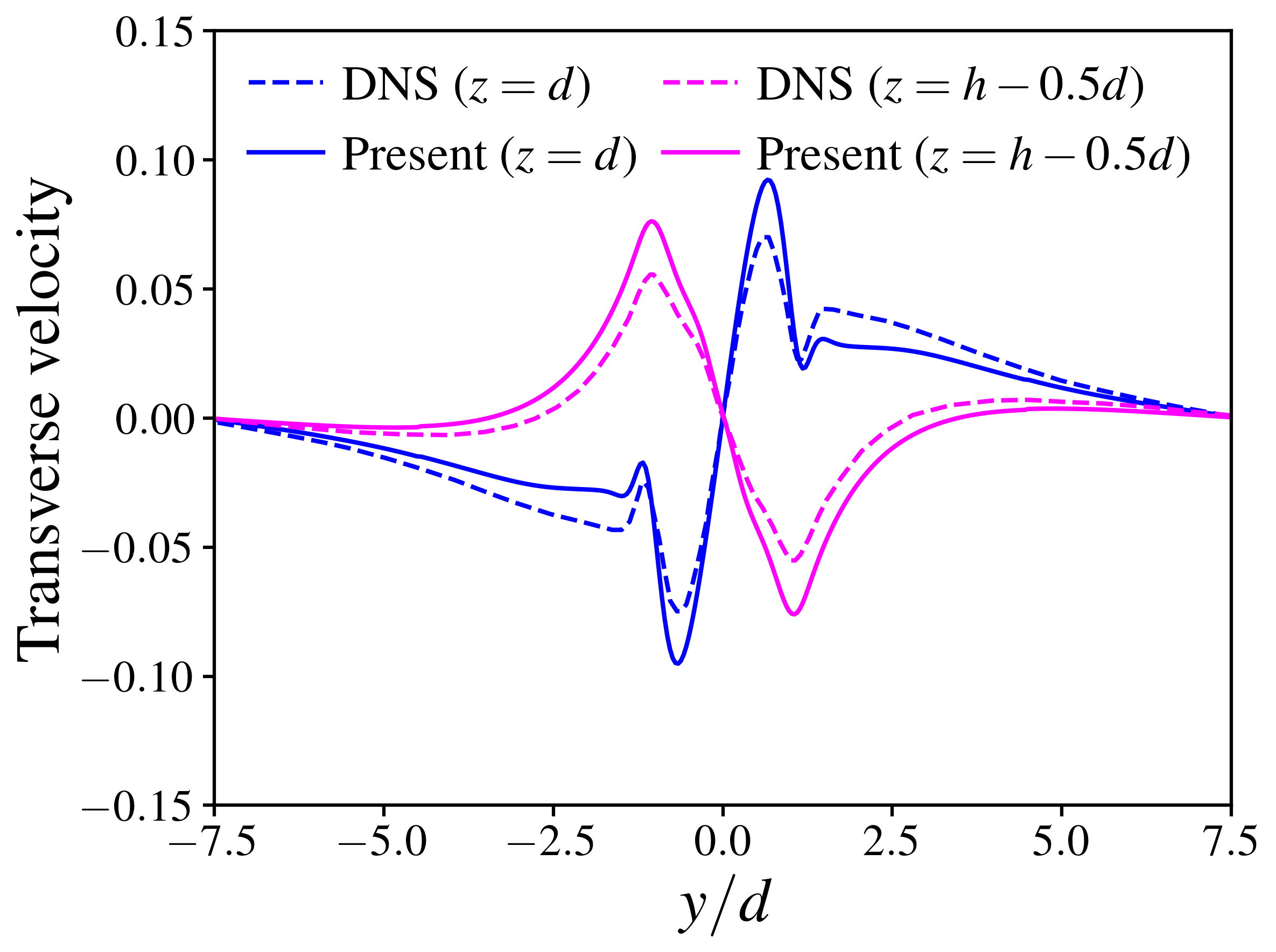}
        \caption{$h/d=4$}
    \end{subfigure}
    \begin{subfigure}[b]{0.32\textwidth}
        \includegraphics[width=\textwidth]{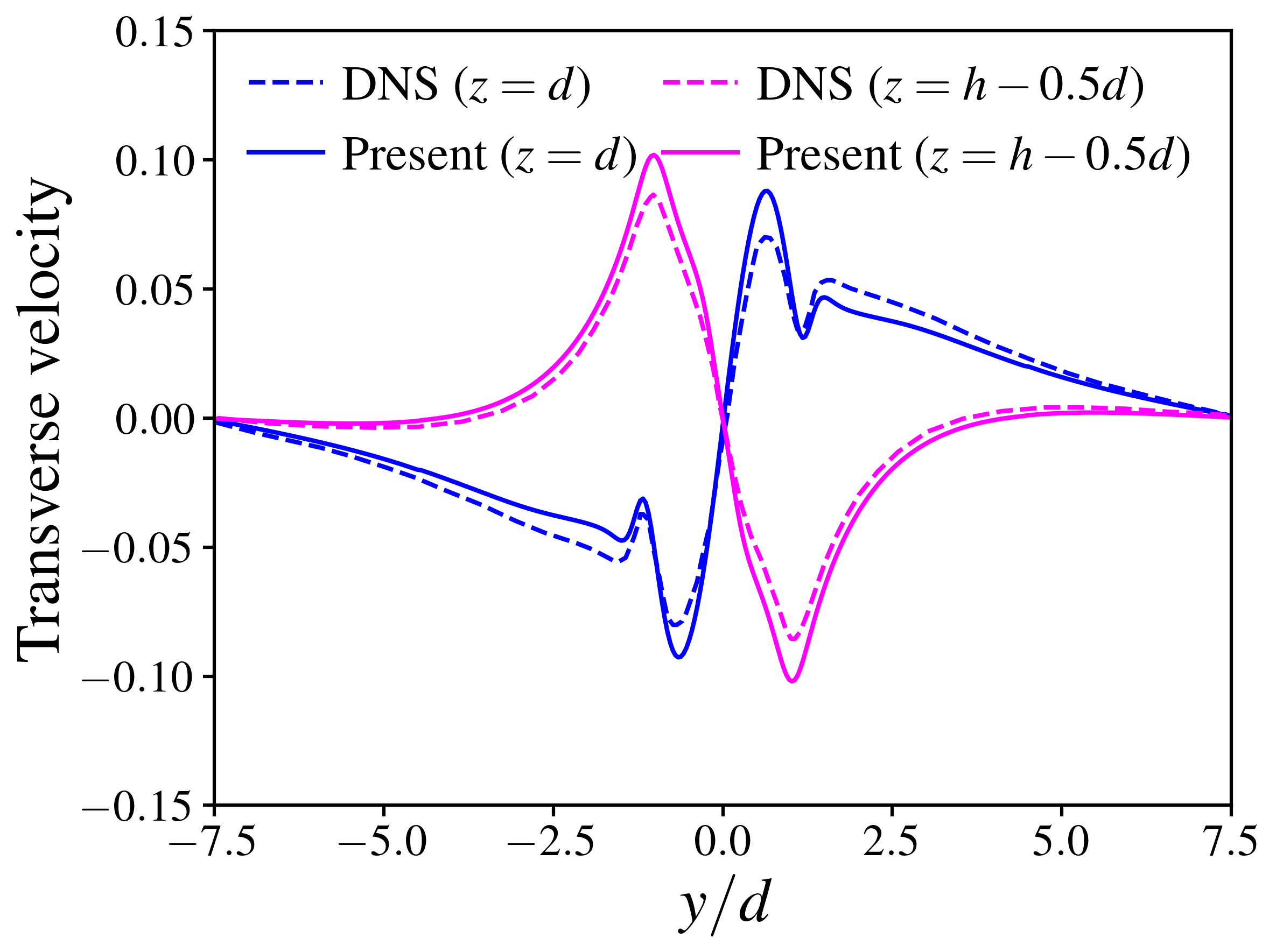}
        \caption{$h/d=5$}
    \end{subfigure}    
    \caption{Time-averaged streamwise and transverse velocity profiles in the transverse direction at $x=3d$, evaluated at $z=d$ and $z=h-0.5d$ for flow past a wall-mounted square cylinder with different aspect ratios $h/d$. Reference DNS results are obtained from \cite{saha2013unsteady}.}\label{fig:SqCyl-transverse}
\end{figure}

Figure~\ref{fig:SqCyl-transverse} compares the time-averaged streamwise and transverse velocity profiles along a transverse line aligned with the $y$-axis at two different heights: one near the bottom and the other near the cylinder top. Flow recirculation is identified by the presence of reverse flow, i.e., negative streamwise velocity. A single recirculation core is observed near the bottom surface ($z=d$), while two separate cores associated with vortex shedding appear near the free end of the cylinder ($z=h-0.5d$). Overall, both the streamwise and transverse velocity profiles show good agreement with the benchmark results. A slight discrepancy is observed in the transverse velocity profile for $h/d=3$ near the bottom, however, the magnitude of the velocity component is very small.

\begin{figure}[h!]
    \captionsetup[subfigure]{justification=centering}
    \centering
    \begin{subfigure}[b]{0.32\textwidth}
        \includegraphics[width=\textwidth]{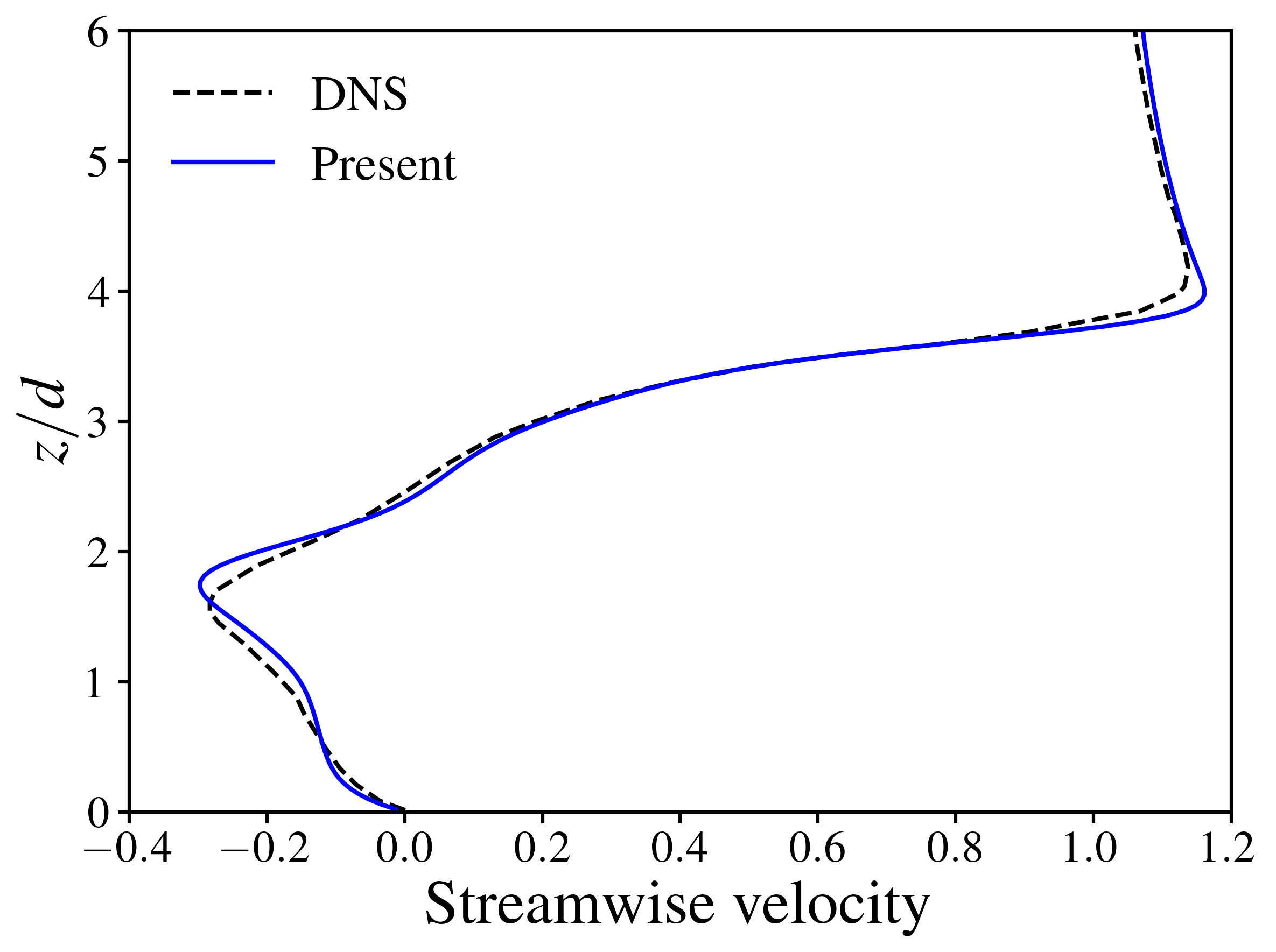}
        \caption{$h/d=3$}
    \end{subfigure}
    \begin{subfigure}[b]{0.32\textwidth}
        \includegraphics[width=\textwidth]{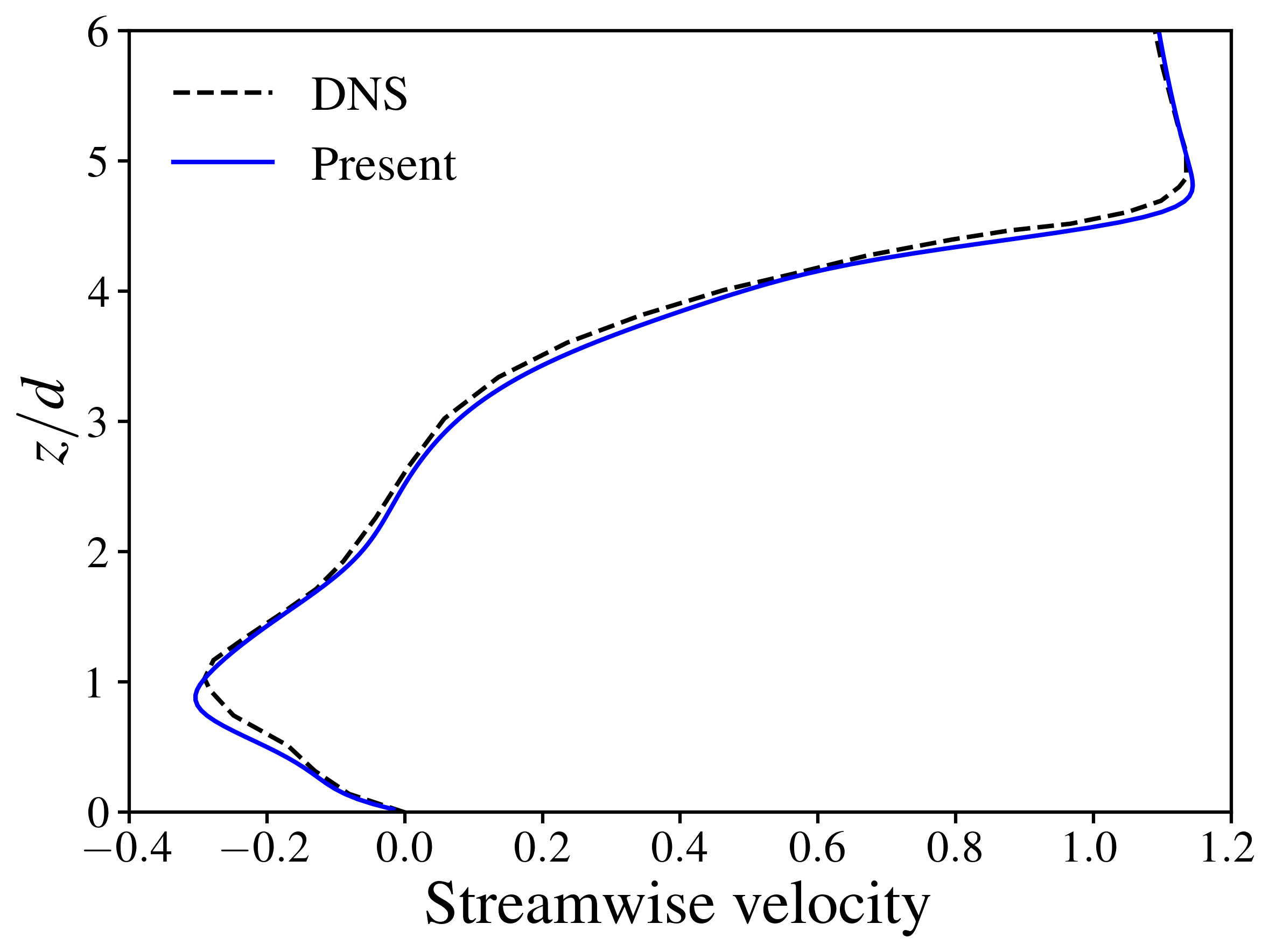}
        \caption{$h/d=4$}
    \end{subfigure}
    \begin{subfigure}[b]{0.32\textwidth}
        \includegraphics[width=\textwidth]{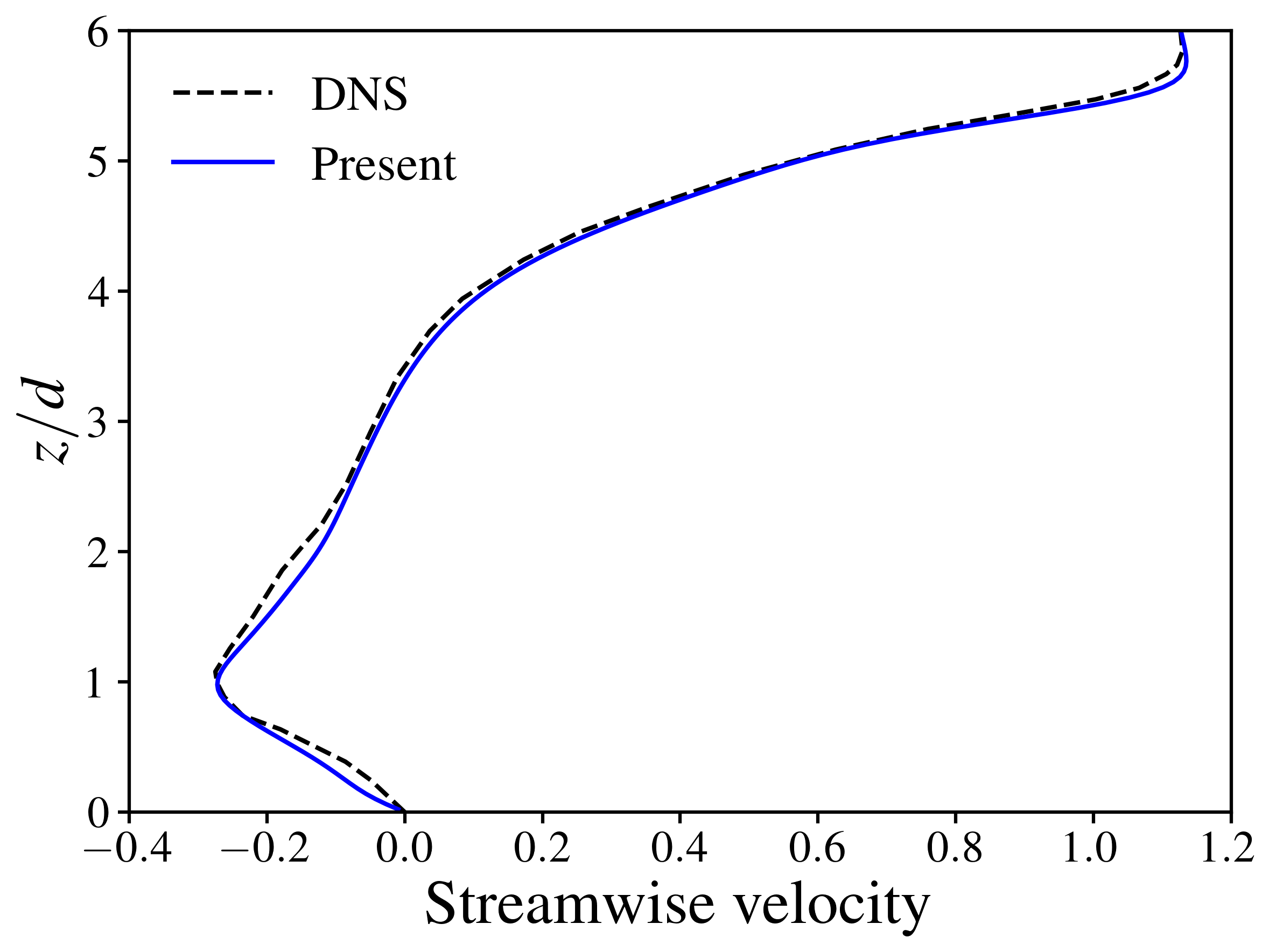}
        \caption{$h/d=5$}
    \end{subfigure}
    \\
    \begin{subfigure}[b]{0.32\textwidth}
        \includegraphics[width=\textwidth]{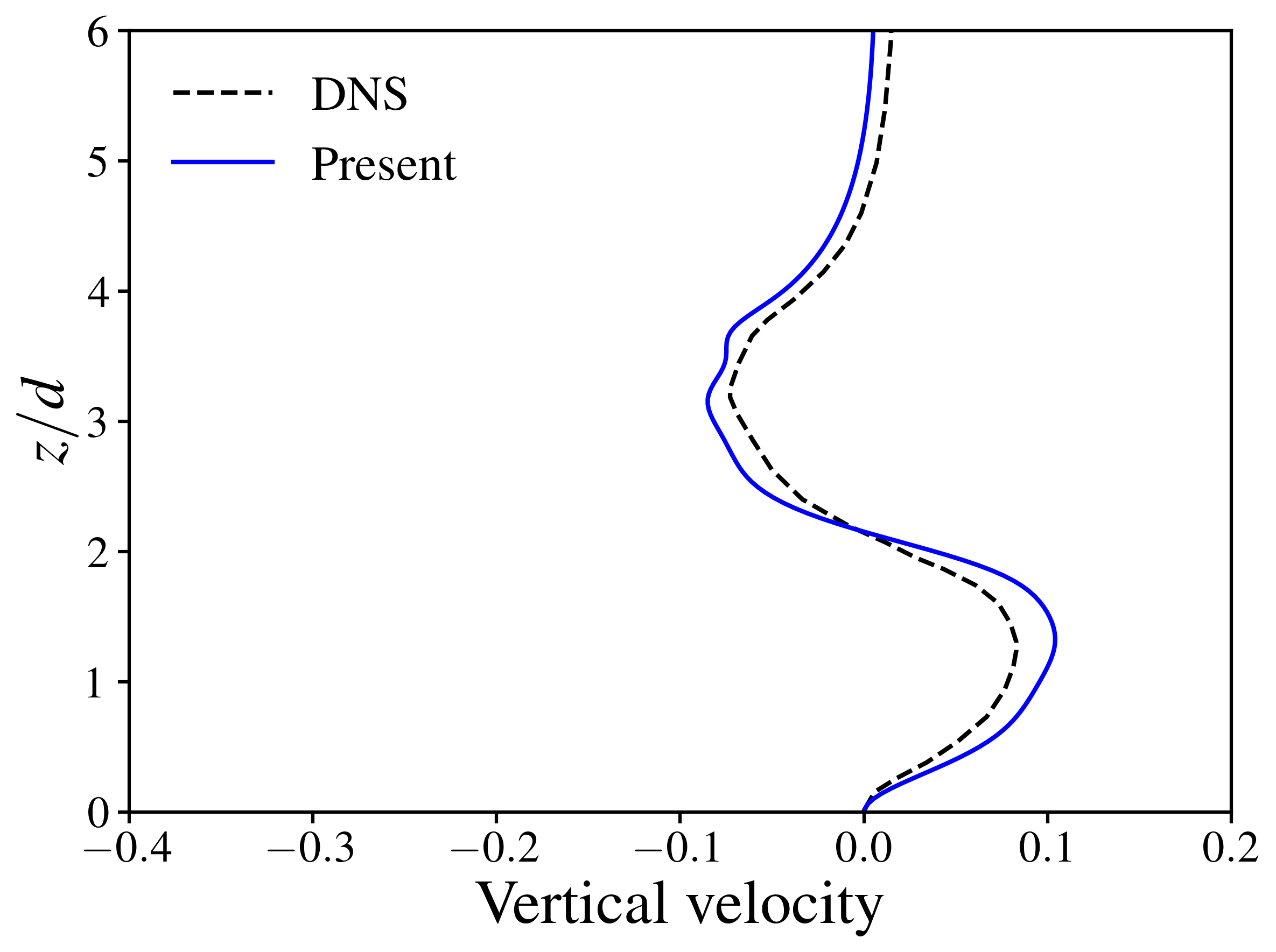}
        \caption{$h/d=3$}
    \end{subfigure}
    \begin{subfigure}[b]{0.32\textwidth}
        \includegraphics[width=\textwidth]{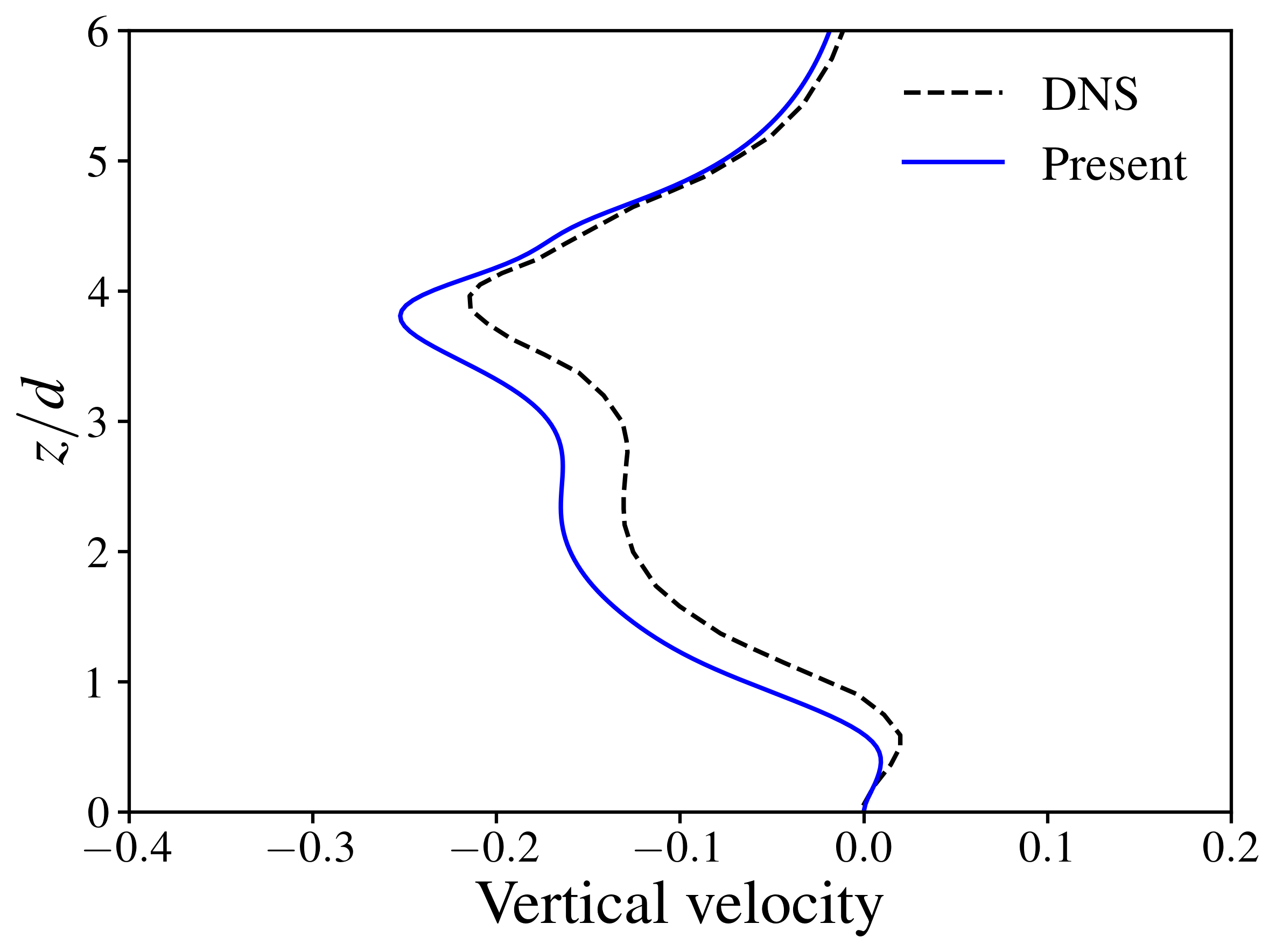}
        \caption{$h/d=4$}
    \end{subfigure}
    \begin{subfigure}[b]{0.32\textwidth}
        \includegraphics[width=\textwidth]{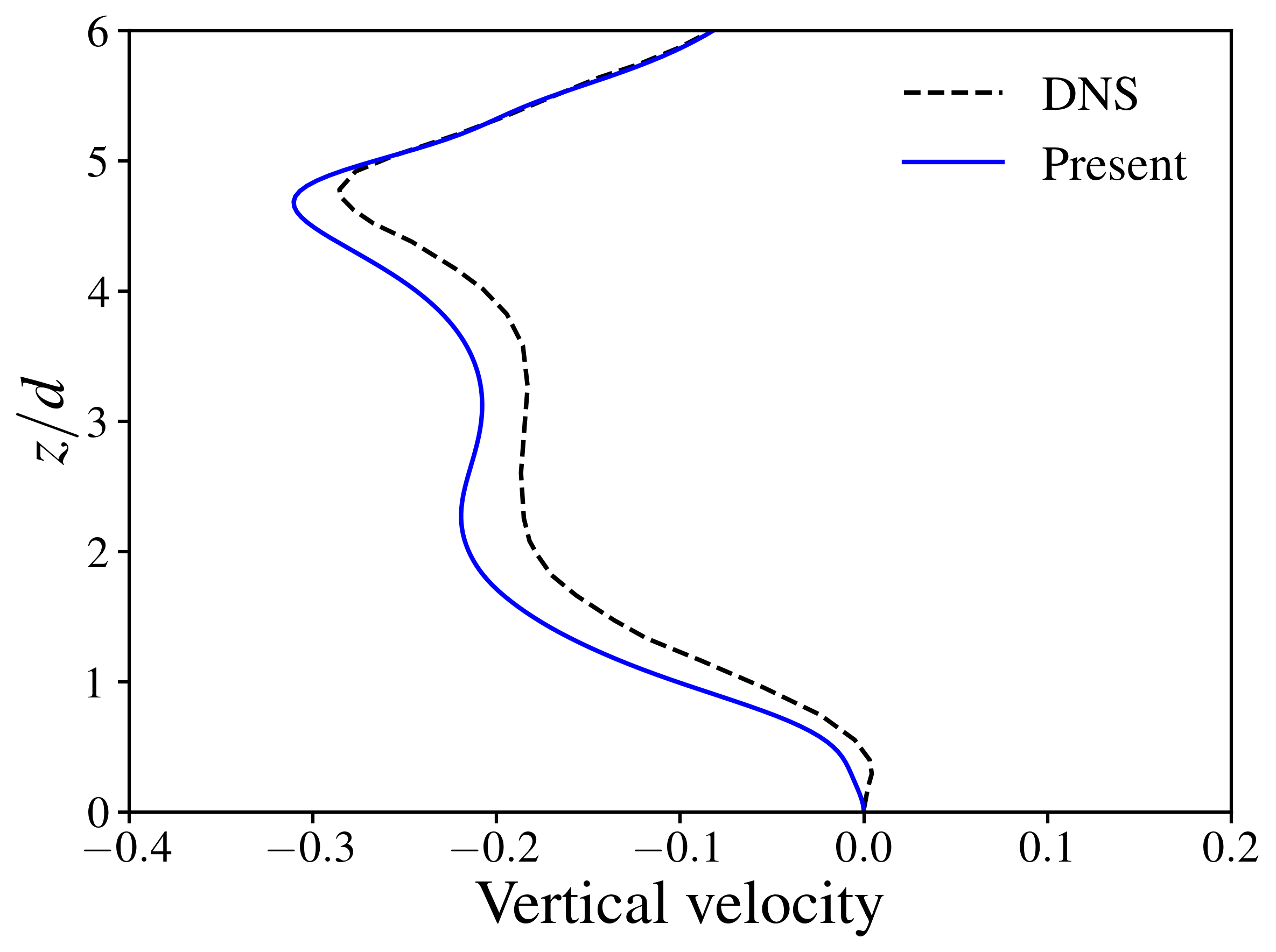}
        \caption{$h/d=5$}
    \end{subfigure}    
    \caption{Time-averaged streamwise and vertical velocity profiles in the vertical direction at $x=3d$ for flow past a wall-mounted square cylinder with different aspect ratios $h/d$. Reference DNS results are obtained from \cite{saha2013unsteady}.}\label{fig:SqCyl-vertical}
\end{figure}

Figure~\ref{fig:SqCyl-vertical} compares the vertical variation of the time-averaged streamwise and vertical velocities. The vertical velocity distributions match well with the benchmark results. Although slight discrepancies are observed in the vertical velocity profiles, the overall patterns and peak locations are accurately captured.

\section{Conclusions}\label{sec:conclusion}

In this paper, we developed an EB method to model atmospheric flows over terrain and buildings using either the fully compressible or anelastic governing equations. We have described the implementation of the EB method on a staggered grid which is widely used in atmospheric modeling. The computational framework adopts block-structured grids with adaptive refinement. The weighted state redistribution scheme is employed to address the small-cell problem near the embedded boundary and enhance numerical stability.

The EB method was validated through comparisons with simulations using terrain-following coordinates or analytical solutions. The set of test cases includes two- and three-dimensional test cases for both compressible and anelastic models. The comparisons demonstrate that the method accurately enforces no-slip and free-slip boundary conditions at embedded terrain surfaces and, consequently, accurately reproduces the flow fields. With this EB capability, the ERF model can accurately and efficiently simulate flows over complex terrain and buildings.

Potential future work includes extending the EB approach to incorporate additional physical parameterizations, including wall boundary layers and subgrid-scale turbulence and microphysics.

\section{Acknowledgements}
This work was performed under the auspices of the U.S.~Department of Energy (DOE) by Lawrence Livermore National Laboratory under Contract DE-AC52-07NA27344 and Lawrence Berkeley National Laboratory under Contract DE-AC02-05CH11231. Kang, Mirocha, and Lundquist were supported by the LLNL LDRD Program under Project No.~24-SI-001. Almgren and Zhang were supported by the U.S. DOE, Office of Science, Office of Advanced Scientific Computing Research, Scientific Discovery through Advanced Computing (SciDAC) Program through the FASTMath Institute under contract No.~DE-AC02-05CH11231. Lattanzi and Natarajan were supported by the U.S. DOE Wind Energy Technologies Office.

\appendix

\newpage

\end{document}